\documentclass[aps,prx,10pt,twocolumn,superscriptaddress,showpacs]{revtex4-1}

\usepackage{amsmath,amssymb,graphics,epsfig,epstopdf,color,verbatim,tabularx}
\usepackage{graphicx}
\usepackage{color}

\DeclareGraphicsExtensions{.pdf,.png,.jpg}

\renewcommand{\Bar}[1]{\mkern 1.5mu\overline{\mkern-1.5mu#1
\mkern-1.5mu}\mkern 1.5mu}

\begin{document}

\title{Dynamical properties of a driven dissipative dimerized $S = 1/2$ chain}

\author{M. Yarmohammadi}
\affiliation{Lehrstuhl f\"ur Theoretische Physik I, Technische Universit\"at 
Dortmund, Otto-Hahn-Strasse 4, 44221 Dortmund, Germany}

\author{C. Meyer}
\affiliation{Institut f\"ur Theoretische Physik, Georg-August-Universit\"at 
G\"ottingen, Friedrich-Hund-Platz 1, 37077 G\"ottingen, Germany}

\author{B. Fauseweh}
\affiliation{Theoretical Division, Los Alamos National Laboratory, Los Alamos, 
NM 87545, USA}

\author{B. Normand}
\affiliation{Paul Scherrer Institute, CH-5232 Villigen PSI, Switzerland}
\affiliation{Lehrstuhl f\"ur Theoretische Physik I, Technische Universit\"at 
Dortmund, Otto-Hahn-Strasse 4, 44221 Dortmund, Germany}
\affiliation{Institute of Physics, Ecole Polytechnique F\'{e}d\'{e}rale 
de Lausanne (EPFL), CH-1015 Lausanne, Switzerland}

\author{G. S. Uhrig}
\affiliation{Lehrstuhl f\"ur Theoretische Physik I, Technische Universit\"at 
Dortmund, Otto-Hahn-Strasse 4, 44221 Dortmund, Germany}

\begin{abstract}
We consider the dynamical properties of a gapped quantum spin system coupled 
to the electric field of a laser, which drives the resonant excitation of 
specific phonon modes that modulate the magnetic interactions. We deduce the 
quantum master equations governing the time-evolution of both the lattice and 
spin sectors, by developing a Lindblad formalism with bath operators providing 
an explicit description of their respective phonon-mediated damping terms. 
We investigate the non-equilibrium steady states (NESS) of the spin system 
established by a continuous driving, delineating parameter regimes in driving 
frequency, damping, and spin-phonon coupling for the establishment of 
physically meaningful NESS and their related non-trivial properties. Focusing 
on the regime of generic weak spin-phonon coupling, we characterize the NESS 
by their frequency and wave-vector content, explore their transient and 
relaxation behavior, and discuss the energy flow, the system temperature, 
and the critical role of the type of bath adopted. Our study lays a foundation 
for the quantitative modelling of experiments currently being designed to 
control coherent many-body spin states in quantum magnetic materials.  
\end{abstract}


\maketitle

\section{Introduction}
\label{sintro}

Both the advent of powerful new laser sources and the increasing demand 
for next-generation magnetic devices, required to power the information 
revolution, are focusing intensive research efforts on time-dependent 
phenomena in condensed matter. On the laser side, x-ray free-electron 
laser sources in the US and Europe now allow the ``ultrafast'' probing of 
materials on the femtosecond timescales of their fundamental electronic and 
magnetic processes. On the device side, the immediate target is designer 
materials for antiferromagnetic (AF) spintronics \cite{Baltz18,Chumak15}, 
to enable the writing, storage, and reading of large-scale classical 
magnetic information with factor-1000 improvements over the current 
levels of speed and power consumption. Already on the horizon, however, 
is the development of magnetic materials as a route to encoding and 
manipulating quantum information, and indeed quantum information processing 
in systems with strong interaction energies would ensure very high-frequency 
operation at the lowest possible dissipation. 

The concept of laser driving generalizes the pump-probe paradigm from simple 
pulse-delay schemes to the imprinting of arbitrary dynamics (within the limits 
of field control). The laser excitation of quantum systems has generated 
theoretical proposals for uniquely out-of-equilibrium states of matter, 
including non-equilibrium steady states (NESS) \cite{rjqq,ru}, non-equilibrium 
topological states \cite{Rudner13}, and many-body localization (MBL) \cite{rdp,
rnh}. To date these ideas have been tested largely on systems of ultracold 
atoms \cite{Chong18,Labouvie16,Schreiber15,rec}, where the laser controls the 
``optical lattice'' on which the atoms reside \cite{rbdz}. The undeniable 
beauty of both the physical concepts and the technological achievements aside, 
these systems are neither very large nor very readily miniaturized.

Laser facilities operating on the energy and ultrafast time scales of 
condensed-matter systems have been deployed recently to observe a wide 
array of novel phenomena in graphene \cite{rg}, superconductors \cite{rcr},  
charge-density-wave materials \cite{Buzzi18,Zong18}, and correlated insulators 
near their metallic transition \cite{Iwai03,Zong19}. Beyond inducing, enhancing,
or destroying a symmetry-broken state, a key focus of these experiments has 
been the high-frequency Floquet regime, where steady laser driving can induce 
new topological states \cite{rkobfd,rlrg}, the ``time crystal'' \cite{rrbs}, 
or more generally allow the ``Floquet engineering'' of the electronic bands 
\cite{Iadecola13,Sentef15}. 

While any material can be laser-driven, the key question is whether this 
driving creates a coherent quantum state \cite{ru}. Some of these new 
phenomena, notably photo-enhanced superconductivity \cite{rcr}, occur 
because the laser drives particular phonon excitations of the lattice 
hosting the electrons. Because strong laser driving can lead to very high 
populations of any targeted mode, exploiting the anharmonic part of the 
lattice restoring force leads to the concept of nonlinear phononics 
\cite{Foerst11,rscg,vonHoegen18}. However, the phonon ensemble determines 
the temperature, and hence heating of the system is a fundamental issue in 
determining whether any of these novel laser-driven phenomena, and 
particularly their quantum nature, can survive beyond the initial ultrafast 
laser pulses.

Among the extensive body of theoretical studies of non-equilibrium quantum 
systems are analyses of short-time transient behavior caused by quenches 
\cite{rlk,rhu,rpfokmm}, including those due to laser pulses \cite{Schwarz20,
rfz}, and of long-time thermalization behavior \cite{rr,rhu}. Ideas 
from (near-)integrable systems include MBL, which is known at least in one 
dimension (1D) \cite{rnh}, and pre-thermalization \cite{rbbw}, while numerous 
studies have explored the Floquet regime \cite{rok}. Of the many numerical 
approaches to quenched or driven models, one of the most successful is 
non-equilibrium dynamical mean-field theory (DMFT) \cite{rakl,Aoki14}, which 
has been applied to many problems in cold atoms \cite{rqh} and condensed 
matter \cite{rew,rmw,rhmew}. However, these studies are largely restricted 
to fermionic systems and focus mostly on leading qualitative effects due to 
intrinsic system dynamics, rather than on the dynamics in the presence of 
dissipation. 

By contrast, a realistic NESS requires a path for outflow of the injected 
energy \cite{ru}. The most general approach to describe a dissipative (open) 
quantum system is the Lindblad formalism \cite{rl}, in which damping is 
provided by bath operators whose Hamiltonian dynamics are not required to 
formulate the equations of motion governing the time-evolution of physical 
observables in the Heisenberg representation \cite{rbp,rw}. Recent studies 
of driven condensed-matter systems have included dissipative effects by using
a phenomenological Gilbert damping \cite{Sato19}, a phenomenological phonon 
damping \cite{Babadi17}, or numerical methods where a thermal bath of phonons 
\cite{rllr,rlar,Peronaci20}, a temperature-independent fermionic bath 
\cite{rmw,Walldorf19}, or both \cite{Murakami17}, form(s) part of the 
system on which calculations are performed. While these studies therefore 
consider NESS implicitly or explicitly, in fact none correspond to the 
problem of an open, driven quantum system subject to Lindblad dissipation 
processes, whose quantitative treatment is the aim of the current work.

For this purpose we will focus on quantum magnetic systems, which 
historically have provided a clean, readily realized, low-dissipation test 
bed for many concepts in condensed-matter and statistical physics. The small 
number and unique behavior of the spin degrees of freedom lead to exact 
solutions including the Heisenberg spin chain, the transverse-field Ising 
model, and the Kitaev model. In non-equilibrium physics, idealized (and 
often integrable) spin-chain models as the Hamiltonian part of a Lindblad 
system have provided the framework for illustrating NESS \cite{rp}, MBL 
\cite{rzpp,ri}, Floquet prethermalization \cite{rwk}, and dynamical quantum 
phase transitions \cite{rzhks}, as well as lending themselves very well to 
numerical investigation. With a view to future device application, single 
spins have long been considered as excellent candidate qubits and the 
application of suitable laser control schemes \cite{ru} has been attempted 
in ensembles of quantum dots \cite{Greilich07,rkea}. The entangled quantum 
many-body states available in magnetic materials present not only a rich 
variety of options for encoding (protected) quantum information, keywords 
including (topological) magnonics \cite{Chumak15,rsmmo,rnkkl,Wang18,Malki19}, 
quantum spin liquids (QSLs) \cite{rbcknns}, and magnetic textures such as 
vortices \cite{Gao17} and skyrmions \cite{rdkl}, but also many routes for 
exploiting intrinsic interactions \cite{Wadley16,Lebrun18,ryba} or extrinsic 
materials-design flexibility \cite{rgkmn} to obtain ``handles'' for 
manipulating magnetism using laser light \cite{Li13,Bossini16,Bossini19}.

The reason why insulating quantum magnets are a relative late-comer to 
the game of laser excitation and pump-probe physics is the weak direct 
coupling of light to spin. In general one may consider four routes for the
creation of magnetic excitations by incident light. (1) The response of 
metallic magnetic systems is usually described in terms of the inverse Faraday 
effect; this mechanism remains present (in the form of virtual electronic 
processes) in insulators and is quadratic in the electric-field strength of 
the light. It was exploited recently \cite{rjbasyzb} to study the coherent 
transport of GHz precession modes of the magnetization over 100 $\mu$m 
distances in ferromagnetic iron garnet films. (2) At the intrinsic frequencies 
of magnetic modes in condensed matter, which are of order 1 THz, processes 
by which one photon creates one magnon depend on anisotropies in the spin 
Hamiltonian. While many forms of spin anisotropy exist, they are in general 
a consequence of spin-orbit coupling and thus they are rather weak in the most 
familiar quantum magnetic materials, whose magnetic ions are 3$d$ transition 
metals. However, they are present in type-II multiferroics and other systems 
with finite magnetoelectric \cite{Manipatruni19} and thermomagnetic coupling 
\cite{Seifert18}, and one such anisotropy was exploited in a recent discussion 
of a laser-pumped spin chain as a test case for a Generalized Gibbs Ensemble 
approach to near-integrable dissipative systems \cite{rllr}. 

(3) The mechanism invoked most commonly in condensed matter emerges from the 
coupling of the electrons to an electromagnetic vector potential described by 
the Peierls substitution. In insulating magnets, the leading-order processes 
are of Raman type, where the scattering of one photon excites two magnon modes 
\cite{rss,rmbe} and thus spin is conserved. For this type of process, the light 
frequency should be a significant fraction of the on-site Coulomb repulsion, 
$U$, of the electrons being excited virtually; because $U$ is of order 5 eV, 
the incident light should be around the visible range. At lowest order, 
incident photons with frequency $\omega$ modify $U$ to $U - \omega$ or $U + 
\omega$, thereby affecting the magnetic (super)exchange interaction. If one 
considers the effect of the electromagnetic field not on the (virtual) 
electronic hopping but on localized electronic energy levels, the interaction 
between two spins localized on sites $i$ and $j$ that have the same energy is 
not changed at linear order by the electric field. By contrast, if the energies 
on $i$ and $j$ are different, the electric field of the light can have a linear 
(albeit weak) influence on the exchange interactions. However, in the common 
situation where the atomic structure ensures a mirror symmetry between ions, 
this interaction vanishes.

(4) The lattice geometry is fundamental to the magnetic interactions, because 
exchange and superexchange processes are very sensitive to the distances and 
angles of the bonds between the ions along the exchange path. Thus the 
selective excitation of specific phonon modes would provide direct control 
of magnetic interactions through a mechanism resonant both between laser and 
phonon and between the selected phonon and the spin sector. By symmetry, the 
phonons must be infrared (IR)-active if they are to be driven directly by the 
light. Once the excitation of a phonon ensures that the atoms in a magnetic 
material are displaced, the modulation of the interactions is in general 
linear in the displacement coordinate; the structural complexity of most 
materials ensures both IR- and Raman-active phonons over a range of 
frequencies, and only for bond paths and displacement directions of 
especially high symmetry do the linear terms vanish. 

We comment for completeness that recent experiments have used nonlinear 
mixing of two driven IR-active phonons to produce excitations at the sum 
and difference frequencies, whose symmetry compositions include Raman-active 
phonons \cite{Melnikov18} and magnetic modes \cite{Giorgianni20}. While this 
mechanism should allow Raman-active phononic modulation of the magnetic 
interactions at quadratic order in the electric-field strength, it is
important to distinguish such ``nonlinear driving with harmonic phonons'' 
\cite{Melnikov18,Giorgianni20} from ``nonlinear phononics'' \cite{Foerst11}. 
The latter depends on anharmonic phonons, and has been exploited to influence 
the electronic properties of correlated many-body states in superconductors 
\cite{Singla15,Kennes17,Sentef17,Murakami17} and Mott insulators 
\cite{Grandi20}. Although nonlinear phononic effects on magnetism have to 
date been considered only in the form of creating effective static magnetic 
fields \cite{Nova17}, more sophisticated protocols could be devised that 
provide a further channel for dynamical driving. Here we restrict our focus 
to the simple case of direct and coherent driving of the spin system by 
single, IR-active phonon modes in the harmonic regime. This situation was 
given the name ``magnetophononics'' by the authors of Ref.~\cite{Fechner18}, 
who performed a theoretical study of classical magnets with phenomenological 
damping, and here we apply the magnetophononic protocol to a quantum spin 
system with quantum dissipation.

\begin{figure}[t]
\includegraphics[width=\columnwidth]{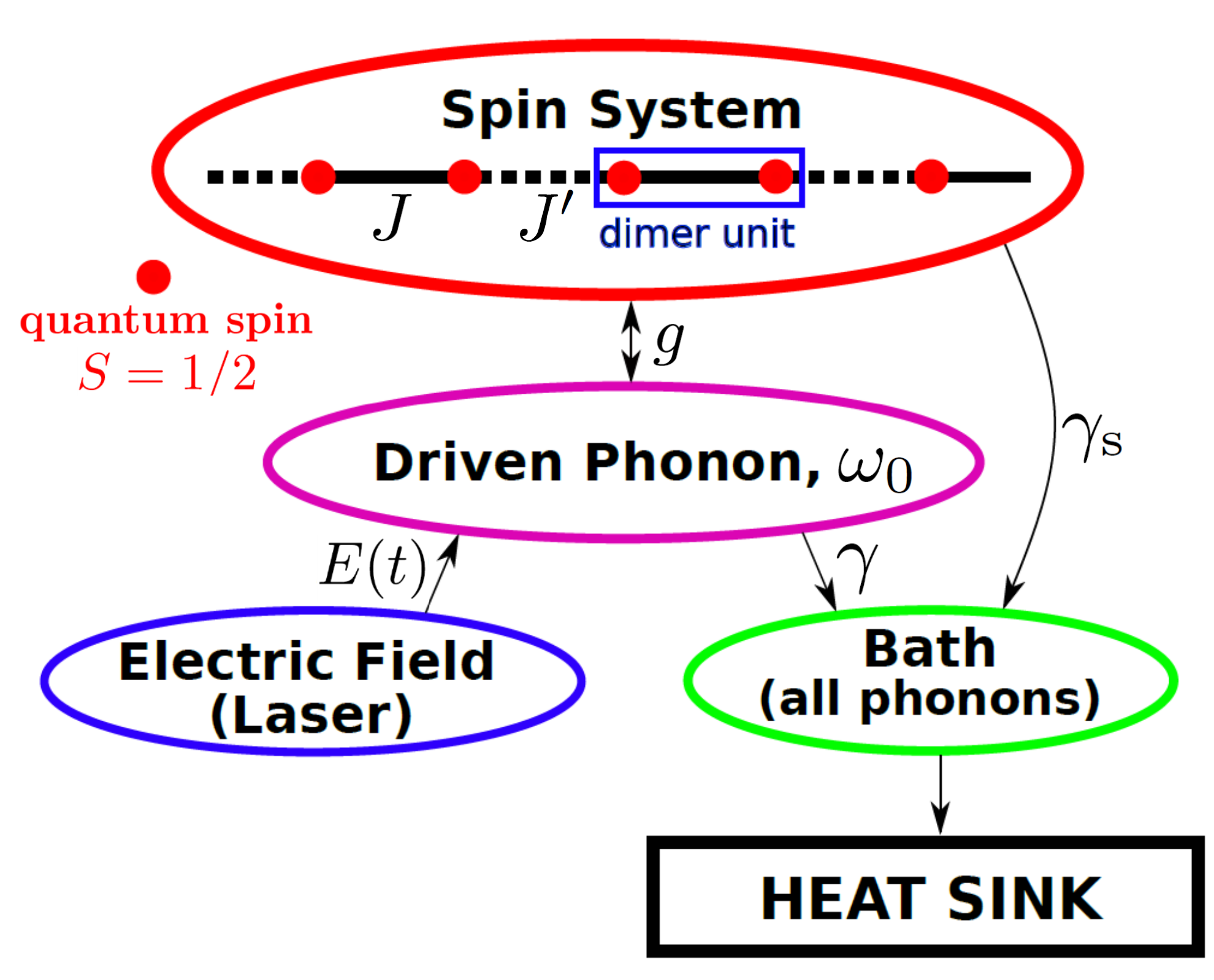}
\caption{Schematic representation of a lattice spin system, here a structurally 
dimerized chain with antiferromagnetic interaction parameters $J > J'$, driven 
by the selective excitation of one specific phonon mode of the lattice. Both 
the driven phonon and the spin system are damped by the ensemble of lattice 
phonons. We envisage an experimental geometry with the sample attached to an 
efficient heat sink for thermal regulation.}
\label{fdcschem}
\end{figure}

To discuss the dynamics of a driven dissipative quantum magnet we use the 
example of the alternating spin chain shown in Fig.~\ref{fdcschem}. The 
driving is effected by laser excitation of an Einstein phonon that couples 
to one of the magnetic interactions in the spin chain and the dissipation is 
modelled in the Lindblad formalism by bath operators that damp both the lattice 
and spin sectors directly. We establish the equations of motion governing 
the basic physics of quantum NESS in this system, in terms of the driving 
frequency, the system parameters, and the response of the separate lattice and 
spin sectors. These equations enable us to discuss the different regimes of 
weak and strong spin-phonon coupling, of weak and strong damping, and all 
the timescales associated with driving, NESS formation, and relaxation. Thus 
our study establishes a foundation for many types of extension, specifically 
to different types of spin system, to different types of bath (characterized 
by whether they conserve spin and momentum), to finite system temperatures 
and thus to driving protocols for the management of heat and of coherence, 
and to quantitative studies of materials and device geometries for practical 
experiments. 

The structure of this article is as follows. In Sec.~\ref{smm} we present 
our model for the quantum spin chain, for the laser-coupled phonon mode 
that drives it, and for the Lindblad bath operators that damp it. We derive 
the equations of motion for the coupled lattice and magnetic sectors and 
comment on their structure. Section \ref{sness} contains a preliminary 
analysis of the content of these master equations, with specific attention 
paid to NESS. We demonstrate numerically that NESS can indeed be established, 
and illustrate how their basic properties are governed by the primary system
parameters, namely the driving power, the driving frequency, the lattice and 
spin damping coefficients, and the spin-lattice coupling. With this basis, in 
Sec.~\ref{sdp} we concentrate on the regime of low spin-phonon coupling to 
perform a complete investigation of the dynamical properties of the NESS in 
the spin sector, characterizing their response by frequency, wave-vector 
components, and spin damping. 

In Sec.~\ref{str} we turn to a different but essential aspect of NESS, namely 
the transient processes occurring as they are established, from the moment the 
laser driving is switched on, and the relaxation processes by which equilibrium 
is restored when the drive is removed. At higher net occupancies of lattice 
and spin excitations we find anomalously slow convergence to NESS, and in 
Sec.~\ref{str}A we apply analytical arguments to discuss the underlying 
physics. Section \ref{str}B extends this analysis to the question of limits 
in parameter space for the existence of NESS within our model framework and 
Sec.~\ref{str}C provides a brief discussion of relaxation and temperature. In 
Sec.~\ref{seh} we analyze the energy flow in the NESS, considering both its 
uptake by the spin system as a function of laser power and frequency and its 
dissipation by the Lindblad terms. This allows us to provide experimentally 
oriented estimates for the rate of temperature increase in the driven system, 
for its control by the heat sink shown in Fig.~\ref{fdcschem}, and the 
resulting timescales for read-out and control processes. In Sec.~\ref{sd} 
we discuss the context of our results from a number of angles, including 
methodology, the influence of the bath model, timescales and heating effects, 
and laser experiments on real materials. Section \ref{ss} consists of a brief 
summary and perspectives for future extensions of the framework established 
in this study. 

\section{Model and Methods}
\label{smm}

We begin by representing the Hamiltonian of the coupled system shown in 
Fig.~\ref{fdcschem} as 
\begin{equation}
H = H_\text{s} + H_\text{sp} + H_\text{p} + H_\text{l},
\label{etsh}
\end{equation}
where the four terms describe respectively the spin system, the spin-phonon 
coupling, the Einstein phonon, and the effect of the laser electric field 
on this phonon. The bath operators damping the spin and lattice systems do 
not enter Eq.~\eqref{etsh} explicitly, but are introduced at the level of 
the Lindblad formalism. While some authors have investigated the strong and 
controllable effects obtained by considering the quantum nature of the light 
field, generally referred to as ``cavity QED'' \cite{rmg,rsrr}, for the
purpose of driving phonon modes we treat the laser light field as classical.

\subsection{Spin system}

We express the Hamiltonian for the structurally dimerized, antiferromagnetic 
spin chain as  
\begin{equation}
H_\text{s} = \sum_i J {\vec S}_{1,i} \! \cdot \! {\vec S}_{2,i} + 
J' {\vec S}_{2,i} \! \cdot \! {\vec S}_{1,i+1}, 
\label{ehs}
\end{equation}
with $J > J' > 0$. For simplicity we consider only Heisenberg interactions 
between the spins and neglect any anisotropy terms; in real materials these 
could be of exchange, Dzyaloshinskii-Moriya, single-ion, $g$-tensor, or 
other origin, and as noted in Sec.~\ref{sintro} are generally weak in 3$d$ 
transition-metal compounds. A representation particularly useful for dimerized 
spin systems is the bond-operator description \cite{rsb,rgrs}, in which the 
Hamiltonian is transformed by expressing the two spin operators on each dimer 
using the identity 
\begin{equation} 
S^{\alpha}_{1,2} = \pm {\textstyle \frac{1}{2}} (s^{\dag} t_{\alpha} + t_{\alpha}^{\dag} 
s) - {\textstyle \frac{1}{2}} i \sum_{\beta \zeta} \epsilon_{\alpha \beta \zeta} 
t_{\beta}^{\dag} t_{\zeta},
\label{ebot} 
\end{equation} 
where $s$ and $t_{\alpha}$ ($\alpha = x,y,z$) are operators for the 
singlet and triplet states of each dimer ($J$) bond. These operators 
have bosonic statistics, required to reproduce the spin algebra of 
$S^{\alpha}_{1,2}$; however, because each dimer may only be in a singlet 
state or one of the three triplets (equivalent to the four possible states 
of two spin-1/2 entities), the bond operators must also obey a local 
hard-core constraint, 
\begin{equation} 
s_{i}^{\dag} s_{i} + \sum_{\alpha} t_{i,\alpha}^{\dag} t_{i,\alpha} = 1, 
\label{eboc} 
\end{equation} 
on each dimer $i$, and hence are hard-core bosons. 

For a system whose magnetic interactions are inversion-symmetric, the minimal 
Hamiltonian of Eq.~(\ref{ehs}) takes the form $H_\text{s} = H_0 + H_2 + H_4$, 
where \cite{rgrs,rmnrs} 
\begin{eqnarray} 
H_0 & = & \sum_{i} - J ({\textstyle \frac{3}{4}} s_{i}^{\dag} s_{i} - 
{\textstyle \frac{1}{4}} t_{i,\alpha}^{\dag} t_{i,\alpha}) \\ & & \;\;\;\;\;\; 
 - \mu_i (s_{i}^{\dag} s_i + t_{i,\alpha}^{\dag} t_{i,\alpha} - 1) , \nonumber
\label{ebosh0} 
\end{eqnarray} 
with summation over the repeated index $\alpha$,
\begin{eqnarray} 
H_2 & = & - {\textstyle \frac{1}{4}} J' \sum_{i,\alpha} \big( t_{i,\alpha}^{\dag} 
t_{i+1,\alpha} s_{i+1}^{\dag} s_i \\ & & \;\;\;\;\;\;\;\;\;\;\;\;\;\;\;\;
 + t_{i,\alpha}^{\dag} t_{i+1,{\alpha}}^{\dag} s_i s_{i+1} + {\rm H. c.} \big) , 
\nonumber
\label{ebosh2} 
\end{eqnarray} 
and 
\begin{eqnarray} 
H_4 & = & {\textstyle \frac{1}{8}} J' \sum_{i,\alpha \ne \beta} \big( 
t_{i,\alpha}^{\dag} t_{i+1,\beta}^{\dag} t_{i+1,\alpha} t_{i,\beta} \\ & & 
\;\;\;\;\;\;\;\;\;\;\;\;\;\;\;\; - t_{i,\alpha}^{\dag} t_{i+1,\alpha}^{\dag} 
t_{i+1,\beta} t_{i,\beta} + {\rm H. c.} \big) . \nonumber
\label{ebosh4} 
\end{eqnarray} 
The second term in $H_0$ enforces the constraint [Eq.~(\ref{eboc})] using 
the Lagrange multipliers $\mu_i$. At zero applied magnetic field, the term 
quadratic in the singlet operators is negative, which ensures a singlet 
condensation and justifies their replacement by a constant, $s_i = \langle 
s_i \rangle$, on each dimer. The ground state of the system is then a 
condensate of singlets with a spin gap to all triplet excitations, whose 
dispersion is specified by the quadratic terms in $H_2$. Here we will not 
consider any spatial gradients (for example in temperature, magnetic field, 
or laser flux) and hence $\langle s_i \rangle = \Bar{s}$ and $\mu_i = 
\mu$; the latter condition enforces the hard-core constraint at a global 
level, but not locally. For the purposes of the present analysis we will 
not consider triplet-triplet interactions, and thus we neglect $H_4$.

We transform the quadratic triplet Hamiltonian $H_0 + H_2$ to reciprocal 
space using 
\begin{equation}
t_{i,\alpha} = \frac{1}{\sqrt{N}} \sum_k t_{k,\alpha} e^{-ikr_i},
\end{equation}
where $N$ is the number of dimers, and express the result in the form  
\begin{eqnarray} 
H_{\rm mf} & = & E_0 + \sum_{k,\alpha} \big[ \big( \textstyle{\frac{1}{4}} J \! -
 \! \mu \big) t_{k,\alpha}^\dag t_{k,\alpha} - \textstyle{\frac{1}{4}} J' \Bar{s}^2 
\cos k \big( t_{k,\alpha}^\dag t_{k,\alpha} \nonumber \\ & & \;\;\;\;\;\;\;\;
\;\; + \, t_{-k,\alpha} t_{-k,\alpha}^\dag + t_{k,\alpha}^\dag t_{-k,\alpha}^\dag + 
t_{-k,\alpha} t_{k,\alpha} \big) \big],
\label{ehmf}
\end{eqnarray}
where 
\begin{equation} 
E_0 = N \big[ (- \textstyle{\frac{3}{4}} J - \mu) \Bar{s}^2
 + \textstyle{\frac{5}{2}} \mu - \textstyle{\frac{3}{8}} J \big]. 
\label{eeo}
\end{equation}
We note that the only terms generated are those coupling operators at wave 
vectors $k$ and $-k$, and there is no mixing of the triplet indices, $\alpha$. 
The conventional approach \cite{rgrs,rmnrs} is to symmetrize the Hamiltonian 
matrix, diagonalize it to obtain a new bosonic quasiparticle, known as the 
triplon \cite{rsu}, form two mean-field equations, and solve these for $\mu$ 
and $\Bar{s}$. By the use of effective quasiparticle statistics, this procedure 
may also be followed at finite temperatures \cite{rrnmnfkgsm,rnr}.  

Here we adopt one further simplification with a view to applying 
equation-of-motion methods. In the ``Holstein-Primakoff'' approximation 
\cite{rmnrs}, the singlet occupation is replaced directly by invoking the 
local constraint [Eq.~(\ref{eboc})], giving 
\begin{equation}
\Bar{s}^2 = 1 - \frac{1}{N} \sum_{k,\alpha} t_{k,\alpha}^\dag t_{k,\alpha}.
\label{esr}
\end{equation}
At quadratic order, this substitution reduces to the approximation 
$\Bar{s} = 1$, $\mu = - \textstyle{\frac{3}{4}} J$, which is clearly 
valid in the limit of a strongly dimerized chain. From extensive studies 
of the spin ladder \cite{rnr}, it is generally recognized that the 
bond-operator description retains semiquantitative validity for interaction 
ratios $J'/J \lesssim 1/2$ at low temperatures. For the present qualitative 
purposes, this approximation has the major advantage of not requiring a 
solution of the self-consistent equations at each time step. 

From the spin Hamiltonian in the ``mean-field'' form 
\begin{eqnarray} 
H_\text{s} & = & \sum_{k,\alpha} \big[ J t_{k,\alpha}^\dag t_{k,\alpha} - 
\textstyle{\frac{1}{4}} J' \cos k \, \big( 2 t_{k,\alpha}^\dag t_{k,\alpha} 
\nonumber \\ & & \;\;\;\;\;\;\;\;\;\; + \, t_{k,\alpha}^\dag t_{-k,\alpha}^\dag
 + t_{k,\alpha} t_{-k,\alpha} \big) \big], 
\label{ehmfhp}
\end{eqnarray}
we diagonalize it by applying the Bogoliubov transformation
\begin{subequations}
\label{ebt} 
\begin{eqnarray} 
t_{k, \alpha} & = & \tilde{t}_{k, \alpha} \cosh \theta_k + \tilde{t}_{-k, 
\alpha}^{\,\dagger} \sinh \theta_{k},  \\ 
t_{k, \alpha}^{\,\dagger} & = & \tilde{t}_{k, \alpha}^{\,\dagger} \cosh \theta_{k}
 + {t}_{-k, \alpha} \sinh \theta_{k} ,
\end{eqnarray}
\end{subequations}
where 
\begin{equation} 
\tanh 2 \theta_{k} = \frac{\lambda \cos k }{2 - \lambda \cos k}
\label{etk}
\end{equation} 
with $\lambda = J'/J$, to obtain 
\begin{equation} 
H_\text{s} = \sum_{k, \alpha} \omega_k \tilde{t}_{k, \alpha}^{\,\dagger} 
\tilde{t}_{k, \alpha}.
\label{ehsd}
\end{equation} 
The operators $\tilde{t}_{k, \alpha}^{\,\dagger}$ and $\tilde{t}_{k, \alpha}$ create 
and destroy the triplon modes of the dimerized chain and have dispersion 
relation 
\begin{equation} 
\omega_k = J \sqrt{1 - \lambda \cos k}. 
\label{ewk}
\end{equation} 

\subsection{Phonon system and spin coupling}

As shown in Fig.~\ref{fdcschem}, we consider a situation in which the 
interaction $J$ is modulated by the oscillations of one specific phonon mode 
on every bond. We take this to be an Einstein phonon with wavevector $q = 0$ 
and a finite energy, $\omega_0$. As noted in Sec.~\ref{sintro}, we focus 
on the situation where this optical phonon is IR-active and hence is driven 
directly by the electric field of the incident light, and do not consider the 
further possibilities offered by high-order phonon excitation processes. We 
assume that the laser illuminates the entire sample, meaning that we treat 
the driving as a bulk effect. In a real material, many different phonon modes 
are present in addition to the driven phonon, and all of them, in particular 
the acoustic phonons, are responsible for the dissipation of energy from both 
the lattice and spin sectors. 

The Hamiltonian terms involving the driven phonon are 
\begin{eqnarray}
H_\text{p} + H_\text{sp} + H_\text{l} & = & 
\sum_j  \big[ \omega_0 b_j^\dag b_j + g (b_j + b_j^\dag) 
\, {\vec S}_{1,j} \! \cdot \! {\vec S}_{2,j} \nonumber \\ 
& & \qquad\qquad\quad + \, E(t) (b_j + b_j^\dag) \big],
\label{ehp}
\end{eqnarray}
where $g$ is the spin-phonon coupling constant and $E(t) = a \cos (\omega t)$ 
is the oscillating electric field of the laser, which we assume to contain a 
single driving frequency, $\omega$; as noted above, for the amplitudes $a$ we 
consider, $E(t)$ may safely be treated as classical field. For our purposes, 
$E(t)$ is an internal field, meaning it is the fraction of the incident laser 
light transmitted into the sample, and we do not concern ourselves with the 
reflected component. The dissipative terms do not enter Eq.~(\ref{ehp}), but 
will be included using the Lindblad formalism in Sec.~\ref{smm}C.  

The transformation of Eq.~(\ref{ehp}) includes single- and triple-operator 
terms. For pedagogical accuracy we take a conventional definition of the 
Fourier transform,
\begin{equation}
b_j = \frac{1}{\sqrt{N}} \sum_q b_q e^{-iqr_j},
\end{equation}
under which the electric-field term becomes 
\begin{subequations}
\begin{align}
\frac{E(t)}{\sqrt{N}} \sum_{j,q} (b_q e^{-iqr_j} + b_q^\dag e^{iqr_j}) & = 
\frac{E(t)}{\sqrt{N}} \sum_q  (b_0 + b_0^\dag), \\
& = N E(t) d,
\label{eeft}
\end{align}
\end{subequations}
where only the $q = 0$ mode is selected, but we express it as an intensive 
quantity summed over $q$, with effective displacement operator $d = 
{\textstyle \frac{1}{\sqrt{N}}} (b_0 + b_0^\dag)$. Even more simply, the 
phonon term becomes $\sum_q \omega_0 b_q^{\,\dag} b_q$. 

Finally, the spin-phonon coupling term becomes 
\begin{eqnarray}
& & \frac{1}{N\sqrt{N}} \sum_{j,q,k,k',\alpha} \big( b_q t_{k,\alpha}^\dag t_{k',\alpha} 
e^{i(k-k'-q)r_j} + \text{H.c.} \big) \nonumber \\ & & =  \frac{1}{\sqrt{N}} 
\sum_{q,k,\alpha} \big( b_q t_{k,\alpha}^\dag t_{k-q,\alpha} + 
b_q^\dag t_{k,\alpha}^\dag t_{k+q,\alpha} \big),   
\label{espt}
\end{eqnarray}
with $q = 0$ as the only relevant phonon mode. At the mean-field level one 
obtains the decoupled terms 
\begin{subequations}
\begin{align}
H_\text{sp} & = H_\text{sp,s} + H_\text{sp,p},
\label{eq:mf-split} \\
H_\text{sp,s} & = g \langle d \rangle \Big[ \sum_{k,\alpha} t_{k,\alpha}^\dag 
t_{k,\alpha} - \langle t_{k,\alpha}^\dag t_{k,\alpha} \rangle_{\rm eq} \Big], \\
H_\text{sp,p} & = g \Big\langle \sum_{k,\alpha} t_{k,\alpha}^\dag t_{k,\alpha} - 
\langle t_{k,\alpha}^\dag t_{k,\alpha} \rangle_{\rm eq} \Big\rangle \, d, 
\label{emfspt}
\end{align}
\end{subequations}
where we have omitted the product of the two expectation values in 
Eq.~\eqref{eq:mf-split} because it has no influence at all on the dynamics 
of the system. Here $H_\text{sp,s}$ contains the operator part acting on the 
spin degrees of freedom, expressed in triplon operators, while $H_\text{sp,p}$ 
contains the operator part acting on the driven phonon. In both terms the 
spin-phonon interaction is expressed by deducting the equilibrium value of 
the triplon occupation, such that it has no effect when the system is not 
driven. While this mean-field decoupling is an approximation, we will show 
in Sec.~\ref{seh} that its quantitative limitations are minor.

To transform $H_\text{sp}$ into the diagonal (triplon) basis of the spin sector, 
we use the identity  
\begin{eqnarray}
t_{k, \alpha}^{\,\dagger} t_{k, \alpha} & = & y_k \big(\tilde{t}_{k, \alpha}^{\,\dagger} 
\tilde{t}_{k, \alpha} + {\textstyle \frac12} \big) - {\textstyle \frac12} 
\nonumber \\ & & \quad + \, {\textstyle \frac12} y'_k \big( 
\tilde{t}_{k, \alpha}^{\,\dagger} \tilde{t}_{-k, \alpha}^{\,\dagger}
 + \tilde{t}_{k, \alpha} \tilde{t}_{-k, \alpha} \big),
\label{ehspt}
\end{eqnarray}
in which 
\begin{subequations}
\begin{eqnarray}
y_k & = & \frac{1 - {\textstyle \frac12} \lambda \cos k}{\sqrt{1 - \lambda 
\cos k}} = \frac{J}{2} \frac{1 + \omega^2_k / J^2}{\omega_k} \;\;\; {\rm and} 
\label{eyk} \\ y'_k & = & \frac{ {\textstyle \frac12} \lambda \cos k}{\sqrt{1
 - \lambda \cos k}} = \frac{J}{2} \frac{1 - \omega^2_k / J^2}{\omega_k}, 
\label{eypk}
\end{eqnarray}
\end{subequations}
to obtain the expression
\begin{eqnarray}
H_\text{sp,s} & = & g \, \langle d \rangle \sum_{k, \alpha} \big[ y_k 
[\tilde{t}_{k, \alpha}^{\,\dagger} \tilde{t}_{k, \alpha} - {n}(\omega_k)] \nonumber 
\\ & & \quad + {\textstyle 
\frac12} y'_k \big( \tilde{t}_{k, \alpha}^{\,\dagger} \tilde{t}_{-k, \alpha}^{\,\dagger}
 + \tilde{t}_{k, \alpha} \tilde{t}_{-k, \alpha} \big) \big].
\end{eqnarray}
Here the bosonic occupation function, ${n}(\omega_k) = [\exp (\hbar\omega_k 
/ k_{\rm B} T) - 1]^{-1}$, provides an accurate value for the equilibrium 
occupancy of the triplon mode with frequency $\omega_k$ [Eq.~\eqref{ewk}] 
at the low temperatures we consider, despite the hard-core nature of these 
modes \cite{rfu}.

We define the operators 
\begin{subequations}
\label{ersv}
\begin{eqnarray}
u_k & = & \sum_\alpha  \tilde{t}_{k,\alpha}^\dag \tilde{t}_{k,\alpha} \;\; {\rm and} 
\\ {\tilde{v}}_k & = & \sum_\alpha  \tilde{t}_{k,\alpha}^\dag 
\tilde{t}_{-k,\alpha}^\dag 
\end{eqnarray}
\end{subequations}
in the triplon sector and denote their expectation values at any given 
time, $t$, by
\begin{subequations}
\label{eq:expect}
\begin{eqnarray}
\label{eq:u-def}
u_k(t) & = & \langle u_k \rangle (t) , \\ 
\tilde v_k(t) & = & \langle {\tilde{v}} \rangle (t);
\end{eqnarray}
\end{subequations}
the expectation value of the product of two annihilation operators is 
manifestly the complex conjugate of $\tilde v_k$,
\begin{equation}
\tilde v_k^*(t) =  \sum_\alpha \langle \tilde{t}_{k,\alpha}
\tilde{t}_{-k,\alpha}\rangle (t).
\end{equation}
We comment that the triplon branch, $\alpha$, is summed over here and, 
because we do not consider an applied magnetic field or any anisotropy in 
the spin Hamiltonian, will not enter our considerations again. 

$u_k(t)$ is clearly a real variable due to the hermiticity of the operator 
on the right-hand side of Eq.~(\ref{ersv}a), while the complex variable 
$\tilde v_k(t)$ is conveniently separated into its real and imaginary parts,
\begin{subequations}
\label{eq:vw-def}
\begin{align}
v_k(t) & = {\rm Re} \, \tilde v_k(t) \\
w_k(t) & = {\rm Im} \, \tilde v_k(t).
\end{align}
\end{subequations}
In the equations of motion to be derived in Sec.~\ref{smm}C, the spin-phonon 
coupling introduces two quantities composed of the above expectation values, 
which we include by defining 
\begin{subequations}
\label{ecuv}
\begin{align}
\mathcal{U}(t) & = \frac{1}{N} \sum_k y_k [u_k(t) - 3 n(\omega_k)], \\
\mathcal{V}(t) & = \frac{1}{N} \sum_k y_k' v_k(t),
\end{align}
\end{subequations}
both of which are real by construction. For the description of the spin 
sector in the driven system, we define the number, $n_\text{x}$, of elementary 
(Bogoliubov, or ``dressed'') triplons per site, 
\begin{equation}
n_\text{x}(t) = \frac{1}{N} \sum_k u_k(t),
\label{enx}
\end{equation}
and it will be helpful to compare this with the number of original (or 
``bare'') triplons per site in the starting basis of Eq.~\eqref{etsh},  
\begin{equation}
n_\text{b} (t) = \frac{1}{N} \sum_{k,\alpha} \langle t^\dag_{k,\alpha} t_{k,\alpha} 
\rangle (t).
\end{equation}
Using Eq.~\eqref{ehspt}, this last definition is equivalent to
\begin{equation}
n_\text{b}(t) = \mathcal{U}(t) + \mathcal{V}(t) + \frac{1}{N} \sum_{k,\alpha} 
\langle t^\dag_{k,\alpha} t_{k,\alpha} \rangle_\text{eq},
\end{equation}
in which the last term is given by
\begin{equation}
\label{enbe}
\frac{1}{N} \sum_{k,\alpha} \langle t^\dag_{k,\alpha} t_{k,\alpha} \rangle_\text{eq}
= \frac{3}{2N} \sum_{k} [ (2n (\omega_k) + 1) y_k - 1].
\end{equation}
At zero temperature and for $\lambda = 1/2$, which will be our test case 
in what follows, the equilibrium expectation value is $n_{\rm b0} = 0.028$. 
This number quantifies the quantum fluctuations in equilibrium and will 
serve as a reference for the extent of modifications to the phonon-driven 
spin state relative to the undriven ground state.

\subsection{Equations of motion}

The time evolution of an open quantum system is specified by adjoint 
quantum master equations \cite{rbp} of the form 
\begin{eqnarray}
\label{eaqme}
&& \frac{d}{dt} A_\text{H}(t) = i [H, A_\text{H}(t)] \\ && \quad + \sum_{l} 
\tilde\gamma_l \big[ A_l^\dag A_\text{H}(t) A_l - {\textstyle \frac12} 
A_\text{H}(t) A_l^\dag A_l - {\textstyle \frac12} A_l^\dag A_l A_\text{H}(t) \big] 
\nonumber
\label{eq:lindblad1}
\end{eqnarray}
for any operator $A_\text{H}(t)$ describing a physical observable. In these 
Heisenberg equations of motion, $H$ is the Hamiltonian of the ``reduced'' 
system under consideration, by which is meant the quantum system with no 
environment. The ``Lindblad'' operators, $\{A_l\}$, are formed from the 
Liouville space of the reduced system to describe its interaction with the 
environment (the ``bath''), which is excluded from explicit consideration. 
It was proven by Lindblad \cite{rl} that Eq.~\eqref{eaqme} is the most 
general form of the dissipation term for a separable (system-bath) Hilbert 
space when $l$ describes a bounded set of operators. The coefficients 
$\tilde\gamma_l$ play the role of damping parameters. 

The driven phonon exemplifies the textbook case \cite{rbp,rw} of the 
Lindblad equations, namely those of the damped harmonic oscillator. 
The Lindblad operators in this case are $A_1 = b_0^\dag$ and $A_2 = b_0$,
with damping rates $\gamma_1$ and $\gamma_2$. For the system to relax back 
to its equilibrium state in the absence of driving, it is known \cite{rbp} 
that the ratio of the two rates must be given by the ratio $n(\omega_0) / 
(1 + n(\omega_0))$, and hence the conventional parameterization is
\begin{subequations}
\begin{align}
\label{eq:excite}
\tilde\gamma_1 & = \gamma n(\omega_0)
\\
\tilde\gamma_2 & = \gamma (1 + n(\omega_0)),
\end{align}
\end{subequations}
leaving only one damping parameter, $\gamma$. For physical transparency we
separate the Lindblad operators into those that excite the system by an 
energy $\omega_l$, which we denote by $B_l$, and those that de-excite, 
which are given by the Hermitian conjugates, $B_l^\dag$. The dissipative
part of Eq.~\eqref{eaqme}, which is the second line, may then be separated 
into the two contributions 
\begin{subequations}
\label{eq:dissipation}
\begin{align}
\label{eq:t1}
T_1 & = {\textstyle \frac12} \sum_{l} \! \gamma_l n(\omega_l) \big\{ \! \big[ 
B_l, [A_\text{H}(t) , B_l^\dag] \big] \! + \! \big[ B_l^\dag , [A_\text{H}(t) , B_l] 
\big] \! \big\}, \\
\label{eq:t2}
T_2 & = {\textstyle \frac12} \sum_{l} \gamma_l
\big\{ [B_l , A_\text{H}(t)] B_l^\dag + B_l [A_\text{H}(t) , B_l^\dag] \big\}.
\end{align}
\end{subequations}
The commutators in these expressions facilitate their rapid evaluation 
in comparison with the expression in Eq.~\eqref{eaqme}; if the observable
and the Lindblad operators are linear bosonic operators, as for the damped
phonon, it can be seen without explicit calculation that the term $T_1$ 
vanishes and hence no dependence on the bosonic occupation, $n(\omega_l)$, 
arises. 

To describe the driven phonon we consider the real variables 
\begin{subequations}
\label{eq:expval-phon}
\begin{eqnarray}
q(t) & = & \big\langle {\textstyle \frac{1}{\sqrt{N}}} (b_0 + b_0^\dag) 
\big\rangle (t), \\ 
p(t) & = & \big\langle {\textstyle \frac{i}{\sqrt{N}}} (b_0^\dag - b_0) 
\big\rangle (t), \\  
n_{\rm ph}(t) & = & \big\langle {\textstyle \frac{1}{N}} b_0^\dag b_0 \big\rangle 
(t), 
\label{erpv}
\end{eqnarray}
\end{subequations}
describing respectively the displacement of the Einstein phonon ($d$ in 
Sec.~\ref{smm}B), the conjugate phonon momentum, and the number operator. 
We recall that, despite the presence of all phonon modes, $b_q^\dag$, in 
$H_\text{p}$, only the operators $b_0$ and $b_0^\dag$ appear elsewhere in the 
Hamiltonian of the reduced system, and hence are candidates for the formation 
of Lindblad operators. The other phonons form the environment and their 
presence gives rise to the damping, which is contained in the single 
parameter $\gamma$.

By evaluating Eqs.~\eqref{eaqme} or \eqref{eq:dissipation} with the 
expectation values from the spin sector [Eqs.~\eqref{ecuv}] in the 
spin-phonon coupling term, one obtains the closed set of equations of 
motion \cite{rbp}
\begin{subequations}
\label{eq:eom-phonon}
\begin{eqnarray}
\frac{d}{dt} q(t) & = & \omega_0 p(t) - {\textstyle {\frac12}} \gamma 
q(t), \label{eompa} \\
\frac{d}{dt} p(t) & = & - \omega_0 q(t) - {\textstyle {\frac12}} \gamma p(t) 
\nonumber \\ & & \;\;\; - 2 [ E(t) + g (\mathcal{U}(t) + \mathcal{V}(t))], 
\label{eompb} \\
\frac{d}{dt} n_{\rm ph}(t) & = & - [E(t) + g (\mathcal{U}(t) + \mathcal{V}(t))] 
p(t) \nonumber \\ & & \;\;\;\;\;\; - \gamma [n_{\rm ph}(t) - {n}(\omega_0)]. 
\label{eompc}
\end{eqnarray}
\end{subequations}
One observes the characteristic structure in Eqs.~\eqref{eompa} and 
\eqref{eompb} of the displacement and momentum serving as conjugate time 
derivatives, but damping themselves through the $\gamma/2$ term. The 
electric-field driving and the spin-system coupling appear only in the 
equation for the phonon momentum [Eq.~\eqref{eompb}]. The number operator 
reflects the driving of the momentum [Eq.~\eqref{eompc}] and also features 
as its own damping term, where $n(\omega_0)$ is the occupation of phonon 
mode $\omega_0$ at thermal equilibrium. Here we do not extend these 
considerations to bilinear Lindblad operators in either the phonon or 
the spin sector.

Turning to the spin degrees of freedom, we consider the real expectation 
values $u_k(t)$, $v_k(t)$, and $w_k(t)$ introduced in Eqs.~\eqref{eq:u-def}
and \eqref{eq:vw-def} to describe the dynamical spin processes diagonal 
and off-diagonal in the triplon number basis. To determine the equations of 
motion, it is expected that the spin sector will be subject to a direct 
damping due both to weak spin-anisotropic terms and to phononic processes 
arising from the many acoustic and optical phonon modes in the Hamiltonian 
of any real material. The specific nature of these damping processes will be 
the subject of more extended discussion in Secs.~\ref{sdp} and \ref{sd}, but 
the available Lindblad operators will in general be linear and bilinear 
combinations of $\tilde{t}_k$ and $\tilde{t}_k^\dag$. In the present analysis 
we focus on linear operators, in order to present the primary phenomena 
associated with the driven dissipative quantum spin chain. The equations of 
motion we will deduce have the analytical advantage of maintaining a simple 
form with transparent physical consequences. However, it is true that such 
one-triplon Lindblad operators are spin non-conserving, meaning that this 
type of bath is appropriate for materials with the non-negligible spin 
anisotropies more commonly associated with systems of 4$d$ and 5$d$ magnetic 
ions. Usually such anisotropic terms are nevertheless corrections to the spin 
Hamiltonian of Eq.~\eqref{ehs}, whereas they may be the leading dissipative 
terms; we discuss this situation in more detail, and comment on the case 
of spin-conserving bath operators, in Sec.~\ref{sd}.

Thus the Lindblad operators, $B_{k}$, that we consider are simply 
$\tilde{t}_{k,\alpha}^\dag$, and have damping coefficient
\begin{equation}
\tilde \gamma_{k} = \gamma_\text{s} n(\omega_k),
\end{equation}
while $B_{k}^\dag$ has damping $\gamma_\text{s} (1 + n(\omega_k))$. We neglect 
a possible dependence of $\gamma_\text{s}$ on the wave vector, $k$, along 
the chains. While one may ask whether this approximation constitutes a severe 
omission, given that energy and momentum conservation allow dissipation only 
for particular combinations of both, we observe that energy conservation as 
contained in Fermi's Golden Rule does not impose a strong constraint when 
one recalls that the 1D chains are embedded in a three-dimensional (3D) 
crystal. Thus $k$-dependent damping coefficients, $\gamma_\text{s}(k)$, are 
averaged over the transverse momentum, $\vec{k}_\perp$, and the assumption 
that energy conservation is satisfied at some value of $\vec{k}_\perp$ is 
fully justified. While some dependence of $\gamma_\text{s}$ on the longitudinal 
momentum, $k$, may indeed remain, we proceed for the purposes of our present 
pedagogical exposition with a single value of $\gamma_\text{s}$ for clarity.

To deduce the equations of motion when $A_\text{H}(t)$ in Eq.~(\ref{eaqme}) 
is one of the bilinear operator combinations in Eq.~(\ref{ersv}), we first 
consider the Hamiltonian parts of the respective expressions, 
\begin{subequations}
\label{eq:eom-sps}
\begin{eqnarray} 
\left[H_\text{s} , u_k  \right] & = & 0, \\ 
\left[H_\text{s} , {\tilde{v}}_k \right] & = & 2 \omega_k {\tilde{v}}_k, \\ 
\left[H_\text{sp,s} , u_k \right] & = & g q(t) y'_k \big( {\tilde{v}}^\dag_k
 - {\tilde{v}}_k \big), \\
\left[H_\text{sp,s} , {\tilde{v}}_k \right] & = & 2 g q(t) \big[ y_k 
{\tilde{v}}_k + y'_k \big( u_k + {\textstyle \frac{3}{2}} \big) \big].
\end{eqnarray}
\end{subequations}
Combining the unitary parts of Eqs.~\eqref{eq:eom-sps} with the dissipative 
part, $T_2$, from Eq.~\eqref{eq:t2}, and taking the appropriate expectation 
values, leads to the final expressions 
\begin{subequations}
\label{eq:eoms}
\begin{eqnarray} 
\frac{d}{dt} u_k(t) & = & 2g q (t) y'_k  w_k (t) - \gamma_\text{s} 
[u_k (t) \! - \! 3 n(\omega_k) ] \label{eomsu} \\
\frac{d}{dt} v_k (t) & = & -2 [\omega_k + g y_k q(t)] w_k (t) 
- \gamma_\text{s} v_k (t) \label{eomsv} \\ 
\frac{d}{dt} w_k (t) & = & 2 [\omega_k + g y_k q(t)] v_k (t) 
\nonumber \\ & & + 2g q (t) y'_k \left[ u_k(t) + {\textstyle \frac{3}{2}} 
\right] - \gamma_\text{s} w_k (t). 
\label{eomsw}
\end{eqnarray}
\end{subequations}
In combination with Eqs.~\eqref{eompa}-\eqref{eompc}, these form the equations 
of motion for the coupled spin-lattice system. Regarding the structure of these 
equations, we comment only that $n_{\rm ph} (t)$ [Eq.~\eqref{eompc}] does not 
have any direct effect on the evolution of the other coupled equations and 
hence it appears that this variable can be neglected for dynamical purposes, 
but we will continue to show $n_{\rm ph}(t)$ as a valuable diagnostic of the 
state of the driven phonon sector. 

Regarding the solution of these equations, in order to study the 
steady-state and dynamical properties of the driven and dissipative 
ensemble of Fig.~\ref{fdcschem}, this will be our task in Secs.~\ref{sness} 
and \ref{sdp}. In the majority of our calculations, we will use a periodic 
chain of $N = 400$ dimers and hence by inversion symmetry will have 201 
independent values of $k$, which we will consider both separately and in 
summed quantities such as Eqs.~\eqref{ecuv} and \eqref{enx}. The 
equations of motion [Eqs.~\eqref{eompa}-\eqref{eompc} and 
Eqs.~\eqref{eomsu}-\eqref{eomsw}] have no lower or upper validity cutoff 
in time, and thus can be applied to discuss the formation, switching, and 
relaxation of quantum spin NESS from $t = 0$ to $\infty$. 

\section{NESS in the phonon-driven spin system}
\label{sness}

We begin by choosing input parameters that establish quantum spin NESS, 
deferring a detailed analysis of the limits to NESS formation until 
Sec.~\ref{str}. Our first aim is a preliminary characterization of the 
response of NESS to the different factors influencing their driving. To 
reduce the space of possible driving parameters, in the present analysis we 
restrict our considerations to resonant excitation of the Einstein phonon 
mode, meaning that we select the laser frequency such that $\omega = \omega_0$ 
and hence $E(t) = a \cos (\omega_0 t)$. From a driving standpoint, for the 
electric-field intensities we wish to study and for a generically weak 
spin-phonon coupling, off-resonant driving is largely just a less efficient 
means, by a factor proportional to $[(\omega - \omega_0)^2 + (\gamma/2)^2]^{-1}$,
of exciting a response at frequency $\omega$. However, in systems with 
stronger spin-phonon coupling, nontrivial phenomena are indeed found by 
pumping and probing at frequencies $\omega \ne \omega_0$. We remind the reader 
that the minimal model of Sec.~\ref{smm}A was not designed to describe driving 
by any of the other physical mechanisms summarized in Sec.~\ref{sintro}, all 
of which are less frequency-selective than phonon driving. It is easy to 
anticipate that the strongest effects of the driven phonon on the spin system 
will be found when $\omega_0$ matches the spectrum of triplon excitations. 

\begin{figure}[t]
\includegraphics[width=\columnwidth]{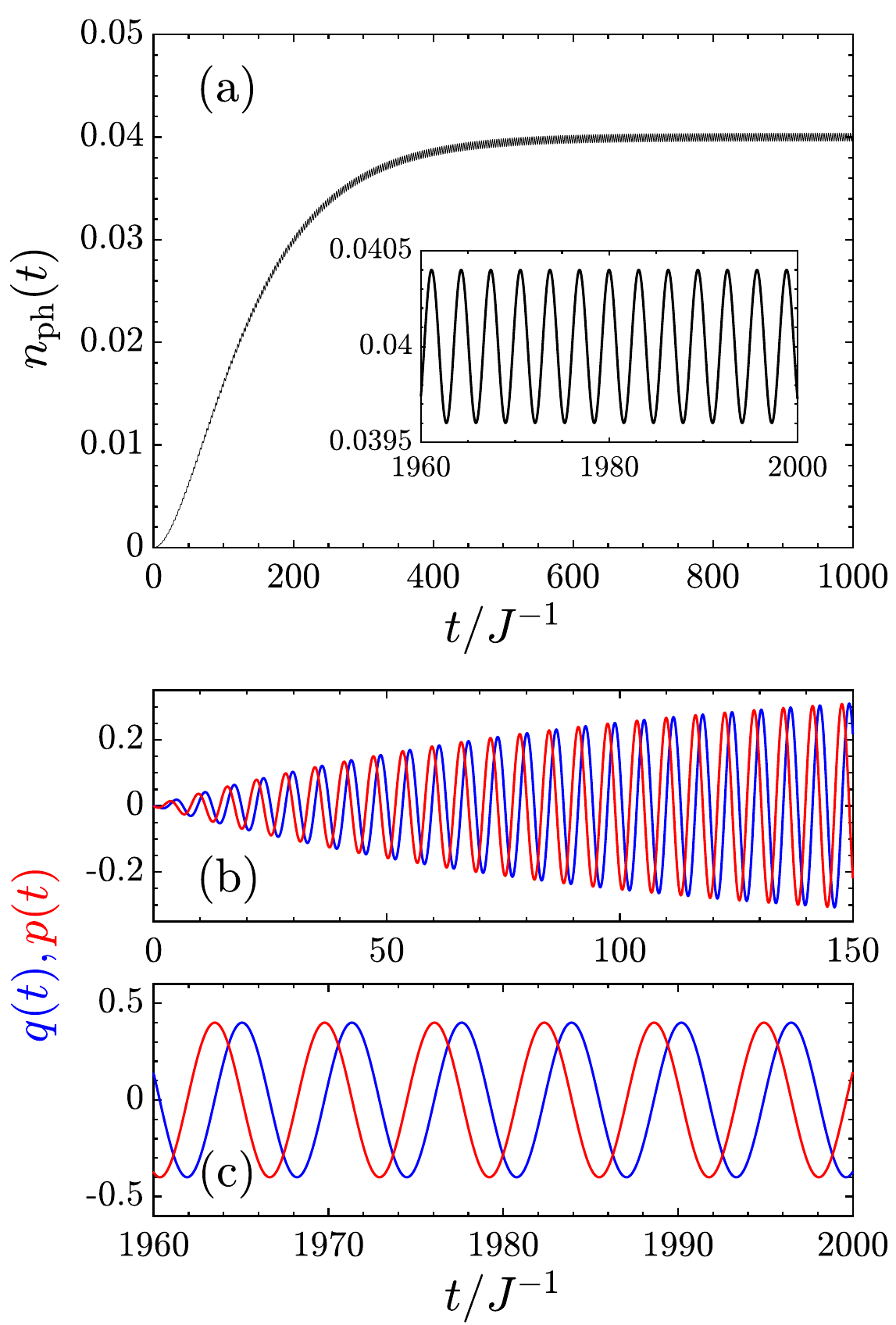}
\caption{Response of the Einstein phonon to a resonant driving field. Here 
$\omega_0/J = 1$, $a = 0.004 J$, $\gamma = 0.02 \omega_0$, and $g = 0$. (a) 
Phonon number, $n_{\rm ph}(t)$, produced by switching on a constant laser 
electric field at $t = 0$. The inset shows the steady state of the driven 
phonon system at long times. (b) Phonon displacement, $q(t)$, and momentum, 
$p(t)$, shown from $t = 0$. (c) $q(t)$ and $p(t)$ at long times.}
\label{fdcphonon}
\end{figure}

We consider first the driven phonon system without coupling to the spin 
chain, meaning with $g = 0$. To represent the phonons of a typical inorganic 
material we choose a damping coefficient $\gamma = 0.02 \omega_0$. From 
Eqs.~(\ref{eompa})--(\ref{eompc}) one observes that, up to a coupling to 
the spin system ($g$) that is typically below 10\%, the phonon has the 
behavior of a classical damped harmonic oscillator with a driving term. This 
is borne out by the time-dependence of the variables $q$, $p$, and $n_{\rm ph}$, 
shown in Fig.~\ref{fdcphonon}. Figure \ref{fdcphonon}(a) illustrates that the 
phonon number is driven up to a finite average value, and the inset that it 
oscillates steadily around this constant value for all later times; this is 
the NESS of the driven phonon system. Figures \ref{fdcphonon}(b) and 
\ref{fdcphonon}(c) show the corresponding behavior of the displacement and 
momentum, which have a relative $\pi/2$ phase difference. 

Several straightforward comments are in order. First, the phonon number 
operator in this laser-pumped steady state has been driven to a non-equilibrium 
average value of approximately 0.04. Although this value appears small, it 
does constitute a macroscopic occupation of a single mode. This driven 
$\omega_0$ phonon is the primary source of lattice excitations in the system, 
and all other phonon modes will have very low occupations at low temperatures. 
In all of the considerations to follow, we maintain the value of $n_{\rm ph}$ 
in this range, both for meaningful comparisons as other parameters are varied 
and for a realistic account of the temperature of the steadily driven system, 
as we discuss in Sec.~\ref{seh}.

Second, the frequency of the oscillations in the driven phonon occupation, 
$n_{\rm ph}(t)$, is twice that of $q(t)$ and $p(t)$, as expected from the 
number of nodes in the displacement cycle; the latter pair can be taken as 
the base frequency of the system, while the former is characteristic of $2 
\omega_0$, reflecting the fact that $n_{\rm ph}$ is essentially the sum of the 
squares of $q$ and $p$. Third, the characteristic timescale for convergence 
of the average of $n_{\rm ph}$ to the phonon NESS is $2/\gamma$ for all three 
quantities [Figs.~\ref{fdcphonon}(a) and (b)]. For $q(t)$ and $p(t)$, 
this is to be expected from the corresponding equations of motion 
[Eqs.~\eqref{eq:eom-phonon}], which contain explicit terms with prefactor
$-\gamma/2$, while for $n_{\rm ph}(t)$ it is the behavior of $p(t)$ on the 
right-hand side of Eq.~\eqref{eompc} that induces the same convergence rate.
Because the convergence is exponential, the actual establishment of a phonon 
NESS depends on the chosen accuracy criterion. 

\begin{figure}[t]
\includegraphics[width=\columnwidth]{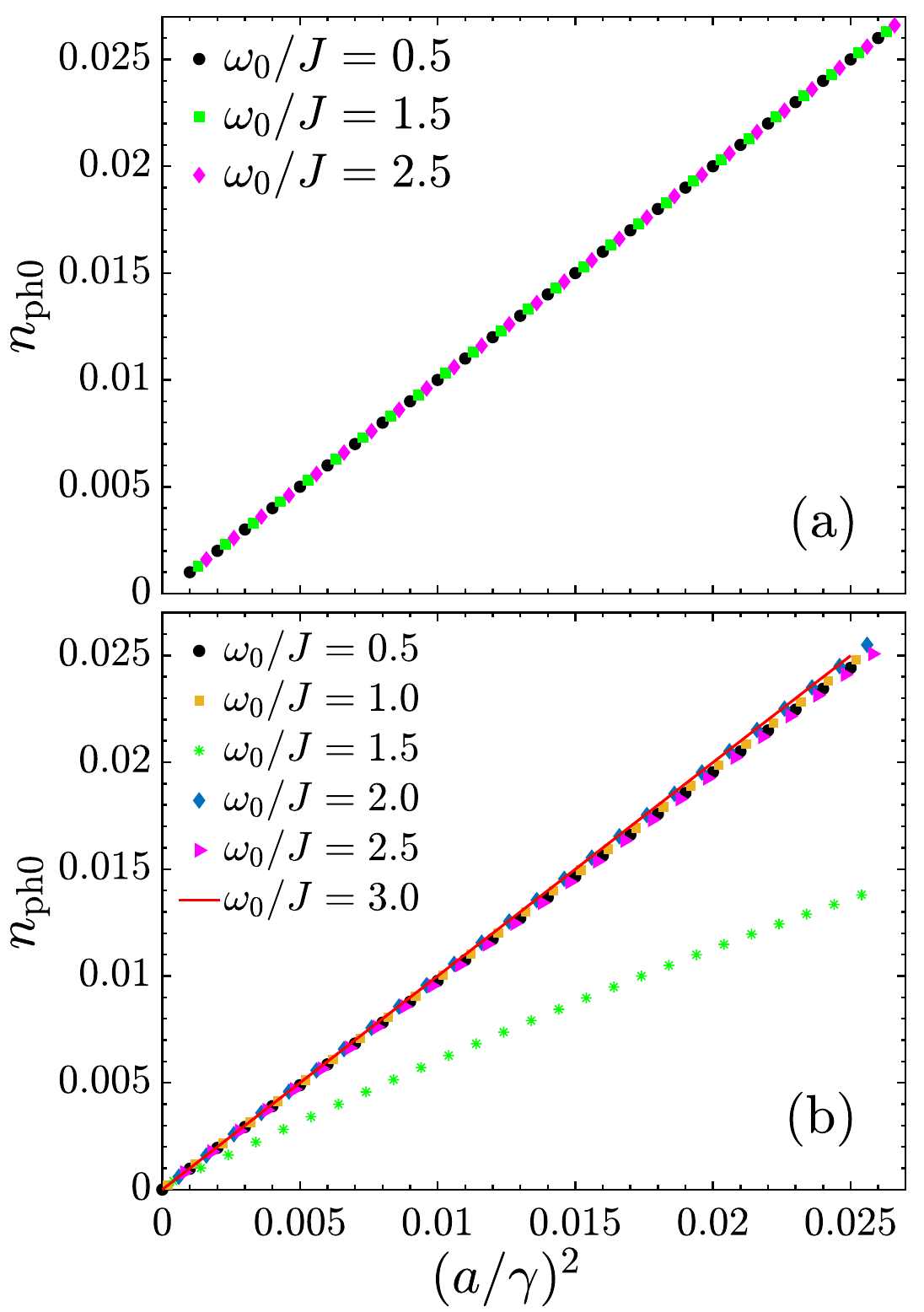}
\caption{Average value of the driven phonon occupation number, $n_{\rm ph0}$, 
in a NESS of the phonon system, displayed as a function of $a^2$ for various 
driving frequencies at fixed $\gamma = 0.02 \omega_0$ and $\gamma_{\rm s} = 
0.01 J$. (a) No coupling to the spin system ($g = 0$). (b) $g = 0.1 J$.}
\label{fdcnph}
\end{figure}

To a good approximation, the phonon number in the NESS [inset, 
Fig.~\ref{fdcphonon}(a)] is given by $n_{\rm ph} (t) = n_{\rm ph0} + n_{\rm ph2} 
\cos (2 \omega_0 t)$; we will investigate the corrections to this situation, 
which arise due to coupling to the spin system, in Sec.~\ref{sdp}. To study the 
quasi-stationary behavior of the NESS, we focus on the mean phonon occupation, 
$n_{\rm ph0}$. Figure \ref{fdcnph}(a) shows that, for all driving frequencies, 
the average energy in the driven phonon mode rises with the driving power, 
which is proportional to the square of the electric-field amplitude. In a 
fully classical treatment of the driven oscillator, $n_{\rm ph0} \propto 
(a/\gamma)^2$, and this result may equally be understood from Fermi's Golden 
Rule, where the flow of energy into the system is proportional to the square 
of the matrix element, and hence to $a^2$. We discuss the topic of energy flow 
in detail in Sec.~\ref{seh}. Because we have chosen for realism to scale 
$\gamma$ to the phonon frequency, $a$ will also be scaled to $\omega_0$ in 
all of the studies to follow, thereby maintaining a constant $(a/\gamma)^2$ 
when $\omega_0$ is varied.  

The spin system is driven by the pumped phonon through the coupling 
parameter $g$. Given that the amplitude of the phonon oscillation, $q(t)$, 
is proportional to $a/\gamma$, it follows that the amplitude of the induced 
driving of the spin system is proportional to $g a/\gamma$. Figure \ref{fdcnph} 
compares the driven phonon system with $g = 0$ to the situation with a finite 
value of $g$. Here we have chosen driving parameters suitable for the 
formation of NESS; those causing the spin system to inhibit NESS formation 
are the explicit focus of Sec.~\ref{str}. We observe in Fig.~\ref{fdcnph}(b) 
that a generic spin-phonon coupling, $g = 0.1 J$, results in only small changes 
being induced by the spin system relative to the isolated driven and damped 
phonons of most frequencies, but that some more significant alterations are 
possible at specific phonon frequencies, for reasons we investigate next. 

Turning now to the response of the driven spin system, it is necessary first 
to establish the nature and characteristic frequencies of the excitations 
created by the driving phonon. Throughout the present study, we will consider 
the dimerized $S = 1/2$ chain of Sec.~\ref{smm}A with an illustrative coupling 
ratio $\lambda = J'/J = 1/2$. Equation \eqref{ewk} states that the triplon 
modes of this chain form one triply degenerate branch dispersing from a value 
of $\omega_{\rm min} = J/\sqrt{2}$ at $k = 0$ to $\omega_{\rm max} = \sqrt{3/2} J$ 
at $k = \pi$. However, by spin conservation it is not possible for a phonon 
to create a single spin excitation, and from the form of $H_\text{sp}$ in 
Eq.~\eqref{ehp} it is evident that one phonon ($b_0^\dag$) couples to two spin 
excitations. One therefore anticipates that the strongest effects of the 
driving phonon on the spin system will be found when $\omega_0$ is chosen to 
lie within the band of two-triplon excitations, namely when $2\omega_{\rm min} 
\le \omega_0 \le 2\omega_{\rm max}$ ($1.414J \le \omega_0 \le 2.449J$). Thus 
an origin for the special behavior of the $\omega_0/J = 1.5$ phonon in 
Fig.~\ref{fdcnph}(b) is apparent immediately, although the detailed mechanism 
will not become clear until Sec.~\ref{str}. 

\begin{figure}[t]
\includegraphics[width=\columnwidth]{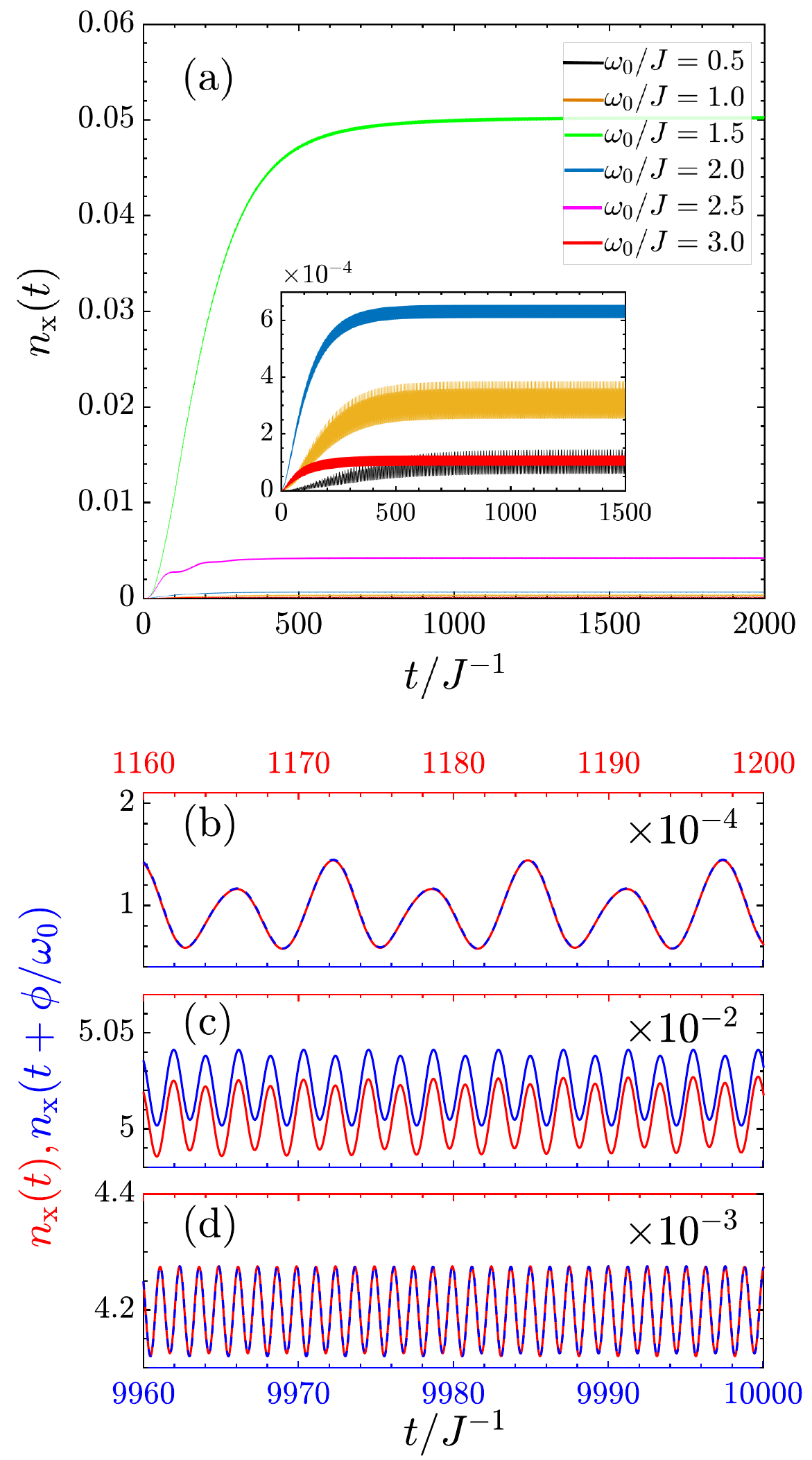}
\caption{(a) Response of the spin system, measured by $n_{\rm x}(t)$ 
[Eq.~\eqref{enx}], to driving phonon frequencies $\omega_0/J = 0.5$, 1.0, 
1.5, 2.0, 2.5, and 3.0. The driving field ensures that $a/\gamma = 0.2$ and 
we set $g = 0.1 J$, $\gamma_{\rm s} = 0.01 J$. $n_{\rm x}(t)$ in the spin NESS is 
shown at (b) $\omega_0/J = 0.5$, (c) $\omega_0/J = 1.5$, and (d) $\omega_0/J
 = 2.5$, where we compare results in the time window $1160 \le t \le 1200$, 
meaning after a small number of spin-system time constants, with those at 
$9960 \le t \le 10000$, meaning at truly long times.}
\label{fdcuw}
\end{figure}

In Fig.~\ref{fdcuw}(a) we choose six driving phonon frequencies below, in, 
and above the two-triplon band, and consider the amplitude of the perturbation 
transferred to the triplon system by the phonon for a spin-phonon coupling 
parameter $g = 0.1J$. We include a direct spin damping, $\gamma_{\rm s} = 0.01 
J$, which we scale to the energy of the spin system; to reflect the observed 
fact that the spin degrees of freedom are in general very weakly damped, we 
also adopt a value that is significantly lower than the phonon damping 
over most of the range of $\omega_0$. Figure \ref{fdcuw}(a) shows that laser 
driving at any frequency does create a response in the spin system that is 
qualitatively similar to that in the phonon system, namely that the spin 
occupation is ``pumped'' to a new average value, about which it oscillates. 
At constant ($a/\gamma$), the average triplon occupation, $n_{\rm x0}$, displays 
a hierarchy of values as the NESS is approached. While at frequencies far from 
the two-triplon band ($\omega_0/J = 0.5$, 1.0, and 3.0) this degree of driving 
produces only a very weak occupation, $n_{\rm x0} < 0.001$ [inset, 
Fig.~\ref{fdcuw}(a)], for frequencies in or near the band we find a state 
with $n_{\rm x0} \simeq 0.05$ at $\omega_0/J = 1.5$, but also one with an 
occupation of only $n_{\rm x0} \simeq 0.0006$ at the band center, 
$\omega_0/J = 2.0$.

Before discussing these occupation amplitudes, we demonstrate that each of 
the driven states is a true NESS. The detailed time structure, $n_{\rm x}(t)$, 
is shown for three selected frequencies in Figs.~\ref{fdcuw}(b) to 
\ref{fdcuw}(d). In each case we compare the triplon occupation in a time 
window near the center of Fig.~\ref{fdcuw}(a) with the long-time limit, for 
which we take the window $9960 \le t \le 10000$; we have shifted all the 
long-time traces by a phase $0 \le \phi < 2 \pi$ to start each cycle at the 
same point. The most important result of Figs.~\ref{fdcuw}(b) to \ref{fdcuw}(d) 
is to prove that the driven model system damped by $\gamma$ and $\gamma_{\rm s}$ 
does indeed host spin NESS, in that identical periodic traces are obtained for 
arbitrarily long times. The subsidiary result is that, for most $\omega_0$ 
values, a good approximation to the NESS is reached already at rather short 
times. Because convergence is exponential, any meaningful accuracy criterion 
will be reached after a single-digit number of time constants, and thus for 
quantitative purposes (Sec.~\ref{seh}), bearing experimental uncertainties 
in mind, we define a NESS to exist using a relative criterion of 2\% 
(corresponding to approximately $4/\gamma_{\rm s}$). According to this 
criterion, the driven state in Fig.~\ref{fdcuw}(c) is not yet a NESS, for 
reasons we will revisit below, but those shown in Figs.~\ref{fdcuw}(b) and 
\ref{fdcuw}(d) are. 

As a benchmark for the meaning of the $n_{\rm x0}$ values in Fig.~\ref{fdcuw}, 
one may compare with the value $n_{\rm b0} = 0.028$ deduced below 
Eq.~\eqref{enbe} (Sec.~\ref{smm}B), which expressed the mixing of dimer 
singlet and triplet states due to the quantum spin fluctuations in the 
pure spin chain. Thus by inspection of the average non-equilibrium triplon 
populations characterized by $n_{\rm x0}$, one may state that the spin NESS 
established at low and high frequencies constitute only a weak perturbation 
of the equilibrium state. This result also implies that in the ``Floquet'' 
regime of frequencies above the two-triplon band, the spin state is not 
altered qualitatively, although it may obtain a nontrivial phase structure. 
By contrast, for some frequencies in and around the two-triplon band, the 
quantum spin NESS can be altered significantly from the equilibrium state, 
and our results for $\omega_0/J = 1.5$ suggest that rather modest phonon 
driving at certain frequencies can create an essentially different type of 
triplon system. We will characterize these qualitatively new states in detail 
in Sec.~\ref{sdp}. 

Here we note that the hard-core nature of the dimer spin states sets an 
absolute upper limit of $n_{\rm x} = 1$ on the triplon occupation, and in fact 
such a situation would represent the most extreme out-of-equilibrium state 
possible, at which many of the approximations in Sec.~\ref{smm}A would no 
longer be valid. Anticipating the discussion of Secs.~\ref{sdp} and 
\ref{str}A, we introduce an operational threshold value of $n_{\rm x} (t)$ 
in the driven spin state, such that our description of the spin sector 
will remain appropriate, and we set this to $n_{\rm x}^{\rm max} = 0.2$.

In addition to the order-of-magnitude differences observed in $n_{\rm x0}$ 
as a consequence of the driving frequency, Fig.~\ref{fdcuw} invites two 
further remarks. First, we observe that the time structure, $n_{\rm x}(t)$, 
of the NESS in Figs.~\ref{fdcuw}(b) to \ref{fdcuw}(d) shows a rather complex 
form, with a definite superposition of different frequency harmonics in 
evidence. We will investigate this harmonic mixing, which appears to be 
strongest at the below-band frequency of Fig.~\ref{fdcuw}(b), in detail in 
Sec.~\ref{sdp}. Second, the timescale over which the spin system reaches its 
NESS appears to be similar at all frequencies, other than $\omega_0/J = 0.5$ 
and 1.5, at $t \approx 400 J^{-1}$. This value corresponds to 4-5 time constants 
of the spin system ($1/\gamma_{\rm s}$). Of the exceptional cases, at 
$\omega_0/J = 0.5$, where $\gamma_{\rm s} = \gamma$, the process is somewhat 
delayed by the phonon ``switch-on'' timescale (Fig.~\ref{fdcphonon}). At 
$\omega_0/J = 1.5$, the process appears to be longer still, with the NESS 
not yet fully established after $t = 1200 J^{-1}$ [Fig.~\ref{fdcuw}(c)]. We 
will investigate the transient behavior of the spin system at switch-on, and 
explain this curiously slow convergence, in Sec.~\ref{str}A. 

\begin{figure}[t]
\includegraphics[width=\columnwidth]{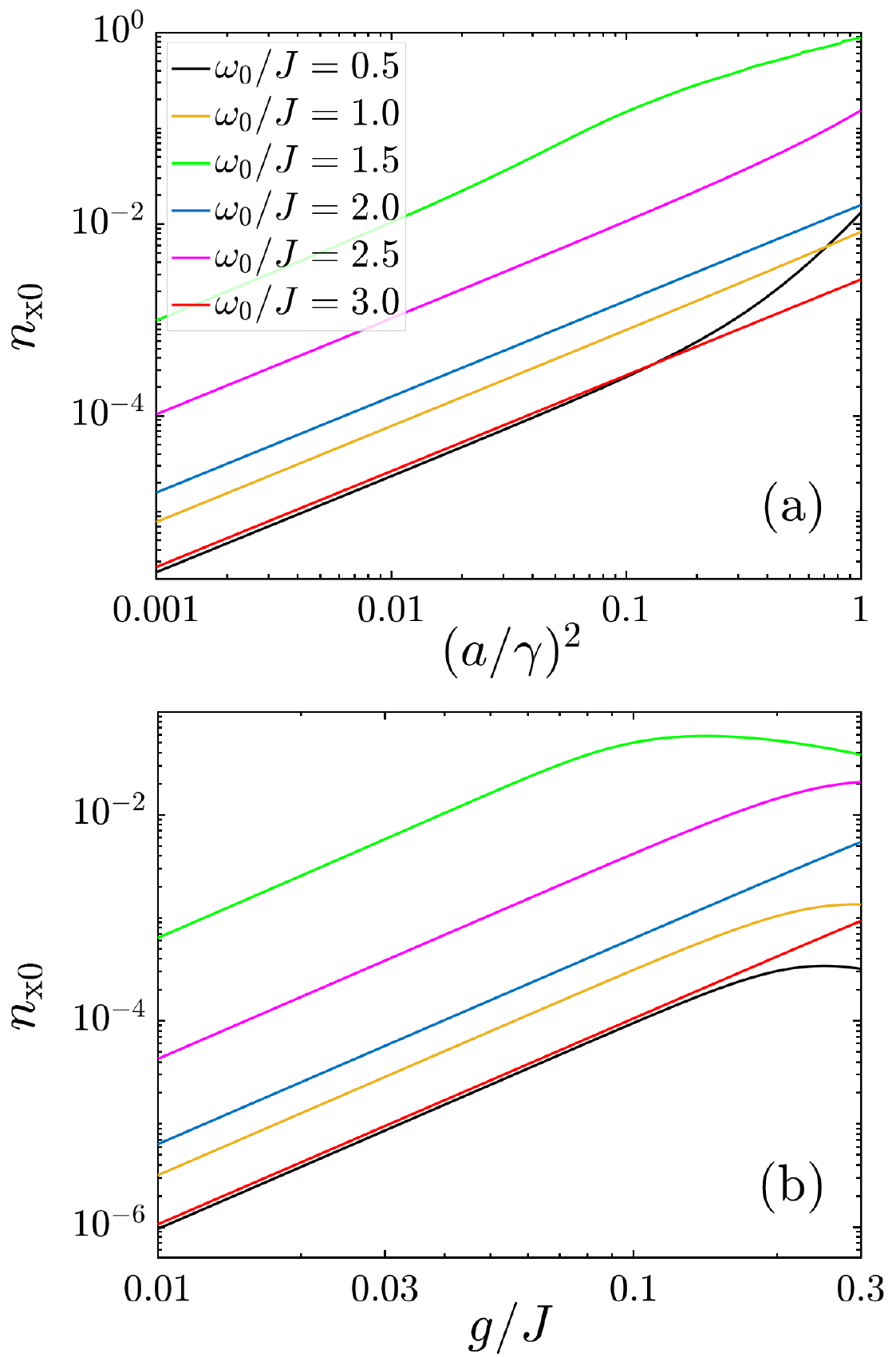}
\caption{(a) Dependence of the average triplon number, $n_{\rm x0}$, in the 
NESS on the fluence, shown as $(a/\gamma)^2$, at driving phonon frequencies 
$\omega_0/J = 0.5$, 1.0, 1.5, 2.0, 2.5, and 3.0. The fixed system parameters 
are $g = 0.1 J$, $\gamma = 0.02 \omega_0$, and $\gamma_{\rm s} = 0.01 J$. Only 
the $\omega_0/J = 0.5$ and 1.5 phonons at very high fluences show deviations 
from a linear form. (b) Dependence of $n_{\rm x0}$ on the spin-phonon coupling 
constant, $g$, for driving phonons of the same six frequencies at fixed 
$a/\gamma = 0.2$. A well-defined $g^2$ dependence at all small couplings 
gives way to a suppression of $n_{\rm x0}$ at larger $g$ values whose onset 
depends on $\omega_0$.} 
\label{fdcfd}
\end{figure}

We conclude our initial survey of spin NESS in response to a driving phonon 
by showing the spin-system analog of Fig.~\ref{fdcnph}. In the analysis of 
experiments, a key quantity in characterizing any phenomenon is its dependence 
on the power, or fluence, of the laser, which is quite straightforward to 
measure. From elementary electrodynamics, the fluence is proportional to the 
squared amplitude of the laser field, and hence in Fig.~\ref{fdcfd}(a) we show 
the dependence on $a^2$ of the average triplon occupation, $n_\text{x0}$, in 
the NESS for the six representative driving frequencies. As for the driven 
phonon, the dependence is clearly linear over the full range of 
$\gamma$-normalized $a^2$ values for all driving frequencies, again except 
for $\omega_0/J = 0.5$ and $\omega_0 /J = 1.5$. The latter shows a saturation 
as $n_\text{x0}$ is driven towards unphysical values at very large $a$, while 
the former shows a crossover to a dependence that it as least quadratic in 
$(a/\gamma)^2$ at strong driving. Next (Sec.~\ref{sdp}) we discuss the 
dynamical properties of the driven NESS, which will allow us to understand 
the origin of this form, after which (Sec.~\ref{str}) we will address the 
issue of limits on $(a/\gamma)^2$ for spin NESS to exist at long driving times. 

In Fig.~\ref{fdcfd}(b) we show the dependence of the driven triplon occupation 
on the spin-phonon coupling, $g$, for the same six driving frequencies. At low 
values of $g$, $n_\text{x0}$ shows a $g^2$ form that is directly analogous to 
its dependence on $a^2$. However, at high $g$ we observe a suppression of 
$n_\text{x0}$ below its expected value, the onset of which occurs at lower 
$g$ for the phonons closest to resonance with the two-triplon band, and find 
that the spin response can even decrease as the coupling is increased. This 
onset of more complex behavior, which is also evident in the response of the 
$\omega_0/J = 1.5$ phonon in Fig.~\ref{fdcnph}(b), allows us to define a regime 
of ``weak'' (or ``linear'') spin-phonon coupling, which terminates around 
$g = 0.08 J$, and a ``strong-coupling'' regime. Most magnetic quantum materials 
do not show strong spin-phonon coupling at equilibrium, and thus for the 
purposes of the present analysis, which is to discuss the properties of a 
generic driven quantum magnet, we will focus on the weak-coupling regime. 
Hence we adopt the value $g = 0.05 J$ to be representative of the class of 
magnetic materials in which to seek linear quantum spin NESS phenomena. 

In this weak-coupling regime, one may exploit the equivalence of $a$ and $g$ 
to define a dimensionless effective driving parameter for the spin system, 
\begin{equation}
\label{eq:p-def}
D = g a / (\gamma J),
\end{equation}
which can be used to simplify the analysis, and we will employ this 
parameterization in Sec.~\ref{str}. However, when working beyond this regime 
it is not possible to avoid studying the full space of $a/\gamma$ and $g$. 
Although we defer the analysis of strong coupling to a later study, we stress 
that all of the treatment in Sec.~\ref{smm} remains fully valid for all the 
$g$ values shown in Fig.~\ref{fdcfd}(b). Nevertheless, as we will mention in 
Sec.~\ref{sd}, values of $g$ up to $0.5 J$ are known in some dimerized-chain 
compounds, and for such extreme spin-phonon coupling one may not exclude the 
possibility of a different type of physics at equilibrium, such as the 
formation of combined phonon-triplon entities; we comment only that the 
formalism of Sec.~\ref{smm} would not be appropriate for such a situation.

\section{Dynamical properties of the quantum spin NESS}
\label{sdp}

We turn now to a quantitative analysis of the dynamics of the spin NESS. It 
is already clear from Sec.~\ref{sness}, and particularly Fig.~\ref{fdcuw}, 
that the superposition of frequencies present in the steady state can be 
complex. For full insight into the harmonic content of the spin NESS, we 
introduce the Fourier transform (FT) of the NESS signal, which we apply to 
$n_{\rm ph}(t)$, to the individual spin components, $u_k(t)$ and $v_k(t)$, and 
to the summed quantities $n_{\rm x}(t)$ [Eq.~\eqref{enx}] and 
\begin{equation}
V(t) = \frac{1}{N} \sum_k v_k(t),
\label{ev}
\end{equation}
which characterize respectively the average triplon occupation and the average 
behavior in the off-diagonal two-triplon sector. The definition of the FT is 
simplied by making use of the results in shown in Fig.~\ref{fdcuw}, where we 
demonstrated that NESS had been achieved at long times. We use one cycle of 
the signal taken from the time window $9960 \le t/J^{-1} \le 10000$ to determine 
the coefficients of the Fourier series
\begin{equation}
X(t) = \sum_m X_m \exp (i m \omega t)
\label{eft}
\end{equation}
for any quantity $X$ appearing in a NESS driven by any frequency $\omega$; 
with this notation, any quantity with an integer subscript ($X_m$) denotes 
a Fourier component, and those with $m = 0$ are all real numbers. Without 
performing a detailed analysis beyond the level of Fig.~\ref{fdcuw}, we 
comment that the system described by the model of Sec.~\ref{smm} does not 
generate any significant dynamics at frequencies other than $m \omega$, 
where $m$ is an integer. We also comment that there are no discernible 
extrinsic features arising in the FT as a consequence of the finite length 
of the chain on which we perform our calculations. 

\begin{figure}[t]
\includegraphics[width=\columnwidth]{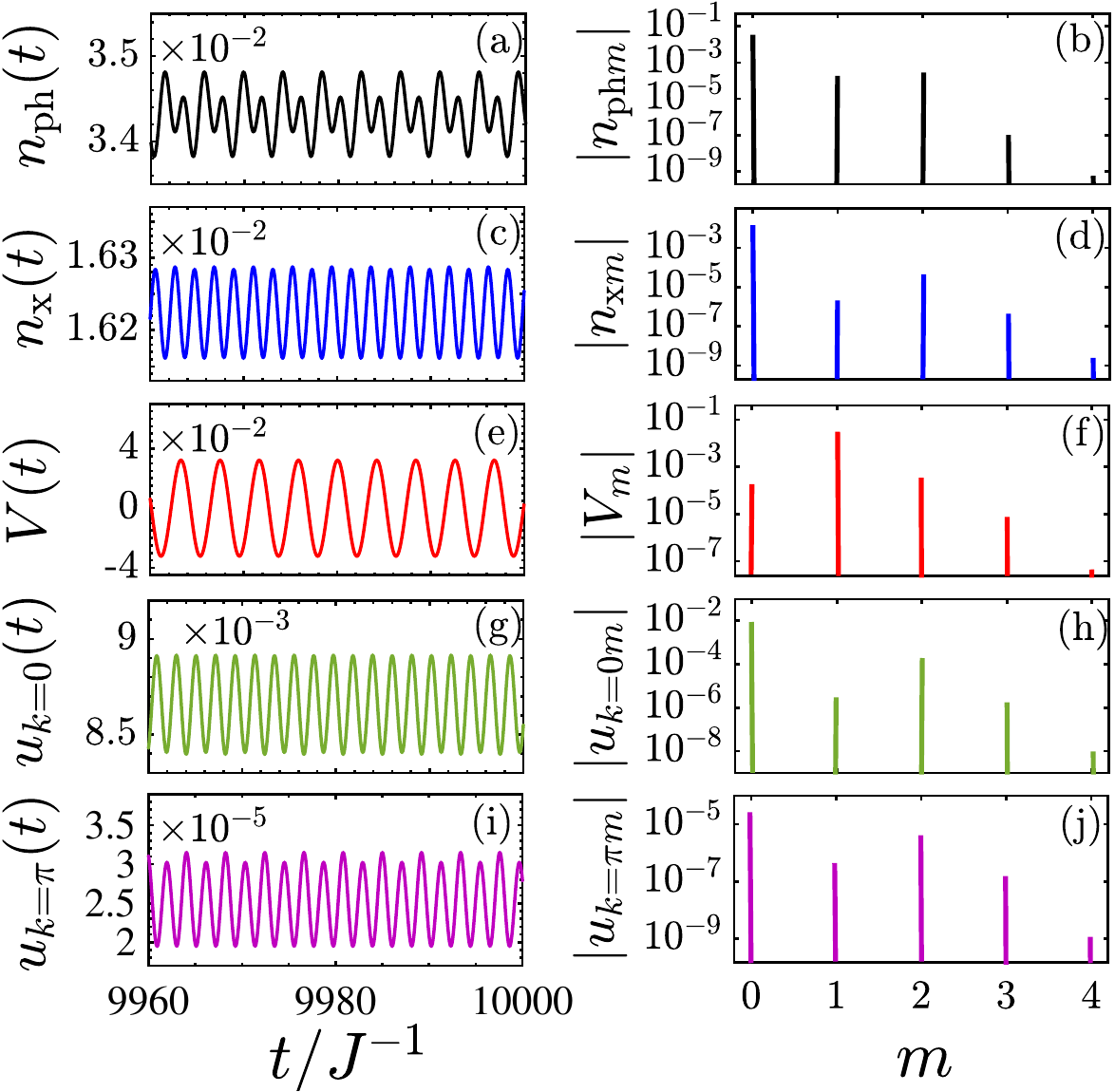}
\caption{Illustration of the quantities (a) $n_{\rm ph}(t)$, (c) $n_{\rm x}(t)$, 
(e) $V(t)$, (g) $u_{k = 0}(t)$, and (i) $u_{k = \pi}(t)$ in the NESS obtained with 
driving parameters $a/\gamma = 0.2$ and $g = 0.05 J$ at frequency $\omega_0/J
 = 1.5$ in the presence of spin damping $\gamma_{\rm s} = 0.01 J$. Panels (b), 
(d), (f), (h), and (j) show the corresponding Fourier decompositions.} 
\label{fdcft}
\end{figure}

Returning to the case of resonant driving ($\omega = \omega_0$), we illustrate 
the FT in Fig.~\ref{fdcft} by showing in the left panels the time structure of 
$n_{\rm ph}(t)$, $n_{\rm x}(t)$, $V(t)$, and the single-$k$ components $u_{k = 0} 
(t)$ and $u_{k = \pi}(t)$ in the NESS of Fig.~\ref{fdcuw} at $\omega_0/J = 1.5$; 
juxtaposed in the right panels are the corresponding harmonic decompositions
determined from Eq.~\eqref{eft}. We have chosen a relatively conventional 
NESS trace [Fig.~\ref{fdcft}(c), similar to Fig.~\ref{fdcuw}(c)], in which 
$n_{\rm x}(t)$ and $u_k(t)$ are dominated by the even Fourier components $m
 = 0$ and 2, while $V(t)$ [and by extension $v_k(t)$] is dominated by $m
 = 1$. This result is quite natural if one considers the equations of motion 
[Eqs.~\eqref{eq:eoms}], taking $q(t)$ to be a sinusoidal driving with small 
amplitude and a frequency $\omega_0$. At leading order in $q(t)$, all 
components $v_k(t)$ and $w_k(t)$ will also oscillate at this same frequency, 
giving a dominant $m = 1$ component, while the leading-order response in 
$u_k(t)$ oscillates at $2\omega_0$ and possesses a constant offset (a zeroth 
harmonic). Because $n_\text{x}(t)$ is the sum over all $u_k(t)$, it therefore 
shows harmonic components primarily at $m = 0$ and 2. All of these features 
are evident in Figs.~\ref{fdcft}(c)-\ref{fdcft}(j). In addition, we observe 
that the different $k$-components of $u_k(t)$ display different harmonic 
contributions, and because $\omega_0/J = 1.5$ excites triplons closer to 
the band minimum, the $m = 0$ and 2 coefficients are larger at $k = 0$ than 
at $k = \pi$; one may verify (data not shown) that the converse is true at 
a driving frequency of $\omega_0/J = 2.5$, and we consider the $k$-dependence 
of the response in more detail below. 

\begin{figure}[t]
\includegraphics[width=\columnwidth]{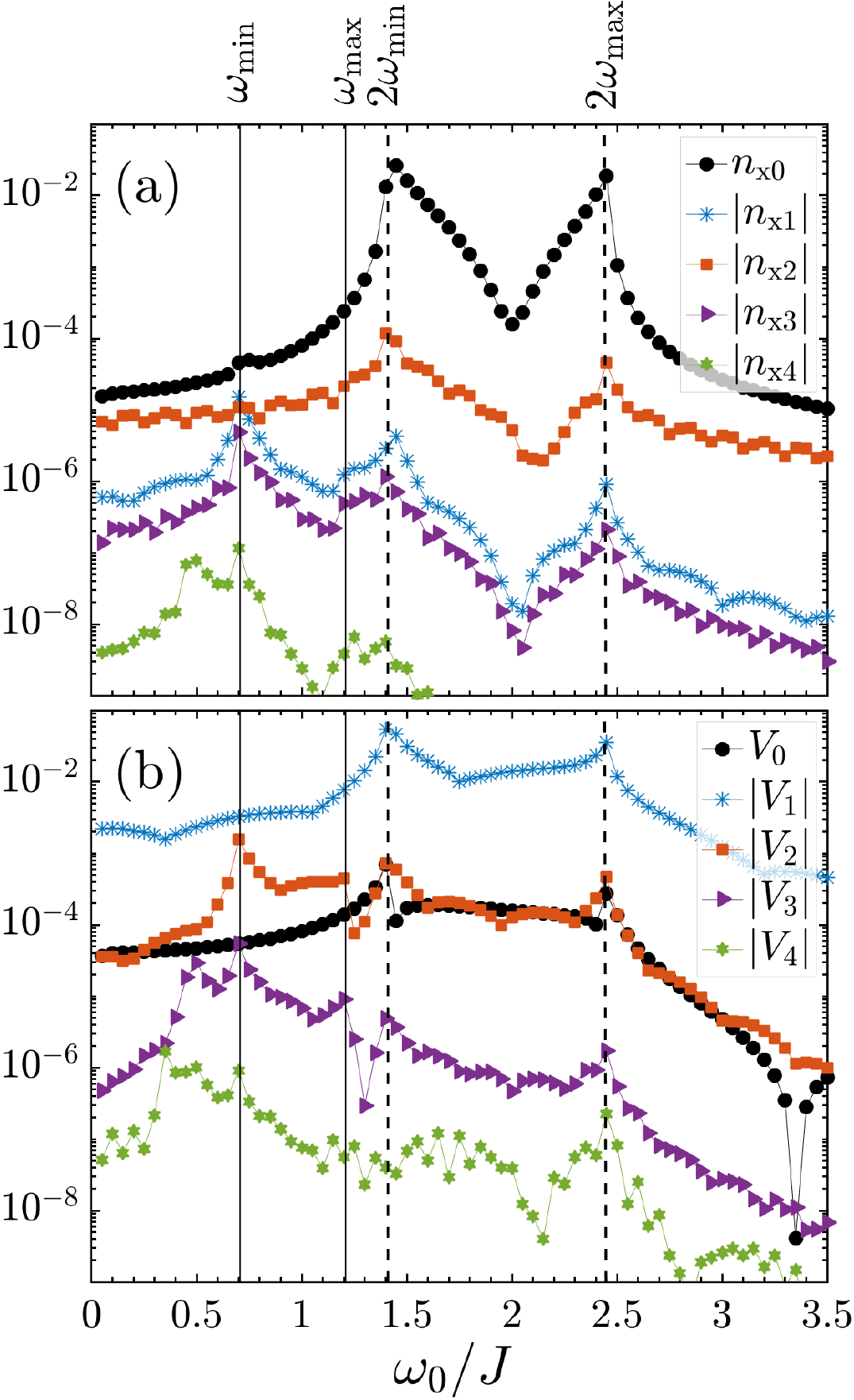}
\caption{Coefficients of the Fourier transforms of (a) $n_{\rm x}(t)$ and 
(b) $V(t)$ in the NESS obtained with driving $a/\gamma = 0.2$ and $g = 0.05 J$, 
shown as a function of the driving phonon frequency, $\omega_0$, for damping 
parameters $\gamma = 0.02 \omega_0$ and $\gamma_s = 0.01 J$.}
\label{fdcftc}
\end{figure}

Given this conventional behavior of the spin NESS, it is somewhat surprising 
to observe the presence of a significant $m = 1$ harmonic in the phonon NESS, 
$n_{\rm ph} (t)$, of Fig.~\ref{fdcft}(a). In fact $n_{\rm ph0}$ is suppressed by 
14\% compared to its $g = 0$ value (Fig.~\ref{fdcphonon}), which is a weaker 
version of the effect visible for the $\omega_0/J = 1.5$ phonon in 
Fig.~\ref{fdcnph}(b). The presence of the $m = 1$ harmonic is another 
consequence of strong feedback from the spin system at this ``resonant'' 
(in-band) frequency, and arises from the term $g \, \mathcal{U}(t) p(t)$ in 
Eq.~\eqref{eompc}, where $\mathcal{U}(t)$ oscillates primarily at $2 \omega_0$ 
and $p(t)$ at $\omega_0$. It is also clear that additional harmonics are 
present in the spin NESS analyzed in Fig.~\ref{fdcft}, including at higher 
multiples of $\omega_0$, and one may anticipate [not least from 
Fig.~\ref{fdcuw}(b)] that for certain frequencies they are significant. 

To investigate the effect of the frequency of the driving phonon, in 
Fig.~\ref{fdcftc} we show the coefficients of $n_x(t)$ and $V(t)$ from $m =
 0$ to 4 as a function of $\omega_0$. Across the full range of frequencies, 
$n_{\rm x}(t)$ is indeed dominated by the $m = 0$ and 2 coefficients 
[Fig.~\ref{fdcftc}(a)] and $V(t)$ by $m = 1$ [Fig.~\ref{fdcftc}(b)], meaning 
that the case study of Fig.~\ref{fdcft}, performed for $\omega_0/J = 1.5$, is 
in fact well representative of the hierarchy of coefficient values, with only 
one significant exception. This is the frequency range around $\omega_0 = 
\omega_{\rm min}$, where a clear peak appears in a number of the harmonic 
components. Although frequencies around $\omega_0/J = 0.7$ are far from a 
direct resonance, their second harmonic ($2\omega_0/J = 1.4 J$) coincides with
the peak density of states at the two-triplon band minimum. Inspection of 
Eqs.~(\ref{eq:eoms}b) and (\ref{eq:eoms}c) reveals that oscillations are 
indeed induced at $2\omega_0$ because $q(t)$ is multiplied by $v_k(t)$ or 
$w_k(t)$. While this process appears at next-to-leading order in $q(t)$, 
it is strongly enhanced when the second harmonic satisfies the resonance 
condition.

The resonantly enhanced second harmonic of $w_k(t)$ in turn induces stronger 
first and third harmonic components in $u_k(t)$, as may be read from 
Eq.~(\ref{eq:eoms}b), in which $w_k(t)$ is multiplied by $q(t)$ where it acts 
as a driving term for $u_k(t)$ at frequencies $(2 \pm 1) \omega_0$. This type 
of harmonic mixing results, for the driving we consider in Fig.~\ref{fdcftc}, 
in the coefficient $|n_{\rm x1}|$ even exceeding $|n_{\rm x2}|$ around $\omega_0
 = \omega_{\rm min}$, where $|n_{\rm x3}|$ is also strongly enhanced. Similarly, 
$|V_2|$ and $|V_3|$ are also enhanced over a wide frequency range around 
$\omega_0 = \omega_{\rm min}$, where at its peak $|V_2|$ approaches $|V_1|$. 
Thus the resonant enhancement of the second harmonic explains why the temporal 
behavior of the spin NESS displays more and different features at below-band 
frequencies around $\omega_0/J = 0.5$ [Fig.~\ref{fdcuw}(b)] than it does for 
the cases $\omega_0/J = 1.5$ [Figs.~\ref{fdcuw}(c) and \ref{fdcft}] and 
$\omega_0/J = 2.5$ [Fig.~\ref{fdcuw}(d)].

We comment briefly on the physical meaning of the ``frequency-doubling'' 
effects that cause the enhancement of so many Fourier components around 
$\omega_0 = \omega_{\rm min}$. First, it is important to stress that the 
response observed at $2 \omega_0$ in $n_{\rm x} (t)$ is not a doubling 
phenomenon; it is merely a consequence of the fact that the triplon number 
is an operator square of the triplon degree of freedom, and in this sense 
the behavior of $u_k(t)$, $v_k(t)$, and $w_k(t)$ is directly analogous to 
that of the driven phonon variables discussed in Sec.~\ref{sness}. By 
contrast, the frequency-doubling observed between phonon driving at 
$\omega_0 = \omega_{\rm min}$ and the strong response of the spin system at 
$2 \omega_{\rm min}$ is a real effect, which at a ``classical'' level can 
be read directly from the equations of motion. At a quantum level, this 
frequency-doubling requires the involvement of two phonons at frequency 
$\omega_0$, taking part in off-shell phonon-triplon processes that are 
allowed in the strongly out-of-equilibrium system. 

We stress again that all physical processes of this type [meaning those 
contained in Eqs.~\eqref{eq:eoms}] do involve multiple driving phonons, as is 
standard in Floquet physics. Our treatment of the lattice system does not allow 
for the creation of phonons with frequencies of $2 \omega_0$, $3 \omega_0$, or 
higher due to anharmonicities in the lattice potential, as was discussed in 
Refs.~\cite{Foerst11,rscg}. Because the factors enhancing multi-phonon response 
and harmonic mixing (Figs.~\ref{fdcft} and \ref{fdcftc}) are the same, it is no 
surprise to find that both phenomena are strongest in the same range of 
frequencies. Quantitatively, the strength of these subdominant signals at 
constant $a/\gamma$ is a product of powers of $g$ with the height of the 
density-of-states peak at $2 \omega_{\rm min}$, and the enhancement can 
exceed an order of magnitude at $\omega_0 = \omega_{\rm min}$.

\begin{figure}[t]
\includegraphics[width=\columnwidth]{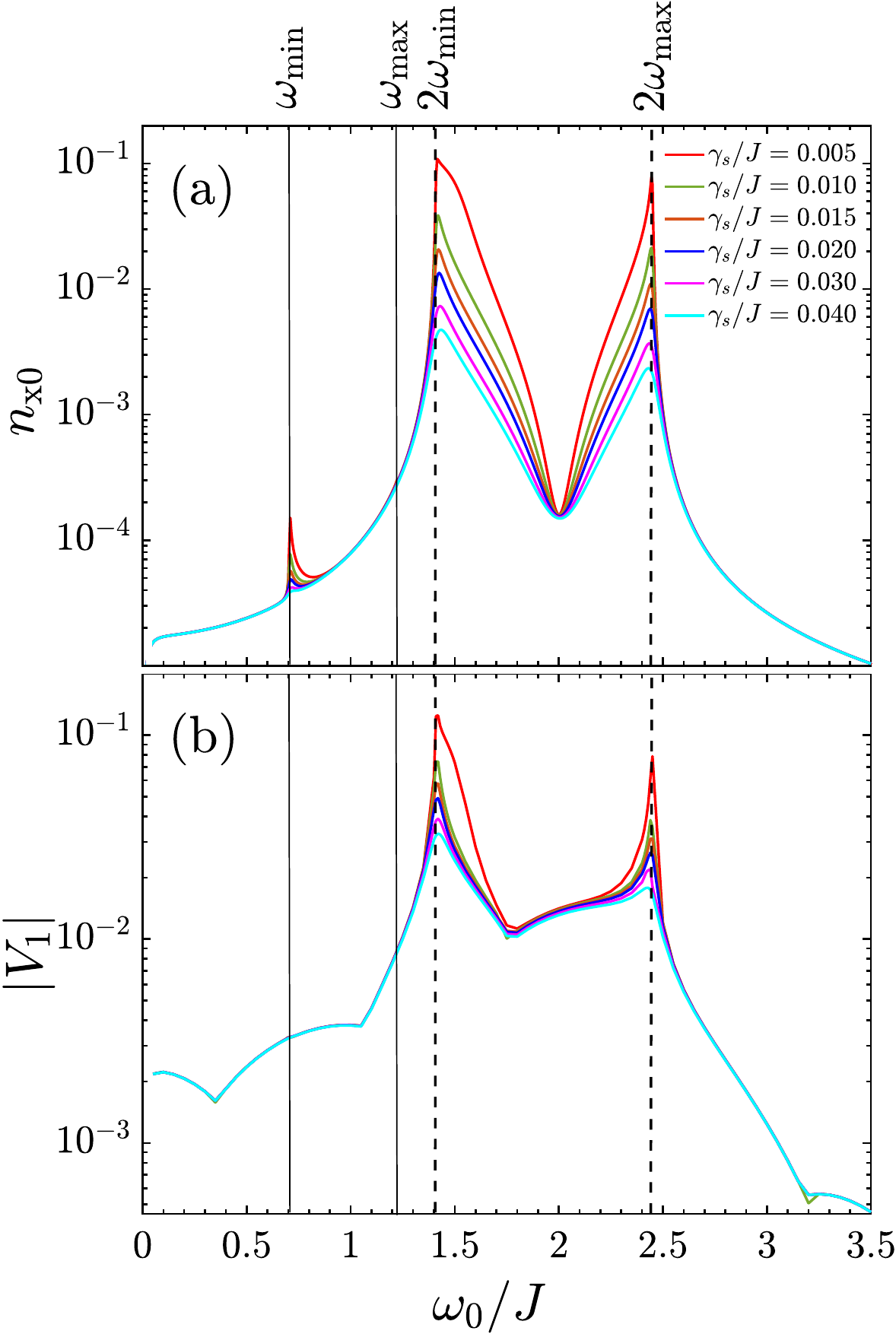}
\caption{(a) Average triplon occupation, $n_{\rm x0}$, in the NESS obtained with 
driving $a/\gamma = 0.2$ and $g = 0.05$, shown on logarithmic axes as a 
function of $\omega_0$ for different values of $\gamma_{\rm s}$. The band-edge 
features become increasingly prominent as $\gamma_{\rm s}$ decreases, as does 
the peak at $\omega_0 = \omega_{\rm min}$, but for most other phonon frequencies 
$n_{\rm x0}$ is quite insensitive to the spin damping. (b) Corresponding 
off-diagonal response, shown by the quantity $|V_1|$.}
\label{fdchpw}
\end{figure}

Turning to the physical quantities characterizing the NESS, we have seen in 
Sec.~\ref{sness}, and see again in Fig.~\ref{fdcftc}, that the response of 
the spin system is very sensitively dependent on the driving frequency, with 
clearly different adiabatic, antiadiabatic, and ``resonant'' (by which is 
meant in-band) forms. However, some in-band frequencies are not particularly 
remarkable, due to small matrix elements or low densities of two-triplon 
states, and some adiabatic frequencies clearly have rather strong anomalous 
(multiphonon) enhancement. For a quantitative visualization of this response, 
in Fig.~\ref{fdchpw}(a) we show the mean amplitude, $n_{\rm x0}$, of the driven 
triplon occupation and in Fig.~\ref{fdchpw}(b) the amplitude of the 
off-diagonal response, which we gauge using $|V_1|$. The rising lines indicate 
decreasing values of $\gamma_{\rm s}$, which we terminate at $\gamma_{\rm s} = 
0.005 J$ to avoid having $n_{\rm x0}$ exceed $n_{\rm x}^{\rm max} = 0.2$, thereby 
allowing NESS formation at all frequencies for the chosen driving parameters. 
At frequencies far from a resonance with the edges of the band, $n_{\rm x0}$ is 
surprisingly insensitive to $\gamma_{\rm s}$ [Fig.~\ref{fdchpw}(a)]. However, 
as $\omega_0$ approaches $2\omega_{\rm min}$ and $2\omega_{\rm max}$, the driven 
$n_{\rm x0}$ varies strongly with $\gamma_{\rm s}$, and the same is true around 
$\omega_0 = \omega_{\rm min}$. 

As Fig.~\ref{fdchpw}(b) makes clear, analogous effects are present at 
$\omega_0 = 2\omega_{\rm min}$ and $2\omega_{\rm max}$ in $|V_1|$, which also 
rises to values of order 0.1 at the lower band edge for $\gamma_{\rm s} = 
0.005 J$ but is essentially independent of $\gamma_{\rm s}$ for driving 
frequencies more than $0.1 J$ outside the two-triplon band. Because we 
have chosen $|V_1|$ as the off-diagonal diagnostic, and this is a primary 
driving term in Eqs.~\eqref{eq:eoms} rather than a driven term, there is no 
$\gamma_{\rm s}$-dependence around $\omega_0 = \omega_{\rm min}$; this response 
of the off-diagonal sector is found rather in the coefficients $|V_2|$ and 
$|V_3|$ in Fig.~\ref{fdcftc}. The other differences in the frequencies of 
characteristic features in $|V_1|$, most notably the in-band minimum occurring 
at $\omega_0/J = 1.7$ rather than 2.0, may be traced to the leading dependence 
in Eq.~\eqref{eomsv} on the coefficient $y_k$ in Eq.~\eqref{eyk} as opposed to 
$y'_k$ in Eq.~\eqref{eypk}.

\begin{figure}[t]
\includegraphics[width=\columnwidth]{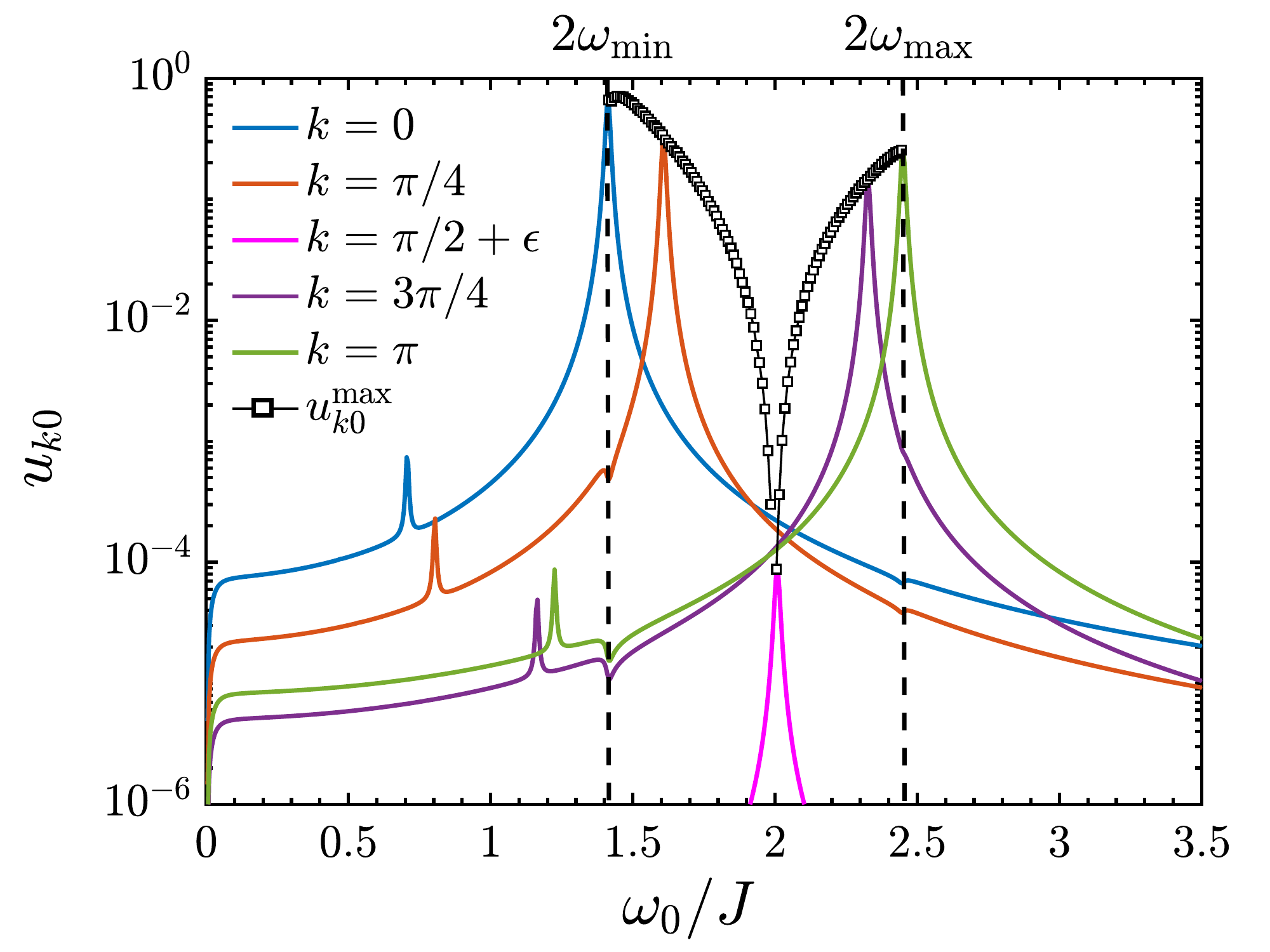}
\caption{Wave-vector-resolved average triplon occupation, $u_{k0}$, in the NESS 
obtained with driving $a/\gamma = 0.2$ and $g = 0.05 J$, shown as a function of 
the driving frequency, $\omega_0$, for $k = 0$, $\pi/4$, $\pi/2 + \epsilon$, 
$3\pi/4$, and $\pi$. Black squares show the maxima, $u_{k0}^{\rm max}$, of the 
$u_{k0}$ functions peaking at different energies across the Brillouin zone, 
which defines the wave vectors $k_{\rm res}$. At all frequencies $\omega_0 < 2 
\omega_{\rm min}$ the strongest peak is found in $u_{k=0}$ and at $\omega_0 > 2 
\omega_{\rm max}$ in $u_{k=\pi}$. $\epsilon = \pi/N$ is an offset from the band 
center, where $u_{k=\pi/2} = 0$.}
\label{fdchpk}
\end{figure}

To understand the degree to which individual $k$-components of the spin 
system are selected by the phonon driving, in Fig.~\ref{fdchpk} we show 
$u_{k0}$ over the full range of driving frequencies for selected values of 
$k$ across the Brillouin zone. For $k = 0$ and $\pi$, it is no surprise 
that the respective $u_{k0}$ functions peak strongly at $\omega_0 = 2 
\omega_{\rm min}$ and $2 \omega_{\rm max}$, because these are the dominant  
available wave-vector components; we note that there is no problem with 
the fact that $u_{k=0}(t)$ exceeds the threshold when the system is driven at 
$\omega_0 = 2 \omega_{\rm min}$, because the triplon occupation is determined 
by the average over all components [Eq.~\eqref{enx}]. For driving frequencies 
within the two-triplon band, one might expect a broad spin response on the 
grounds that triplon pairs from a wide range of wave vectors may contribute. 
However, the response at each frequency remains dominated by the resonance 
condition $\omega_0 = 2 \omega_k$, and thus the components $u_{k0}$ for $k = 
\pi/4$ and $3\pi/4$ continue to show sharp peaks (which fall by one order of 
magnitude over an energy range of 10\% of the band width). Thus $k$-selection 
on the basis of the driving energy is rather accurate and it is well justified 
to introduce a ``resonant'' wave vector, $k_{\rm res}$, selected by each 
$\omega_0$. The black squares in Fig.~\ref{fdchpk} show the maxima, 
$u_{k0}^{\rm max}$, of a sequence of $u_{k_{\rm res}}(t)$ functions selected in 
this way. 

In addition to this characteristic frequency, each $u_{k0}$ shows a pronounced 
below-band two-phonon process, visible at one half of the peak frequency, 
and it is only the act of averaging over all the $k$-components that disguises 
these features in our figures showing $n_{\rm x0}$. For the driving and damping 
parameters used in Fig.~\ref{fdchpk}, no three-phonon processes are discernible 
in the individual $k$-components. Nevertheless, a wealth of structure is 
revealed by considering the FT of the different $k$-components on logarithmic 
axes for a range of frequencies (analogies of Fig.~\ref{fdcft}, data not 
shown). The differential response of different $k$-components is also clearly 
visible when the drive is switched on, leading to complex envelope oscillations 
at initial times, and we will touch on these phenomena in Sec.~\ref{str}A. We 
remind the reader that the structure of our model ensures no interactions 
between triplons at different $k$, and so all $u_k(t)$ components evolve 
independently in time. 

We close our discussion of dynamical phenomena in the NESS by commenting on 
the possibility of new dynamical modes emerging in the driven system, for 
example where the pumped phonon is strongly dressed by triplons. Excitations 
with combined phononic and spin character are known in a number of materials, 
including manganites and ``spin-Peierls'' chains. In general these are a 
property of the equilibrium system arising for strong $g$ and, as noted at 
the end of Sec.~\ref{sness}, their inclusion would require an extension of 
the present treatment. While this treatment does reveal unconventional 
dynamical processes in the driven system, specifically those involving 
multiple phonons, it is not designed to capture the formation of bound 
states of these excitations at equilibrium. 

\section{Transient and relaxation processes}
\label{str}

Although the primary aim of our present study is to discuss NESS themselves, 
clearly their short-time (transient) behavior on ``start-up'' is a key to 
measurement windows, as well as to analyzing switching processes of the 
type one may wish to use in logic operations. Despite the clear presence 
of the timescales set by the lattice and spin dampings, respectively $1/
\gamma$ and $1/\gamma_{\rm s}$, we have already observed in Figs.~\ref{fdcuw}(a) 
and \ref{fdcuw}(c) that curiously slow convergence to a NESS can take place. 
To shed light on this result, we first analyze the convergence process and 
identify a further effective timescale arising from the driving. This allows 
us to illustrate the nature of convergence within the spin system, given the 
narrow resonance regimes of all the different $k$-components shown in 
Sec.~\ref{sdp}. We then discuss the consequences of this relationship between 
driving and convergence for the possibilities, both theoretical and practical, 
that NESS may not be reached at all because the system is driven too strongly. 
Finally, the long-time behavior of the NESS in the absence of driving has both 
important benchmarking properties for theoretical purposes and a key role in 
thermal control for experimental implementations. As noted in Sec.~\ref{smm}C, 
the formalism we derived there has no lower or upper cutoff in time, and thus 
can be applied to address every aspect of switching on and off a quantum 
spin NESS.

\subsection{Transients at switch-on}

In the introduction to NESS in our model (Sec.~\ref{sness}), we showed
in Figs.~\ref{fdcphonon}(a) and \ref{fdcphonon}(b) how the phonon variables 
are ``pumped up'' on application of the electric field, with $n_{\rm ph}(t)$ 
approaching its steady state, and thus becoming a steady drive for the spin 
system, after a time of approximately $4/\gamma$. In Fig.~\ref{fdcuw}(a) we 
showed how the spin system reacts to this oscillatory driving, with $n_{\rm x} 
(t)$ approaching its steady state after a time of approximately $4/\gamma_{\rm 
s}$ at most of the driving frequencies in Fig.~\ref{fdcuw}(a). However, it 
was clear from the $n_{\rm x}(t)$ curve at $\omega_0/J = 1.5$, which took 
significantly longer to reach its NESS [Fig.~\ref{fdcuw}(c)], that this 
reasoning alone does not explain every aspect of the spin response at the 
onset of driving. 

For a quantitative analysis of the convergence timescale, we focus first on 
in-band driving ($2 \omega_{\rm min} \le \omega_0 \le 2 \omega_{\rm max}$) and 
consider the process by which $n_{\rm x}(t)$ is ``pumped up'' by the driving 
phonon [Fig.~\ref{fdcuw}(a)]. We simplify the analysis by using the fact 
that, in this range of $\omega_0$, the driven phonon approaches its plateau 
of constant $n_{\rm ph0}$ more quickly than the spin system, because $1/\gamma
 < 1/\gamma_{\rm s}$. Thus we take the phonon oscillations as sinusoidal with 
a fixed amplitude [Fig.~\ref{fdcphonon}(c)], which to match the unit slope 
of Fig.~\ref{fdcnph} is given by 
\begin{equation}
\label{eq:driving}
q(t) = 2 \frac{a}{\gamma} \sin(\omega_0 t) \; = \; 2 \frac{D J}{g} 
\sin(\omega_0 t),
\end{equation}
where we reintroduce the driving parameter, $D$ [Eq.~\eqref{eq:p-def}], of 
the weak-coupling regime. For any selected $k$-value we define the function 
\begin{equation}
f_k(t) = 4 D J \sin(\omega_0 t) y'_k [u_k (t) + 3/2],
\end{equation}
which appears as an inhomogeneous term in the linear differential equation
of Eq.~\eqref{eomsw}. We combine Eq.~\eqref{eomsw} with Eq.~\eqref{eomsv} 
by defining the variable $z_k(t) = v_k(t) + i w_k(t)$, which then obeys the 
inhomogeneous differential equation
\begin{equation}
\label{eq:z-diff}
\frac{dz_k}{dt} = 2 i [\omega_k + 2 D J \sin(\omega_0 t) y_k] z_k
 - \gamma_\text{s} z_k + i f_k(t).
\end{equation}
A suitable primitive of the prefactor of the first term on the right-hand side 
is
\begin{subequations}
\begin{align}
h_k (t) & = 2 \int [\omega_k + 2 D J y_k \sin(\omega_0 t)] dt \\
& = 2 \omega_k t - \frac{4 D J y_k}{\omega_0} \cos(\omega_0 t),
\end{align}
\end{subequations}
allowing the solution of Eq.~\eqref{eq:z-diff} to be expressed in the form
\begin{equation}
\label{eq:z-solution}
z_k(t) = i e^{ih_k(t) - \gamma_\text{s} t} \int_0^t f_k (t') e^{-ih_k(t') + \gamma_\text{s} t'} 
dt'.
\end{equation}
This is not yet an explicit expression, because the right-hand side depends 
on $u_k(t)$, which remains unknown, but can be related to $z_k(t)$ by an 
expression based on Eq.~\eqref{eomsu},
\begin{subequations}
\begin{align}
{\tilde u}_k (t) & = u_k (t) e^{\gamma_\text{s} t} \\ & \!\!\!\!\!\! = - 2 i D J 
y'_k \!\! \int_0^t \!\! \sin(\omega_0 t') [z_k(t') - z_k^*(t')] e^{\gamma_\text{s} t'} 
dt' \!,
\label{eq:q-def-b}
\end{align}
\end{subequations}
where $z^*(t')$ denotes the complex conjugate. While this general expression 
still does not represent an explicit function, it can be used to identify the 
primary trends in the response of the spin NESS. 

We focus on the slowly varying component of $n_\text{x}(t)$, and not on the 
rapidly oscillating ones. For this it is sufficient to consider the slowly 
varying parts of each mode occupation, $u_k(t)$, as may be verified by 
numerical integration of Eqs.~\eqref{eq:eoms}. Figure \ref{fdchpk} indicates 
that the dominant term will be the one at the resonant momentum, $k_\text{res}$, 
which is determined from the driving frequency by $2 \omega_{k_\text{res}} = 
\omega_0$. The behavior of $k$-components away from 
$k_\text{res}$ is discussed in App.~\ref{app:detuned} and the results are 
summarized below. Henceforth we omit the subscript $k_\text{res}$. The slowly 
varying component of the right-hand side of Eq.~\eqref{eq:z-solution} is 
obtained by averaging over one period, $T_0 = 2\pi/\omega_0$, giving 
\begin{equation}
\label{eq:bessel}
\frac{1}{T_0} \! \int_0^{T_0} \!\! \sin(\omega_0 t) e^{-i h(t)} dt = 
J_1(\beta)/\beta,
\end{equation}
in which $\beta = 4 D J y/\omega_0$ and $J_1(\beta)$ is the Bessel 
function of the first kind. Replacing $\sin(\omega_0 t') \exp [-i h(t')]$ in 
the integrand of Eq.~\eqref{eq:z-solution} by its average taken from 
Eq.~\eqref{eq:bessel} leads to
\begin{subequations}
\begin{equation}
\label{eq:z-approx}
z(t) = i \frac{y'}{y} \omega_0 J_1(\beta) e^{ih(t) - \gamma_\text{s} t} F(t)
\end{equation}
with
\begin{equation}
\label{eq:F-def}
F(t) = \int_0^t \big( {\tilde u} (t') + {\textstyle \frac{3}{2}} 
e^{\gamma_\text{s} t'} \big) dt',
\end{equation}
\end{subequations}
which is a real quantity. We stress that the approximations leading to this 
result are well justified because the driving oscillations are much faster 
than the build-up in the triplon expectation values [Fig.~\ref{fdcuw}(a)].
By inserting Eq.~\eqref{eq:z-approx} into Eq.~\eqref{eq:q-def-b} we obtain
\begin{subequations}
\begin{align}
\label{eq:q-eq1}
{\tilde u}(t) & = \frac{(2y')^2 D J \omega_0 J_1(\beta)}{y} {\rm Re} 
\Big[ \! \int_0^t \!\! \sin(\omega_0 t') e^{ih(t')} F(t') dt' \Big] \\
& = \left( \frac{y' \omega_0 J_1(\beta)}{y} \right)^2 \int_0^t F(t') dt',
\label{eq:q-reson}
\end{align}
\end{subequations}
where again we have used Eq.~\eqref{eq:bessel} to obtain the last expression. 
Taking the second derivative yields
\begin{equation}
\label{eq:q-dgl-reson}
\frac{d^2 {\tilde u}}{dt^2} = \Gamma^2 \big( {\tilde u}(t) + {\textstyle 
\frac{3}{2}} e^{\gamma_\text{s} t} \big),
\end{equation}
in which we have defined
\begin{equation}
\label{eq:Gamma-def}
\Gamma = \left| \frac{y'\omega_0}{y} J_1 \left( \frac{4 D J y}{\omega_0} 
\right) \right|.
\end{equation}
The differential equation is readily solved with the relevant initial 
conditions, ${\tilde u}(0) = 0$ and $d {\tilde u}/dt(0) = 0$, to give the 
final expression for $u(t)$ as 
\begin{equation}
\label{eq:result-reson}
u(t) = \frac{3\Gamma}{4} \left( \frac{1 - e^{-(\gamma_\text{s} - \Gamma)t}}
{\gamma_\text{s} - \Gamma} - \frac{1 - e^{-(\gamma_\text{s} + \Gamma)t}}{\gamma_\text{s}
 + \Gamma} \right).
\end{equation}

This result makes the essential feature clear immediately. At the level of 
the present analysis, the true convergence rate is given by the quantity 
\begin{equation}
\label{eq:tgams}
\tilde\gamma_\text{s} = \gamma_\text{s} - \Gamma,
\end{equation}
which can become arbitrarily small when $\Gamma$ approaches 
$\gamma_\text{s}$. The qualitative situation is quite intuitive: 
$\gamma_\text{s}$ decribes the rate of relaxation of the system back to a 
state with zero triplons at zero temperature, which is the case considered 
here (and discussed in Sec.~\ref{str}C); the phonon driving acts in the 
opposite direction by creating pairs of triplons, and thus strong driving 
changes the effective relaxation (damping) timescale. In fact it is clear 
that Eq.~\eqref{eq:tgams} also specifies a regime where $\Gamma$ exceeds 
$\gamma_{\rm s}$, so that triplon creation outweighs the relaxation term and
Eq.~\eqref{eq:result-reson} specifies that the resonant triplon occupation, 
$u(t)$, will undergo an exponential divergence. This situation will be the 
focus of our attention in Sec.~\ref{str}B. Quantitatively, in most 
circumstances the argument of $J_1$ will be non-negative and smaller than 
1.84, which is where the function has its first maximum. In this interval, 
$J_1$ is a monotonically increasing function of the driving strength, $D$, 
and thus one expects that $\Gamma$ can indeed be raised to values on the 
order of $\gamma_\text{s}$.

Before computing $\Gamma$ as a function of $\omega_0$, we make two further 
general remarks. First, in the qualitative view of $\Gamma$ as a driving rate, 
or excitation rate, that competes with the relaxation rate, $\gamma_\text{s}$, 
one is tempted to interpret $\Gamma$ in terms of Fermi's Golden Rule. However, 
a conventional application of the Golden Rule gives a rate proportional to the 
square of the matrix element, whereas in the present analysis $\Gamma = 2DJy'$ 
at small driving ($D \to 0$), meaning that $\Gamma$ is linearly proportional 
to the driving amplitude. In more detail, the value of $u(t)$ in the NESS is 
given from the long-time limit of Eq.~(\ref{eq:result-reson}) by $3 \Gamma^2/ 
[2(\gamma_\text{s}^2 - \Gamma^2)]$ and thus is indeed proportional to 
$\Gamma^2$, and hence to $D^2$, in accordance with the Golden Rule. However, 
the timescale of the transient behavior as the system approaches the NESS is 
governed by a different coherent mechanism that yields $\Gamma \propto D$.

Second, for quantitative purposes it is necessary to consider the effect of 
driving at frequency $\omega_0$ on the modes at $k \neq k_{\rm res}$, meaning the 
action of the driving phonon as a ``detuned'' pump of all other triplon modes. 
The algebra of the detuned case is presented in App.~\ref{app:detuned} and we 
summarize the results as follows. As a function of a detuning parameter we 
define as 
\begin{equation}
\label{eq:detun-def}
\delta = 2\omega - \omega_0,
\end{equation}
there are two possible regimes. If $|\delta| < \Gamma$, it is convenient 
to define the quantity 
\begin{equation}
\label{eq:tGam-def}
\widetilde\Gamma = \sqrt{\Gamma^2 - \delta^2}, 
\end{equation}
in terms of which 
\begin{equation}
\label{eq:u-regimeI}
u(t) = \frac{3 \Gamma^2}{2(\gamma_\text{s}^2 \! - \! \widetilde\Gamma^2)}
\bigg[ 1 \! - \! e^{-\gamma_\text{s}t} \Big( \cosh(\widetilde\Gamma t) + 
\frac{\gamma_\text{s}}{\widetilde\Gamma} \sinh(\widetilde\Gamma t) \Big) \bigg].
\end{equation}
Thus from the behavior of the hyperbolic function, $\widetilde\Gamma$ adopts 
the role of $\Gamma$ in Eq.~\eqref{eq:tgams} and the relevant convergence 
rate becomes $\tilde\gamma_\text{s} = \gamma_\text{s} - \widetilde\Gamma$.
By contrast, when $|\delta| > \Gamma$, so that the detuning of the driving 
frequency exceeds the driving threshold, it is convenient to define the 
quantity 
\begin{equation}
\label{eq:tGam-def2}
\tilde\delta = \sqrt{\delta^2 - \Gamma^2},
\end{equation}
in terms of which 
\begin{equation}
\label{eq:u-regimeII}
u(t) = \frac{3 \Gamma^2}{2(\gamma_\text{s}^2 + \tilde\delta^2)} \bigg[ 1
 - e^{-\gamma_\text{s}t} \Big( \cos(\tilde\delta t) + \frac{\gamma_\text{s}}
{\tilde\delta} \sin(\tilde\delta t) \Big) \bigg].
\end{equation}
Because all the hyperbolic functions become trigonometric, the sole remaining 
exponential convergence is governed by $\gamma_{\rm s}$, leading to the result 
that the convergence is conventional. We note in this case that slow 
oscillations arise at frequency $\tilde\delta$, which may cause the triplon 
number to overshoot before it converges to its NESS limit (example data not 
shown).

\begin{figure}[t]
\centering
\vspace{-0.5cm}
\includegraphics[width=\columnwidth]{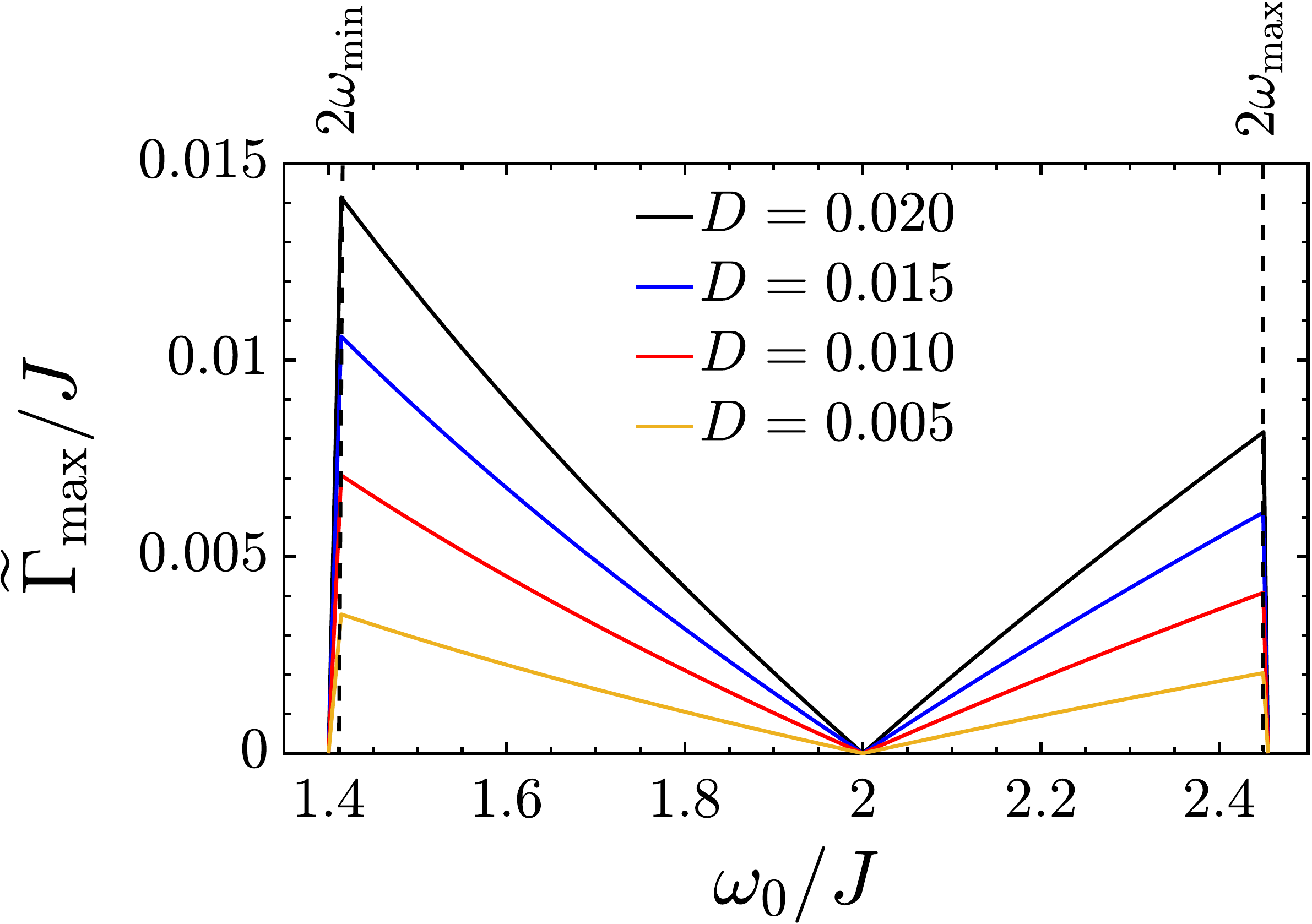}
\vspace{-0.5cm}
\caption{Dependence of the inverse driving timescale, $\widetilde 
\Gamma_{\rm max}$, on the frequency, $\omega_0$, of the driving phonon, 
shown for four values of the driving strength, $D$.} 
\label{fdcGamma}
\end{figure}

Although one might assume that the resonant case, $2\omega_k = \omega_0$,  
described by Eq.~(\ref{eq:result-reson}) will provide the highest threshold 
value, making $\widetilde\Gamma_{\rm max} = \Gamma$, the complicated dependence 
of $\Gamma$ on $k$ [Eq.~(\ref{eq:Gamma-def})] makes it possible that, for a 
given $\omega_0$, a slightly detuned mode at $k \neq k_\text{res}$ yields a 
higher $\widetilde\Gamma$. In particular, for frequencies close to but outside 
the two-triplon band, detuned driving will be of primary importance. To capture 
these possible effects, we compute $\widetilde\Gamma_{\rm max}$ by variation of 
$k$ at each fixed $\omega_0$, and the results are shown in Fig.~\ref{fdcGamma}.

Clearly $\widetilde\Gamma_{\rm max}$ is finite throughout the in-band regime, 
although it drops to zero at the band center ($\omega_0 = 2 J$) due to a 
matrix-element effect ($y'_k|_{k = \pi/2} = 0$). Although the dependence of 
$\widetilde\Gamma_{\rm max}$ on $\omega_0$ is both direct and indirect, 
occurring both through proximity to the resonance condition ($2\omega_{k_{\rm res}} 
 = \omega_0$) and through the momentum-dependence of $y_k$ and $y'_k$, it shows 
an almost linear rise with frequency towards the two band edges. Because we 
consider the linear driving regime of Sec.~\ref{sness}, it is also a linear 
function of $D$. Importantly, $\widetilde\Gamma_{\rm max}$ is also finite 
outside the two-triplon band as a consequence of detuned driving, although 
for the parameters in Fig.~\ref{fdcGamma} it falls rapidly (in fact over a 
frequency window of order $D J$) beyond the band edges. For any given $D$, 
the function $\widetilde\Gamma_{\rm max} (\omega_0)$ indicates the values of 
the triplon damping, $\gamma_{\rm s}$, for which unconventional convergence 
can occur, and it is no surprise to find that in-band frequencies near the 
two band edges are the most likely candidates [Fig.~\ref{fdcuw}(a)]. From 
Fig.~\ref{fdcGamma}, and specifically from the value of $\widetilde 
\Gamma_{\rm max}$ at $\omega_0 = 2 \omega_{\rm min}$, one may read that, at the 
level of our analysis, the value of $\gamma_\text{s}$ ensuring conventional 
or slow convergence at all frequencies for driving $D = 0.01$ (Sec.~\ref{sdp}) 
is approximately 0.007. We comment on the minor discrepancy with our numerical 
findings in Fig.~\ref{fdchpw}, where NESS formation was verified at all 
$\omega_0$ with $\gamma_\text{s} = 0.005$, in Sec.~\ref{str}B. 

Here we make three quantitative side remarks to this analysis. First, the 
effective driving timescale, $\widetilde\Gamma_{\rm max}$, is not easily read 
from the external driving parameters, because it depends crucially on the 
phonon amplitude. Even in the weak-coupling regime, meaning small $g$ as 
defined in Sec.~\ref{sness}, we have seen that the oscillations of the 
driven phonon are not entirely independent of the spin system for in-band 
driving frequencies. Second, we do not consider the additional complexity 
of a $k$-dependent $\gamma_{\rm s}$, although the framework developed here 
could be used without alteration. Third, the effect of non-linear processes 
occurring at multiples of $\omega_0$, is not included in our discussion of 
$\Gamma$ and $\widetilde\Gamma$, although it could be incorporated by 
considering a very weak effective $D$.

\begin{figure}[t]
\includegraphics[width=\columnwidth]{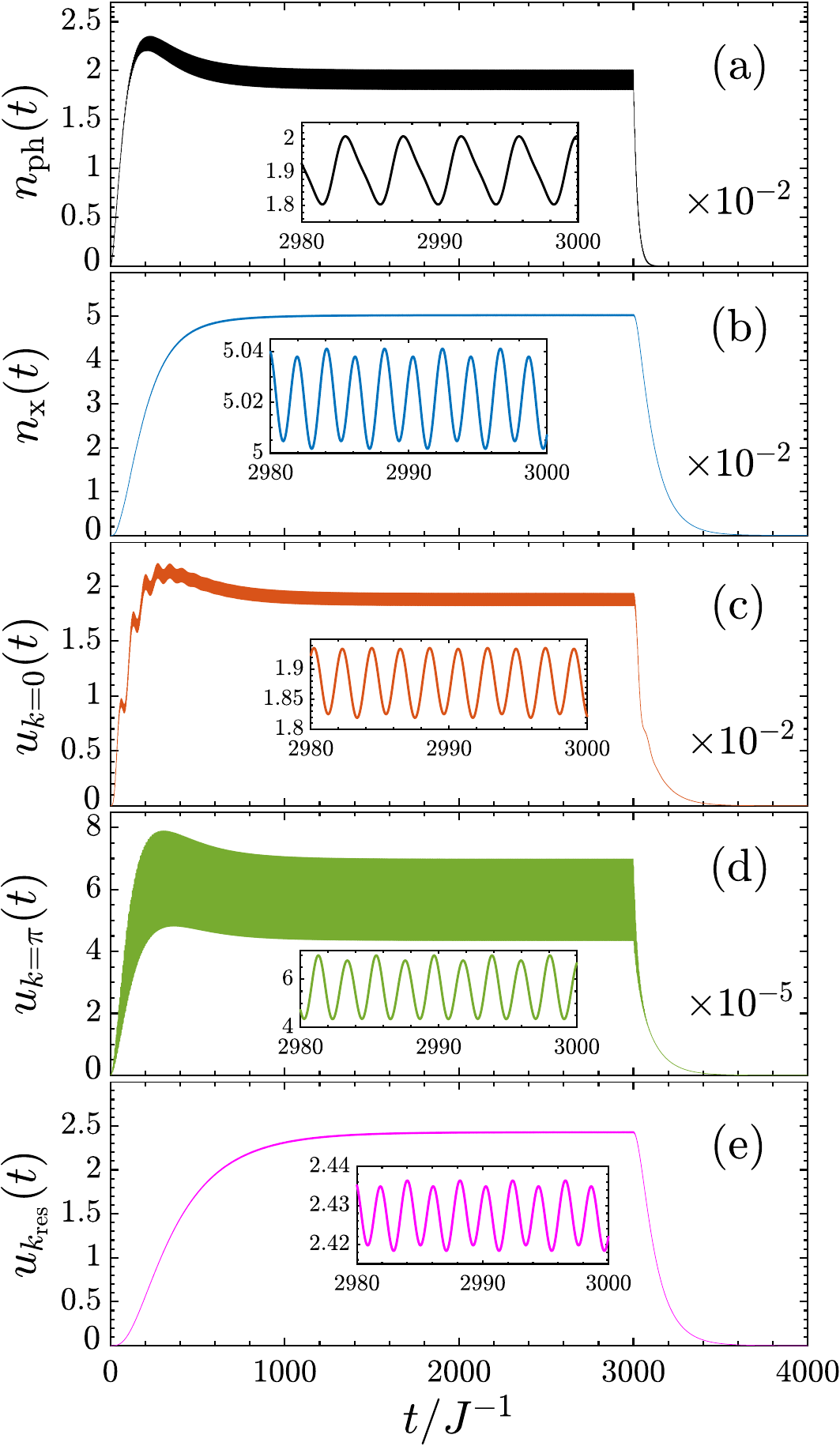}
\caption{Creation of the NESS established with $a/\gamma = 0.2$ and $g = 0.1 
J$ (driving parameter $D = 0.02$) for driving frequency $\omega_0/J = 1.5$ and 
spin damping $\gamma_{\rm s} = 0.01 J$; these are the parameters of the green 
line in Fig.~\ref{fdcuw}(a). (a) $n_{\rm ph}(t)$. (b) $n_{\rm x}(t)$. (c) $u_{k=0}$. 
(d) $u_{k=\pi}$. (e) $u_{k_{\rm res}}$. Also shown is the relaxation of each variable 
when the driving is removed after 3000 time steps.}
\label{fdcsc}
\end{figure}

To illustrate the phenomenon of slow convergence at switch-on, we consider 
driving field $a/\gamma = 0.2$ and $g = 0.1 J$, which is the situation in 
Fig.~\ref{fdcuw}(a). For the in-band driving frequency $\omega_0 = 1.5 J$ 
and spin damping $\gamma_{\rm s} = 0.01 J$, we show in Figs.~\ref{fdcsc}(a) and 
\ref{fdcsc}(b) the driven phonon and triplon numbers. The driving strength 
is the same as that in Fig.~\ref{fdcphonon}(a), and thus $n_{\rm ph}(t)$ first 
rises towards the plateau value of 0.04 in a time dictated by $1/\gamma$, 
but is pulled down again to an average value $n_{\rm ph0} \simeq 0.02$ 
[Fig.~\ref{fdcnph}(b)] in a time dictated by $1/{\tilde \gamma}_{\rm s}$. This 
is a direct reflection of the ``inertia'' of the spin system as it begins to 
absorb some of the input phonon energy, a topic we analyze in more detail in 
Sec.~\ref{seh}A. The values of $\gamma_{\rm s}$ and $\widetilde\Gamma$ 
(Fig.~\ref{fdcGamma}) place the system very close to the threshold 
specified by Eq.~\eqref{eq:tgams}, with the result that the spin NESS 
[Fig.~\ref{fdcsc}(b)] is reached only after approximately 1200 time steps 
[Fig.~\ref{fdcuw}(c)], indicating that $1/{\tilde \gamma_{\rm s}} \approx 
3/\gamma_{\rm s}$. For a more quantitative understanding of the transient 
phenomena in this regime, in Figs.~\ref{fdcsc}(c) to \ref{fdcsc}(e) we show 
the $k = 0$-, $\pi$-, and $k_{\rm res}$-components of $u_k(t)$. It is not a 
surprise to confirm that the majority of the slow-convergence behavior is 
indeed concentrated in $u_{k_{\rm res}}(t)$ [Fig.~\ref{fdcsc}(e)], which is both 
the largest and the most slowly converging component, apparently requiring 
50\% longer than $n_{\rm x0}(t)$ to converge within 2\% of its final value. 
However, it is somewhat surprising to find that the non-resonant $u_{k}(t)$ 
components actually rise above their NESS values (on a timescale dictated 
by $\gamma_{\rm s}$) before falling again as the driving phonon amplitude 
reaches its final NESS value [on the timescale dictated by $u_{k_{\rm res}}(t)$].

\subsection{Existence of NESS}

In Secs.~\ref{sness}, \ref{sdp}, and \ref{str}A, we have used parameters 
allowing the formation of NESS in order to analyze their response to the 
driving parameters and their internal dynamical properties. Having established 
this foundation, we now discuss the crucial issue of whether a NESS can exist 
at all for strong driving over long driving times. Clearly, unlimited driving 
would lead to heating of the system on a finite timescale, and we defer a 
discussion of this topic until Sec.~\ref{seh}; here we continue to assume 
that the heat sink represented in Fig.~\ref{fdcschem} maintains a steady, low 
system temperature despite the injection of energy from the laser. The focus 
of our present discussion is the possibility that the lattice or spin system 
could be driven so strongly that it breaks down rather than converge to a NESS. 

The integrity of the driven lattice is easy to establish. A straightforward 
application of the Lindemann criterion, whose details we present in 
App.~\ref{app:lc}, leads to the result that lattice melting due to phonon 
driving would become an issue for average phonon mode occupancies on the 
order of $n_{\rm ph0} = 3$. Thus the driving parameters we consider here, and 
the resultant $n_{\rm ph0}$ values, pose no threat to the periodic lattice. By 
contrast, based on the discussion of Sec.~\ref{str}A, one might expect that 
Eq.~\eqref{eq:tgams} represents a threshold of driving strength ($D$) beyond 
which triplon creation exceeds their relaxation and $n_{\rm x}$ should diverge 
exponentially, meaning that NESS formation is impossible. Here we discuss two 
criteria for the loss of NESS. The first is breaching of the condition on the 
triplon occupation, $n_{\rm x}(t) < n_{\rm x}^{\rm max} = 0.2$ (Sec.~\ref{sness}), 
beyond which the formalism of Sec.~\ref{smm} can no longer be applied to the 
spin system. The second is breaching of the positivity of ${\tilde 
\gamma}_{\rm s}$ as defined in Eq.~\eqref{eq:tgams}. 

\begin{figure}[t]
\includegraphics[width=\columnwidth]{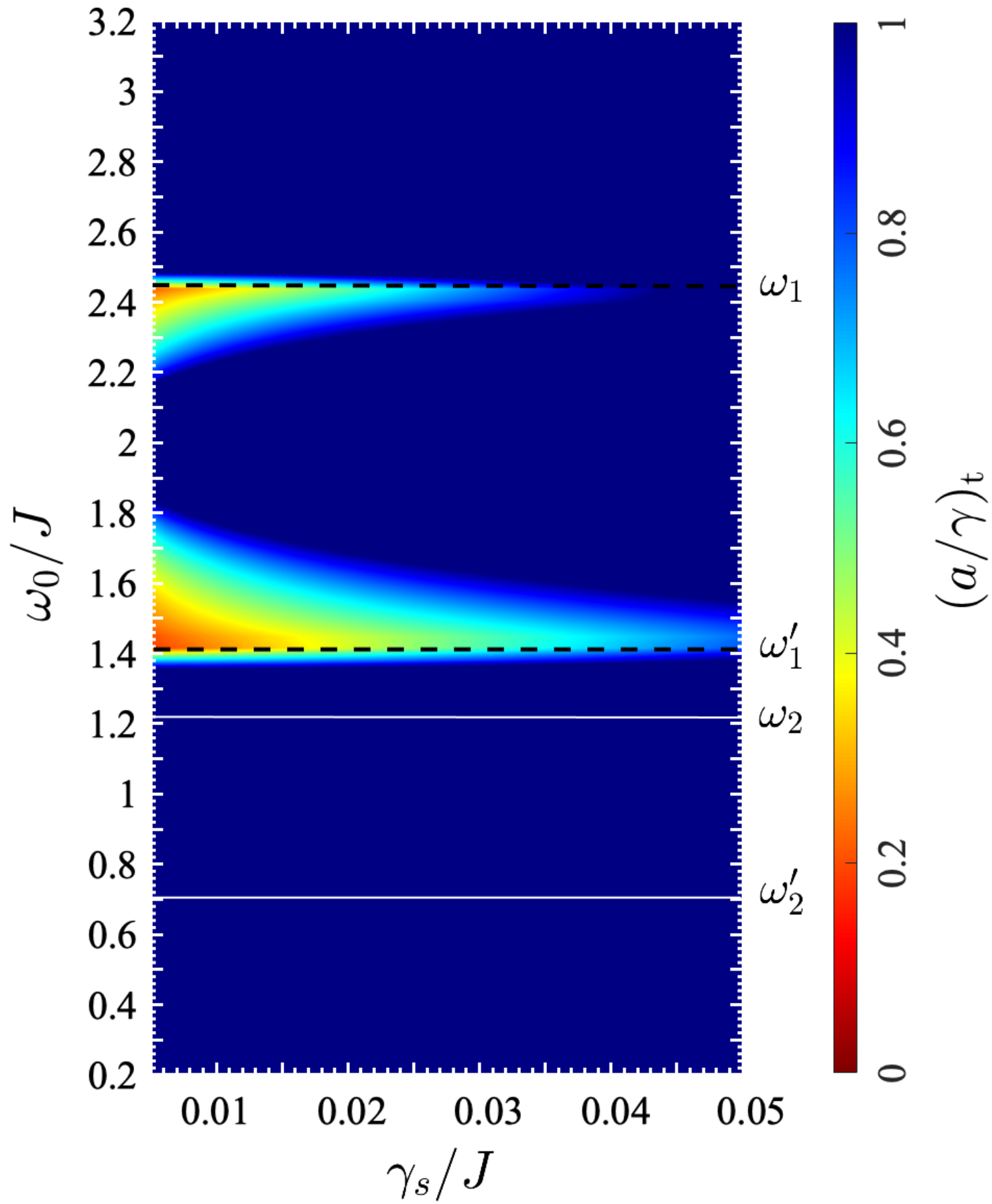}
\caption{Threshold value, $(a/\gamma)_{\rm t}$, of the normalized laser 
electric field strength required to achieve the maximum steady-state triplon 
occupation of $n_{\rm x} = 0.2$, shown as a function of $\gamma_{\rm s}$ and 
$\omega_0$ for fixed $g = 0.05 J$ and $\gamma = 0.02 \omega_0$. We draw 
attention to the 3 regimes of behavior demarcated by $\omega_1^\prime = 2 
\omega_{\rm min}$ and $\omega_1 = 2 \omega_{\rm max}$. Also marked are the 
frequencies $\omega_2^\prime = \omega_{\rm min}$ and $\omega_2 = \omega_{\rm max}$, 
where unlike Fig.~\ref{fdchpw}(a) no additional structure is visible in 
$(a/\gamma)_{\rm t}$.}
\label{fdchpa}
\end{figure}

Considering first the maximum triplon occupation, in Fig.~\ref{fdchpa} we show 
the threshold value of the driving strength, $(a/\gamma)_{\rm t}$, required to 
drive the triplon occupation of the spin NESS above $n_{\rm x}^{\rm max}$. Red 
colors are chosen to represent regions of small $(a/\gamma)_{\rm t}$, because 
this indicates efficient triplon occupation, and these are found at driving 
frequencies corresponding to the lower and upper edges of the two-triplon 
band, intensifying as $\gamma_{\rm s}$ becomes smaller (Sec.~\ref{sdp}). As 
in Fig.~\ref{fdchpw}(a), it is evident that the system does not respond as 
efficiently for in-band frequencies near the band center, and that very 
strong driving is required when $\omega_0$ lies above the two-triplon band. 
In contrast to Fig.~\ref{fdchpw}(a), $(a/\gamma)_{\rm t}$ does not reflect 
the presence of the two-phonon response feature at and above $\omega_0 = 
\omega_{\rm min}$, underlining that the $n_{\rm x0}$ values arising due to these 
processes are indeed small. 

Nevertheless, one may worry that $n_{\rm x}^{\rm max} = 0.2$ is an arbitrary 
criterion, which would have no relevance if a more sophisticated treatment 
of the spin sector were implemented, and thus that the driving criterion 
should give a more rigorous statement on the existence of NESS. However, 
here we encounter departures from the idealized analytical discussion of 
Sec.~\ref{str}A. There we also remarked on the observation in our numerical 
results that the driven phonon does not reach the average occupation, 
$n_{\rm ph0}$, expected from Fig.~\ref{fdcnph}(a) because of the effects of 
the spin system that it drives. While this result was visible at $g = 0.1 J$ 
for the $\omega_0/J = 1.5$ phonon in Fig.~\ref{fdcnph}(b), it is not absent 
in what we have called the weak-$g$ regime, as one may observe from the value 
of $n_{\rm ph0}$ in Fig.~\ref{fdcft}(a). This is one example of a feedback 
effect between the lattice and spin systems, which we will encounter again 
in Sec.~\ref{seh}A. Its consequence for the analysis in Sec.~\ref{str}A is 
that one may no longer assume a fixed driving strength, $D$, but because of 
this downwards renormalization one should work with an effective driving, 
$D_{\rm eff}$. The deviation between $D$ and $D_{\rm eff}$ becomes larger as $D$ 
is increased. We have also encountered strong feedback effects as $\gamma
_{\rm s}$ becomes very small and as $n_{\rm ph0}$ becomes large enough to alter 
the dimerization, $\lambda$, of the spin system. What all of these feedback 
effects have in common is that they are significant only when the triplon 
occupation is large, meaning $n_{\rm x} > n_{\rm x}^{\rm max}$, and thus the 
constraint $n_{\rm x}^{\rm max} = 0.2$ adopts additional significance. Feedback 
is a complex topic that will be important in discussing driven systems with 
strong $g$, but lies beyond the scope of our current exposition. 

\begin{figure}[p]
\includegraphics[width=\columnwidth]{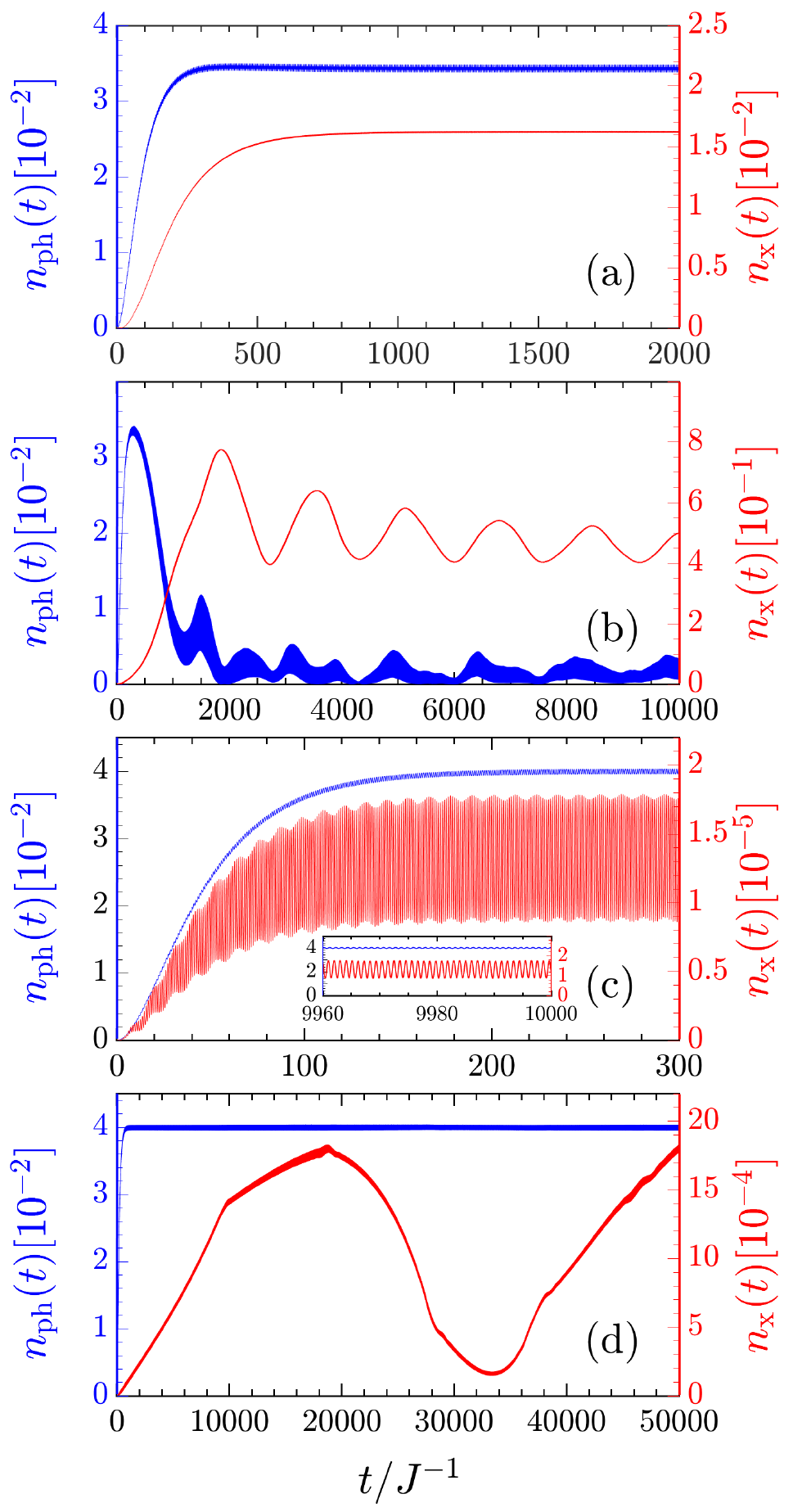}
\caption{Time-dependence of the phonon number, $n_{\rm ph}(t)$, and triplon 
number, $n_{\rm x}(t)$, shown with $a/\gamma = 0.2$ and $g = 0.05 J$ ($D = 
0.01$). (a) When $\omega_0/J = 1.5$ and $\gamma_\text{s} = 0.01 J$ (the 
parameters of Fig.~\ref{fdcft}), the system converges to a NESS on a 
conventional timescale. 
(b) When $\omega_0/J = 1.5$ and $\gamma_\text{s} = 0$, $n_{\rm x}(t)$ increases 
rapidly beyond $n_{\rm x}^{\rm max}$, destabilizing the phonon occupation. (c) 
When $\omega_0/J = 3.0$, the driving frequency lies sufficiently far above 
the two-triplon band that NESS exist even when $\gamma_\text{s} = 0$. (d) When 
$\omega_0/J = 0.75$, the driving frequency lies well below the two-triplon 
band but the second harmonic, $2\omega_0$, lies within it. In this case, 
when $\gamma_\text{s} = 0$ the lattice approximates a NESS, but with this 
near-constant driving of the spin system a NESS cannot be formed.}
\label{fdcexist}
\end{figure}

Even without a driving criterion, it is nonetheless instructive to ask how 
the breakdown of NESS occurs. We perform only a brief and numerical examination 
of how the model of Sec.~\ref{smm} behaves when NESS formation is precluded, 
for which we set $\gamma_{\rm s} = 0$. In Fig.~\ref{fdcexist} we depict the 
time-evolution of the lattice and spin systems for different driving 
frequencies with a fixed driving strength, $D = 0.01$. For in-band 
driving at $\omega_0/J = 1.5$ and a spin damping $\gamma_{\rm s} = 0.01 J$, 
Fig.~\ref{fdcexist}(a) shows the NESS of Fig.~\ref{fdcft}. However, when 
$\gamma_{\rm s} = 0$, Fig.~\ref{fdcexist}(b) shows how the triplon number 
is driven rapidly to a regime well beyond $n_{\rm x}^{\rm max}$, which in turn 
causes the phonon occupation to become unstable and creates complex, aperiodic 
feedback phenomena.

When $\omega_0$ lies above the two-triplon band, one may read from the detuning 
discussion, and also directly from Fig.~\ref{fdchpa}, that two possibilities 
exist. If $\omega_0$ is sufficiently far beyond $2 \omega_{\rm max}$, as shown in 
Fig.~\ref{fdcexist}(c) for the case $\omega_0/J = 3.0$, a NESS can be formed 
even with $\gamma_{\rm s} = 0$. In this case, the phonon driving cannot cause 
the direct occupation of triplons and the steadily driven state of the spin 
sector remains only very weakly excited. Any feedback from the spin to the 
lattice sector under these circumstances is negligible, and thus the latter is 
also unaffected by the value of $g$. The beating envelope in $n_{\rm x} (t)$ in 
Fig.~\ref{fdcexist}(c) is a consequence of transient signals in individual 
components of $u_k(t)$ that are never damped with $\gamma_\text{s} = 0$. The 
second possibility arises when $\omega_0$ lies above but very close to 
$2 \omega_{\rm max}$, in which case the driving phonon acts as a detuned pump 
of the spin response at the upper band edge and the physics is that of 
Figs.~\ref{fdcexist}(a) and \ref{fdcexist}(b).

Finally, the situation for driving frequencies below the lower two-triplon 
band edge is somewhat more complicated. Once again there is a regime of 
potentially divergent behavior due to detuned driving when $\omega_0$ lies 
slightly below $2 \omega_{\rm min}$ (Fig.~\ref{fdchpa}). [This phenomenon also 
allows one to understand why the lower and upper two-triplon band edges do not 
create extremely sharp features, or even discontinuities, as a function of 
$\omega_0$ in the response observed in Figs.~\ref{fdcftc} and \ref{fdchpw}.] 
At frequencies below the detuned regime, the generic situation is that 
illustrated in Fig.~\ref{fdcexist}(d) for a frequency $\omega_0 = 0.75J$. 
The phonondoes indeed approach a NESS, but this essentially steady driving 
does not create a spin NESS because high-order processes always exist that 
pump the undamped spin system on some potentially very long timescale. In 
Fig.~\ref{fdcexist}(d)
the higher-order process involves the second harmonic and it is necessary both 
to follow the spin dynamics to multiples of 10$^4$ time steps and to use long 
chains (here $N = 3000$) to verify the situation. In general, driving of the 
system by a multi-phonon process can be captured by the same analytical 
arguments applied for in-band frequencies, although the effective value of 
$D$ should be replaced by the amplitude of the relevant higher harmonic. As 
a result, the qualitative situation at arbitrary below-band frequencies is 
that of Fig.~\ref{fdcexist}(d), but quantitatively the required timescale 
may extend to millions of steps. We conclude this analysis by stressing 
again that NESS formation is the most natural behavior in the model of 
Fig.~\ref{fdcschem} at all frequencies for realistic values of $a$ and 
$\gamma_{\rm s}$, as shown in Fig.~\ref{fdcexist}(a), as well as throughout 
Sec.~\ref{sness}.

\subsection{Relaxation at switch-off}

The process of relaxation of a system with Lindblad damping is the recovery 
of thermal equilibrium in the absence of the drive. In our analysis, the system 
started at temperature $T = 0$ before the drive was switched on, and thus it 
relaxes back to this state. The present analysis is readily extended to 
finite $T$ by including (i) a thermal phonon occupation, (ii) a more 
sophisticated treatment of the spin system \cite{rnr}, and (iii) the thermal 
factors in the definition of the bath properties that are already contained 
in the Lindblad formalism (Sec.~\ref{smm}C). However, this extension would 
not account for the fact that the driving introduces energy to the system, 
and hence causes heating; for this we appeal to the heat sink represented 
in Fig.~\ref{fdcschem}, which corresponds to the cooling apparatus in any 
condensed-matter experiment. We will discuss the issues associated with the 
system temperature, particularly in the presence of the laser drive, more 
deeply in Secs.~\ref{seh} and \ref{sd}. 

For the purposes of this subsection, in Fig.~\ref{fdcsc} we have switched 
off the phonon drive after 3000 time steps. It is clearly visible in all 
cases that the phonon sector [characterized by $n_{\rm ph}(t)$] relaxes to its 
equilibrium, $n_{\rm ph} = 0$, over a timescale governed by $1/\gamma$ and the 
spin sector [characterized by $n_{\rm x}(t)$] over a timescale governed by 
$1/\gamma_{\rm s}$. This behavior is independent of the value of $\omega_0$ at 
which the system was being driven and of the amplitude of the driving (data 
not shown). In this sense, relaxation can be considered as similar to the 
process of ``pumping up'' the NESS with a very low drive, so that the system 
remains far from the driving-induced timescale obtained in Sec.~\ref{str}A. 
Thus one may conclude that unconventional transient processes appear only 
when the system is driven, and indeed driven near its band edges, whereas 
relaxation dynamics are straightforward. 

\begin{figure}[t]
\includegraphics[width=0.9\columnwidth]{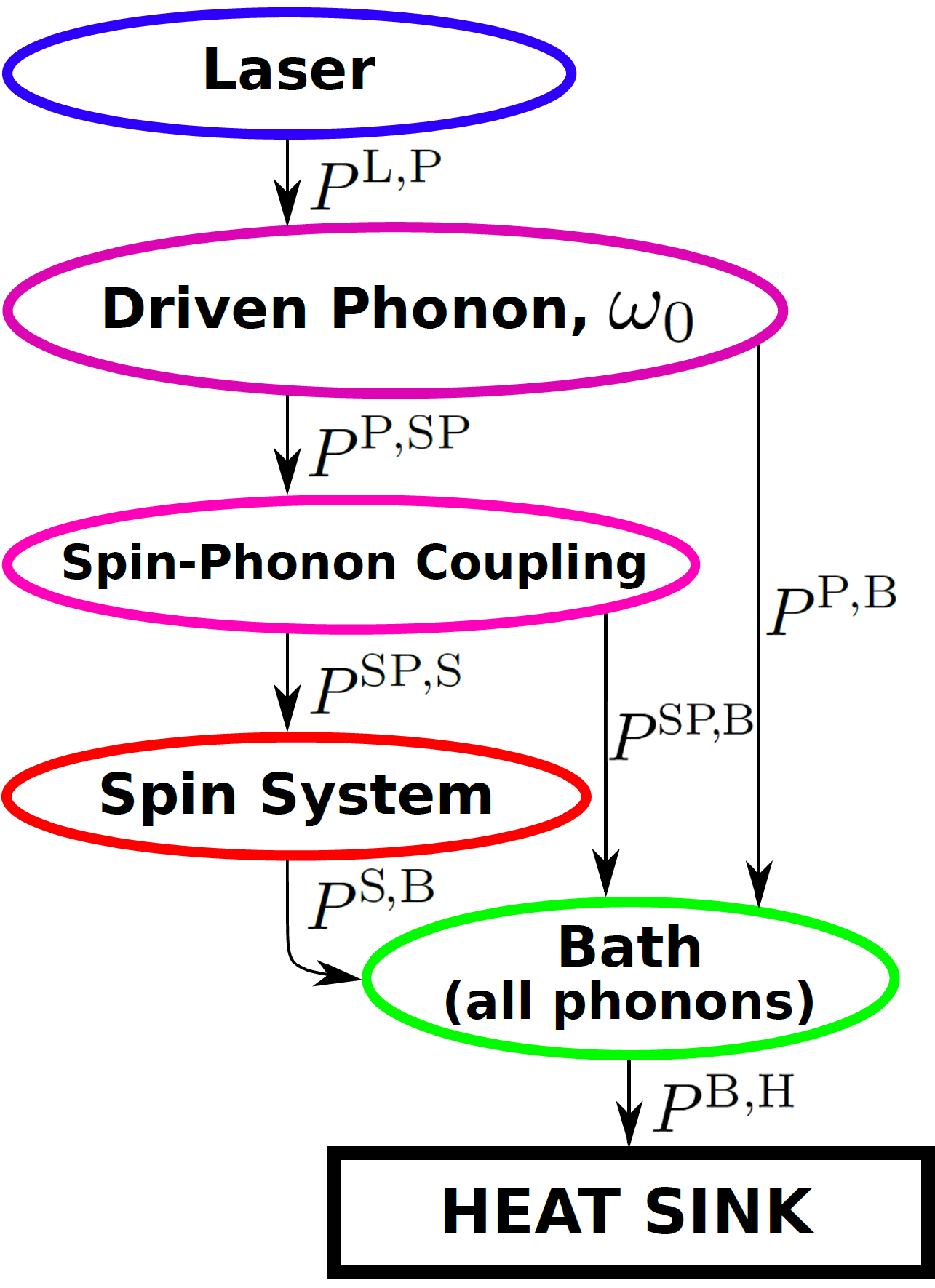}
\caption{Schematic representation of energy flow into and out of the NESS 
of the combined lattice and spin system.}
\label{fdcefschem}
\end{figure}

\section{Energy flow and system heating}
\label{seh}

\subsection{Energy flow}

Particularly valuable for both conceptual and practical purposes is to 
consider the energy flow through the spin-lattice system. For a true NESS, 
the rate of energy throughput should be constant from the driving to the 
final stage of dissipation. Figure \ref{fdcefschem} provides a schematic 
representation of the situation, which we characterize using seven separate 
stages of the flow process. The energy flow (energy per unit time) is a 
power and is defined to be positive in the direction of the arrows in 
Fig.~\ref{fdcefschem}. Clearly the input power is the uptake of laser 
energy by the driven phonon, part of which also drives the spin sector 
through the effect of the spin-phonon coupling. Energy absorption by the 
lattice, which is also the bath, will determine the temperature of the 
system; while this is limited by possible melting of the crystal, the 
loss of coherence in the quantum spin states is a much more stringent 
criterion. To avoid a monotonic rise in the lattice, or bath, temperature, 
we model the system as being connected to a large and efficiently conducting 
heat sink. 

To compute the different energy flows in Fig.~\ref{fdcefschem}, the following 
two relations may be read directly from the differential equation of 
Eq.~\eqref{eompc}, which describes the time evolution of the number of 
energy quanta in the driven phonon. The energy flow from the laser into 
this phonon, normalized to the number of dimers, is given by 
\begin{equation}
\label{eq:p-lp}
P^\text{L,P}(t) = - E(t) \omega_0 p(t),
\end{equation}
while the energy flowing from it and directly to the bath is given by
\begin{equation}
\label{eq:p-pb}
P^\text{P,B}(t) = \gamma \omega_0 [n_\text{ph}(t) - n(\omega_0)].
\end{equation}
The energy flowing out of the driven phonon due to the presence of the 
spin-phonon coupling is given by using the same equation to obtain 
\begin{equation}
\label{eq:p-psp}
P^\text{P,SP}(t) = g \omega_0 [\mathcal{U}(t) + \mathcal{V}(t)] p(t).
\end{equation}
By considering energy conservation for the phonon we obtain the sum rule 
\begin{equation}
\label{esr1}
P^\text{L,P}_0 = P^\text{P,SP}_0 + P^\text{P,B}_0 
\end{equation}
for the temporal averages of each power. We remark that all of the powers in 
Fig.~\ref{fdcefschem} have a temporal oscillation in the NESS at multiples 
of the driving frequency, in the same way as all the other quantites discussed 
in Secs.~\ref{sness} and \ref{sdp}. However, these oscillations are not very 
relevant to the overall energy flow or system temperature and we focus on 
their average values, which are the $m = 0$ harmonics of Sec.~\ref{sdp}, so 
we denote them by $P_0^\text{X,Y}$.

Turning to the spin sector, Eq.~\eqref{eomsu} gives the energy flow into 
the spin system as 
\begin{equation}
\label{eq:p-sps}
P^\text{SP,S}(t) = 2 g q(t) \frac{1}{N} \sum_k y_k' \omega_k w_k(t).
\end{equation}
The same equation also states that the energy flow from the spin system into 
the bath is given by the decay rate of all the triplons, which yields
\begin{equation}
\label{eq:p-sb}
P^\text{S,B}(t) = \frac{\gamma_\text{s}}{N} \sum_k \omega_k [u_k(t)
 - 3 n(\omega_k)].
\end{equation}
Once again, energy conservation within the spin system enforces the sum rule
\begin{equation}
\label{esr2}
P^\text{SP,S}_0 = P^\text{S,B}_0
\end{equation}
on the time-averaged values. However, if one considers Eq.~\eqref{eq:p-psp} 
as the work done by the phonon on the spin system and Eq.~\eqref{eq:p-sps} 
as the work received by the spin system due to the phonon, it is evident 
that there is no mathematical reason for these two quantities to be equal. 
To obtain the physical sum rule, it is necessary to consider in detail the 
spin-phonon coupling term, $H_\text{sp}$ in Eqs.~\eqref{etsh} and 
\eqref{eq:mf-split}. In the mean-field approximation, we have by construction
\begin{equation}
{\textstyle \frac{1}{N}} \langle H_\text{sp} \rangle (t) = g q(t) (\mathcal{U} 
 + \mathcal{V})(t),
\end{equation}
and hence the time derivative 
\begin{equation}
{\textstyle \frac{1}{N}} \partial_t \langle H_\text{sp} \rangle = g [ (\partial_t
q) (\mathcal{U} + \mathcal{V}) + q \partial_t (\mathcal{U} + \mathcal{V}) ].
\end{equation}
Using Eqs.~\eqref{eompa}, \eqref{eomsu}, and \eqref{eomsv} to evaluate the 
partial derivatives on the right-hand side yields
\begin{subequations}
\begin{align}
\label{eq:line1}
& {\textstyle \frac{1}{N}} \partial_t \langle H_\text{sp} \rangle = 
g \omega_0 (\mathcal{U} + \mathcal{V}) p - {\textstyle \frac12} g \gamma 
q (\mathcal{U} + \mathcal{V}) \\
\label{eq:line2}
& \quad + g q \frac{1}{N} \sum_k y_k \big[ 2 g q y'_k  w_k - \gamma_\text{s} 
\big( u_k - 3 n (\omega_k) \big) \big] \\
\label{eq:line3}
& \quad - g q \frac{1}{N} \sum_k y'_k [2 (g q y_k + \omega_k) w_k
 + \gamma_\text{s} v_k ].
\end{align}
\end{subequations}
By inspection, the first term in Eq.~\eqref{eq:line1} is $P^\text{P,SP}(t)$ and 
the second term in Eq.~\eqref{eq:line3} is $P^\text{SP,S}(t)$, while the first 
terms in Eqs.~\eqref{eq:line2} and \eqref{eq:line3} cancel, as a result of 
which the expression takes the form 
\begin{subequations}
\begin{align}
\label{eq:bilanz1}
{\textstyle \frac{1}{N}} \partial_t \langle H_\text{sp} \rangle &= 
P^\text{P,SP}(t) - P^\text{SP,S}(t) \\
& \quad - g q(t) ( {\textstyle \frac12} \gamma + \gamma_\text{s}) (\mathcal{U}
 + \mathcal{V})(t).
\end{align}
\end{subequations}
The second line suggests rather strongly the definition 
\begin{subequations}
\begin{align}
\label{eq:def-flow}
P^\text{SP,B}(t) & = g q(t) ({\textstyle \frac12} \gamma + \gamma_\text{s}) 
(\mathcal{U} + \mathcal{V})(t) \\ & = ({\textstyle \frac12} \gamma + 
\gamma_\text{s}) \langle H_\text{sp} \rangle (t),
\end{align}
\end{subequations}
where $- P^\text{SP,B}(t)$ describes a relaxation of $\langle H_\text{sp} \rangle 
(t)$ towards zero. This quantity corresponds to a flow of energy from the 
spin-phonon coupling towards the bath and its temporal average completes the 
balance
\begin{equation}
\label{eq:bal-sp}
P^\text{P,SP}_0 = P^\text{SP,S}_0 + P^\text{SP,B}_0,
\end{equation}
which results from the fact that the time-average of the derivative $\partial_t 
\langle H_\text{sp}\rangle$ must vanish in a NESS. Thus the definition of 
Eq.~\eqref{eq:def-flow} and the additional sum rule of Eq.~\eqref{eq:bal-sp} 
provide the appropriate linkage to describe energy conservation in the 
coupled spin-phonon system. 

\begin{figure}[t]
\centering    
\includegraphics[width=\columnwidth]{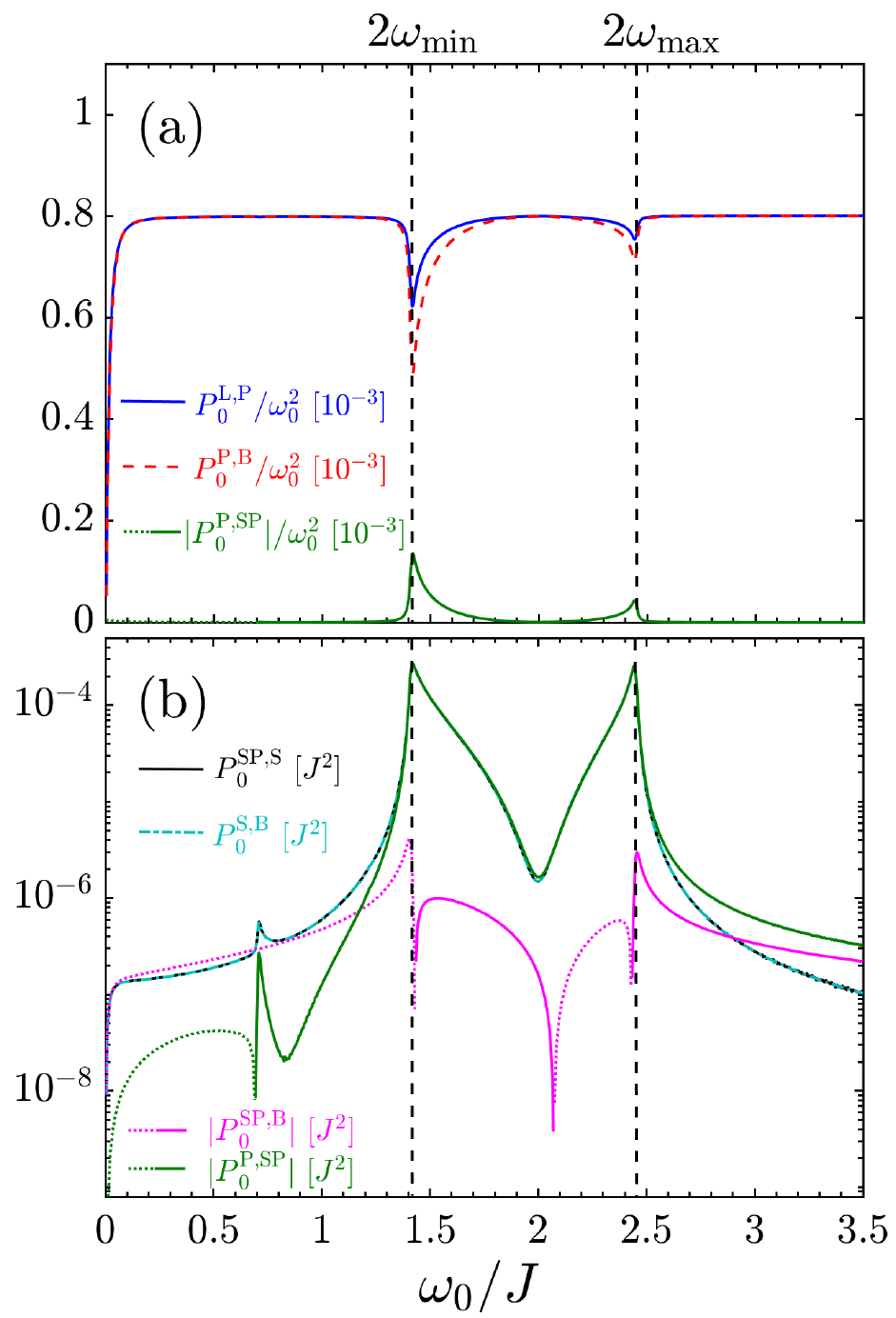}
\caption{Average energy flows, $P^{\rm X,Y}_0$, through the combined lattice and 
spin system depicted in Fig.~\ref{fdcefschem}, normalized to the number of 
dimers and shown as a function of $\omega_0$ for $\gamma = 0.02J$, $a/\gamma
 = 0.2$, $g = 0.05 J$, and $\gamma_{\rm s} = 0.01 J$. (a) Power delivered to the 
driving phonon ($P^{\rm L,P}_0$) and power dissipated directly to the bath from 
this phonon ($P^{\rm P,B}_0$), whose difference is the power transferred towards 
the spin system from this phonon ($P^{\rm P,SP}_0$). For clarity we show 
$P^{\rm X,Y}_0$ normalized by $\omega_0^2$. (b) Work done by the phonon system 
due to the spin system ($P^{\rm P,SP}_0$) and power delivered to the spin system 
($P^{\rm SP,S}_0$), with their difference ($P^{\rm P,SP}_0$) represented on the 
logarithmic scale as a modulus with a solid (dashed) line for a positive 
(negative) power. $P^{\rm SP,S}_0$ is identically equal to the power dissipated 
by the effect of the bath on the spin system ($P^{\rm S,B}_0$).}
\label{fdcef}
\end{figure}

In Fig.~\ref{fdcef} we show how the energy flows of Eqs.~\eqref{eq:p-lp}, 
\eqref{eq:p-pb}, \eqref{eq:p-psp}, \eqref{eq:p-sps}, \eqref{eq:p-sb}, and 
\eqref{eq:def-flow} depend on the driving frequency. All of the powers we 
compute obey the steady-state sum rules of Eqs.~\eqref{esr1}, \eqref{esr2}, 
and \eqref{eq:bal-sp}, which describe every stage of the process. This is 
clearest when considering the spin system, shown in Fig.~\ref{fdcef}(b), 
where the forms of $P_0^\text{SP,S}$ and $P_0^\text{S,B}$ reflect the 
exponential increase in its sensitivity to driving at frequencies 
near the edges of the two-triplon band, which we have seen already in 
Secs.~\ref{sdp} and \ref{str}B. This is particularly true in Fig.~\ref{fdchpa}, 
which can in fact be understood as a graph of energy absorption by the spin 
system (red colors being high values). Although both energy flows bear a close 
resemblance to Fig.~\ref{fdchpw}(a), we note from the latter that the quantity 
summed to obtain the net power contains an additional weighting factor of 
$\omega_k$; among other effects, this acts to make the heights of the peaks at 
$2\omega_{\rm min}$ and $2 \omega_{\rm max}$ more symmetrical in Fig.~\ref{fdcef}(b)
than in Sec.~\ref{sdp}.

Turning to the phonon system, in Fig.~\ref{fdcef}(a) we show the input energy 
flow from the laser, $P_0^\text{L,P}$, the energy flowing out of the driven 
phonon due to the spin system, $P_0^\text{P,SP}$, and the output flow directly 
from this phonon to the bath, $P_0^\text{P,B}$. Our first observation is that, 
in the regime of weak spin-phonon coupling considered here, the majority of 
the laser energy flows directly to the bath, while the quantity central to 
our analysis, $P_0^\text{P,SP}$, is always relatively small. Next we observe 
that it peaks around $\omega_0 = 2 \omega_{\rm min}$ and $2 \omega_{\rm max}$, as 
anticipated from Sec.~\ref{sdp}. To illustrate the relative importance of the 
energy in the spin-phonon coupling term, $H_{\rm sp}$, we show $P_0^\text{P,SP}$ 
once more as the green line in Fig.~\ref{fdcef}(b) for comparison with 
$P_0^\text{SP,S}$. Their difference, $P^\text{SP,B}$, remains at the percent 
level for all driving frequencies within the two-triplon band, indicating 
that $H_{\rm sp}$ does not act to store significant energy, but in essence 
transmits it from the phonon to the spin system as expected physically. At 
very high and very low frequencies, $|P^\text{SP,B}|$ becomes a more significant 
fraction of the energy in the spin system, but this energy is in turn a very 
small fraction of the total (laser) energy flowing through the system. We take 
these results as evidence that treating the spin-phonon term as a perturbation 
in the mean-field approach is well justified, and by extension that the neglect 
of higher spin-phonon correlations is appropriate for the relevant driving 
frequencies.

We comment in passing that $P^\text{SP,B}$ can in fact have a negative sign, 
implying a small energy flow from the bath due to the spin-phonon coupling 
term. While this may at first appear counterintuitive, we stress that the 
splitting of the system Hamiltonian into the three parts $H_\text{p}$, 
$H_\text{s}$, and $H_\text{sp}$ [Eq.~\eqref{etsh}] is somewhat arbitrary, and 
combining $H_\text{s}$ and $H_\text{sp}$ would remove this feature. In total, 
there is no violation of the fact that energy flows from the combined 
spin-phonon system into the bath, and indeed one may compute this net power, 
$P_0^\text{P,B} + P_0^\text{SP,B} + P_0^\text{S,B}$, which by the sum rules at 
each step of Fig.~\ref{fdcefschem} matches $P_0^\text{L,P}$. We do not calculate 
$P_0^\text{B,H}$, assuming simply that it matches the power flowing into the 
bath. 

\begin{figure}[t]
\centering    
\includegraphics[width=\columnwidth]{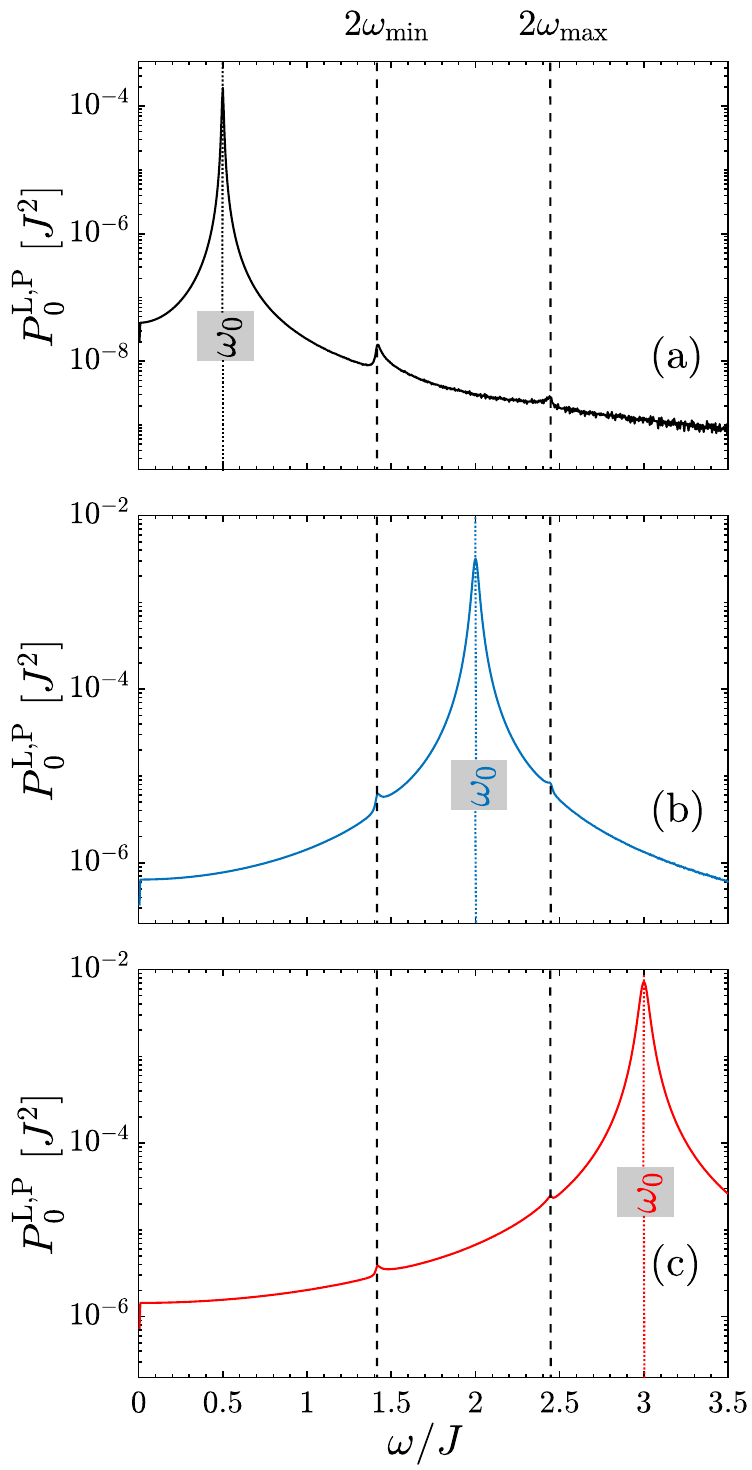}
\caption{$P^{\rm L,P}_0$ shown as a function of $\omega$ for $\gamma = 0.02 
\omega_0$, $a/\gamma = 0.2$, $g = 0.1 J$, and $\gamma_{\rm s} = 0.01 J$, for 
systems with one phonon (coupled to the $J$ bond as in Sec.~\ref{smm}) at a 
frequency (a) $\omega_0/J = 1.0$, (b) $\omega_0/J = 2.0$, and (c) $\omega_0/J
 = 3.0$.}
\label{fdces}
\end{figure}

Because $P^{\rm L,P}_0$ is the average power, or fluence, taken up by the 
combined spin-lattice system, it is closely related to quantities that 
might be measured in an absorption experiment. To make contact with 
experimental methods it is necessary to generalize our treatment. In 
comparison with a conventional pump-probe procedure, we have considered 
only the pumping step, because in a NESS there is no concept of a delay 
time before probing. Further, we have considered pumping only at the 
frequency of one hypothetical phonon, lying at any value of $\omega_0$, 
which we have varied to probe the behavior of the spin system. By contrast, 
in a real material there is only one, or a small number of, phonon(s) coupled 
strongly to the primary magnetic bonds, but it is relatively straightforward 
to pump the system at all frequencies $\omega \ne \omega_0$. Thus in 
Fig.~\ref{fdces} we depart from the conventions used so far in our study 
and adjust the frequency of the driving laser in order to illustrate the 
fluence as a function of $\omega$ for systems with one strongly-coupled 
phonon, which lies at a frequency below, in, or above the two-triplon band. 

In this type of experiment it is clear that the resonant phonon at $\omega_0$ 
dominates the absorption. However, the fingerprints of the two-triplon band 
are visible as, for the parameters chosen, 0.1\%-level effects across the full 
frequency range of the band. This non-resonant absorption is naturally stronger 
in a material where the relevant phonon lies close to the frequencies $2 
\omega_{\rm min}$ and $2 \omega_{\rm max}$. We recall that the net fluence at the 
phonon peak increases with $\omega_0^2$, and thus that pumping higher-lying 
phonons may result in a stronger signal if these are suitably coupled to 
the spin system. 

However, for driving a phonon that lies very close to a resonant frequency 
of the spin system, we draw attention to an additional phenomenon. The blue 
line in Fig.~\ref{fdcef}(a) shows that the absorption peak at $\omega = 
\omega_0$ is actually suppressed when $\omega_0$ lies in the spin band, most 
strongly so for phonons resonant with $2 \omega_{\rm min}$ and $2 \omega_{\rm max}$.
This ``self-blocking'' effect appears initially to be counterintuitive, as one 
might expect stronger absorption when more system degrees of freedom are at 
resonance with the incoming laser. However, the spin system is not coupled 
directly to the light, being excited only by the driven phonon, and this 
situation suggests a heuristic image of the spin system as an extra 
``inertia'' that the driven phonon must move. While we also used this word 
Sec.~\ref{str}A, a rather more specific description of the physics can be 
read from the prefactor of $p(t)$ in Eq.~\eqref{eompc}, where one observes 
that the spin system acts against $E(t)$, making it more difficult for the 
phonon to draw energy from the laser electric field by oscillating maximally. 
Once this self-blocking effect is taken into account, the difference between 
the blue and red dashed lines in Fig.~\ref{fdcef}(a) shows the additional 
absorption of the incoming fluence actually taken up by the spin system 
[shown again in more familiar form in Fig.~\ref{fdcef}(b)]. 

\subsection{Heating}

Both conceptually and experimentally, extended continuous driving must 
inevitably lead to heating, which without remediation would destroy the 
coherence of the system, and later the system itself. Throughout this work 
we have assumed that the heat sink represented in Figs.~\ref{fdcschem} and 
\ref{fdcefschem} will be able to maintain a constant, low system temperature 
despite the steady drive, and our brief analysis of relaxation to equilibrium 
in Sec.~\ref{str}C was predicated on this assumption. We now turn to a 
quantitative investigation of the reality of the situation in driven 
condensed-matter systems. 

We comment first on the physical meaning of the Lindblad bath model. Because 
the energy flowing directly from the driven phonon to the bath, $P^\text{P,B}$ 
in Eq.~\eqref{eq:p-pb}, is directly proportional to the phonon damping and 
phonon occupation, $\gamma n_{\rm ph0}/\omega_0$ of the phonon energy is 
transferred to the bath in every period. Thus for the parameters we use, 
an energy of $\omega_0$ per dimer is dissipated after approximately 1500 
cycles; we recall that in our model the Einstein phonon modes are present on 
every bond in the system, meaning that the laser driving is a bulk effect. To 
introduce some typical numbers for quantum magnetic materials, we consider 
the inorganic compound CuGeO$_3$, which forms a quasi-1D spin-1/2 system and 
has been well characterized in the context of quantum magnetism at equilibrium. 
In fact CuGeO$_3$ was studied in detail \cite{rhtu,rwgb} due to its spin-Peierls
behavior, by which is meant that it shows a lattice transition from a uniform 
to an alternating chain that is driven by reducing the energy in the magnetic 
sector. This type of transition is a ground-state phenomenon not related to 
the phonon driving we consider here, and our present analysis would in 
principle be applicable to the distorted state; however, CuGeO$_3$ also 
possesses second-neighbor and interchain interactions, and thus the nature 
of the transition has remained the subject of some debate. Although CuGeO$_3$ 
is known for its strong spin-phonon coupling, meaning that its $g$ value lies 
outside the weak-coupling regime we consider here (for analyzing driven, 
out-of-equilibrium physical properties), we borrow its thermal parameters 
to analyze energy transfer and heating.

CuGeO$_3$ has leading magnetic exchange constants of approximately 10 meV 
\cite{uhrig97a}, and so we use this value for illustrative energy estimates. 
Taking the triplon band width into account, we will assume that a phonon 
driving the system near $2\omega_{\rm min}$ also has $\hbar \omega_0 = 10$ meV 
$\approx 2.4$ THz $\approx 80$ cm$^{-1}$. The phonon spectrum of CuGeO$_3$ 
contains Raman and IR-active modes over a wide range of frequencies 
\cite{Popovic95}, with the lowest-lying IR-active modes at 6.0, 11.9, and 
16.4 meV. Because all of these normal modes involve some motion of all the 
atoms in the unit cell, a quantitative model of the type we consider would 
have $J$ and $J'$ modulated simultaneously by the driven phonon. Proceeding 
with the illustrative phonon frequency of 10 meV, from Eq.~\eqref{eq:p-pb} 
we obtain for the parameters of Secs.~\ref{sness} and \ref{sdp} ($n_{\rm ph0}
 = 0.04$ and $\gamma = 0.02 \omega_0$) that the energy deposited in the bath 
by the driving phonon is 
\begin{eqnarray}
\label{eedb}
P^\text{P,B}_0 & = & 1.95 \! \times \! 10^{-11} {\rm ~Js}^{-1} {\rm ~per~dimer} 
\nonumber \\ & = & 5.89 \! \times \! 10^{12} {\rm ~W~per~mole~of~spins.}
\end{eqnarray}

As noted in App.~\ref{app:lc}, this value of $n_{\rm ph0}$ poses no risk of 
melting the lattice. However, to understand the effect of the energy in 
Eq.~\eqref{eedb} on the lattice temperature, we use the result \cite{Lorenz97}
that the low-temperature specific heat of CuGeO$_3$ is given approximately by 
the standard pure-phonon form, $C = \beta T^3$, with prefactor $\beta \approx 
0.3$ mJ/(mol K$^4$). Thus the time required for the driven system to reach a 
temperature $T_{\rm max}$ in the absence of any cooling apparatus would be
\begin{equation}
t_{\rm h} = \frac{\beta}{4P_0^\text{P,B}} \, [T^4_{\rm max} - T^4_{\rm init}] 
\approx 1.2 \! \times \! 10^{-17} \, [T^4_{\rm max} - T^4_{\rm init}] \: 
\frac{\text{s}}{\text{K}^4}. 
\end{equation}
Starting at $T_{\rm init} = 0$ or 2 K, the time required to reach a temperature 
$T_{\rm max} = 20$ K is $t_{\rm h} = 2.04 \! \times \! 10^{-12}$ s. We assume that 
this $T_{\rm max}$ is a realistic estimate of the temperature where one could no 
longer argue for quantum coherence of spin processes taking place in a triplon 
band whose minimum lies at 5 meV. The resulting $t_{\rm h}$ corresponds to only 
5 cycles of the driving phonon and is clearly too short by a factor of several 
hundred when compared with the results of Secs.~\ref{sness} to \ref{str}. 

Even allowing for considerable latitude with system parameters and materials 
choices, it is clear that the study of spin NESS in a quantum magnet is not 
realistic without an efficient heat sink attached to the sample 
(Figs.~\ref{fdcschem} and \ref{fdcefschem}). To address the effect of 
the heat sink, it is necessary to introduce further materials parameters, 
specifically for sample dimensions and the thermal conductivities removing 
heat from the sample. As will shortly become clear, there are two reasons 
why an experiment of the type we analyze is relevant for a very thin sample, 
and thus we illustrate the heat flow for a thickness of 20 nm. Using that the 
mass of one mole of spins in CuGeO$_3$ is 184 g and the density is 5.11 
g$\,$cm$^{-3}$ \cite{Ecolivet99}, a sample of area $A = 1$ mm$^2$ would be 
$5.54 \times 10^{-10}$ moles of CuGeO$_3$, meaning from Eq.~\eqref{eedb} that 
a laser power 
\begin{equation}
P_{\rm laser} = 3.26 \text{~kW},
\end{equation}
should be transported through this area to the heat sink. First for the sample 
itself, the thermal conductivity of CuGeO$_3$ at low temperatures is neither 
constant nor isotropic, but an approximate value for the cross-chain ($b$-axis) 
direction is $\kappa = 0.1$ W/(K cm) \cite{Hofmann02}. For the rate at which 
heat leaves the sample, we compute
\begin{equation}
P_{\kappa} = \kappa A \Delta T /\Delta l = 9.0 \text{~kW},
\end{equation}
where we have set $\Delta T = T_{\rm max} - 2 = 18$ K as the temperature 
difference across the sample ($\Delta l = 20$ nm). Thus the qualitative 
conclusion from this crude estimate is that the thermal conductivity of the 
sample can match the power to be dissipated if a sufficiently thin film can 
be prepared. In slightly more detail, an energy-flow balance would dictate 
that $\Delta T$ should stabilize around 8.5 K. 

This worked example illustrates that $P_{\rm laser}$ is directly proportional 
to $\Delta l$ and $P_{\kappa}$ inversely proportional, making the film thickness 
a crucial parameter. Nevertheless, the penetration depth of light into 
insulating matter is not well characterized for frequencies where the light 
is resonant with phonon excitations, and further with the spin sector. As a 
consequence, a thin film is indeed the most reliable geometry for ensuring 
that the bulk is uniformly irradiated by the incident laser beam. Materials 
that are difficult to prepare as thin films therefore suffer the twin 
disadvantages that their thermal conductivity becomes a bottleneck in the 
energy-flow process and that attenuation of the laser electric field inside 
the sample becomes a concern. While a significantly more detailed and 
materials-specific analysis is required for planning an experiment, our 
considerations indicate that it it always possible to study spin NESS in 
thin-film systems. 

Certainly an optimized cooling system is a prerequisite for such studies, 
even at the nominally weak driving strengths ($n_{\rm ph0} = 0.04$) we have 
considered in Secs.~\ref{sness} to \ref{str}. The heat sink should be a 
highly conducting metal able to remove the input power efficiently, and 
thus no bottleneck should arise due to its thermal contact or the thermal 
conductivity. However, metals are not known to have a high heat capacity, 
and thus we estimate the thermal energy that could be taken up by a metal 
block. We consider high-quality Al (residual resistivity ratio ${\rm RRR}
 = 30$) and note first that $\kappa_{\rm Al} = 1$ W/(K cm) \cite{Duthil14}, 
which is well in excess of the value in CuGeO$_3$. The specific heat has the 
form $C = \gamma_{\rm Al} T$ with $\gamma_\text{Al} \approx 0.05$ J/(kg K$^2$) 
at low temperatures \cite{Duthil14}, and hence for a block with area 1 cm$^2$ 
and thickness 5 mm (giving a mass $m = 1.35$ g \cite{Duthil14}), the energy 
absorbed by increasing the temperature from $T_{\rm init} = 2$ K to $T_{\rm hs}$ is 
\begin{equation}
\Delta E = {\textstyle \frac12} \gamma_\text{Al} m [T_{\rm hs}^2 - T_{\rm init}^2] 
\; = \; 7.1 \times \! 10^{-4} \: \text{J}  
\end{equation}
if the temperature of the heat sink is limited to 5 K. With an input power 
of 3.26 kW, the time to overheating of the block is 
\begin{equation}
t_{\rm sink} = 2.2 \! \times \! 10^{-7} \: \text{s},
\end{equation}
which corresponds to over half a million cycles of the 2.4 THz driving phonon. 
Thus an Al heat sink has plenty of reserve capacity for the purposes of a NESS 
experiment. 

Returning to the beginning of the experimental process depicted in 
Figs.~\ref{fdcschem} and \ref{fdcefschem}, we have assumed that the parameter 
$a$ is freely variable. The maximum electric field in a modern THz laser 
source is approximately $E = 3 \! \times \! 10^8$ Vm$^{-1}$ \cite{Giorgianni20}. 
If one assumes that the field acts on an oxygen ion, one obtains
\begin{equation}
a = {\textstyle \frac{1}{\sqrt{2}}} 2 e E q_\text{osc} = 4.8 \text{~meV}
\end{equation}
where $q_\text{osc} = 1.1 \! \times \! 10^{-11}$ m when computed using $M_{\rm O}$.
For comparison, the value we have used to ensure $n_{\rm ph} = 0.04$ in 
Secs.~\ref{sness} to \ref{str} corresponds with $\hbar \omega_0 = 10$ meV 
to $a = 0.16$ meV. Thus even for predominantly reflective surfaces, values 
of $a$ suitable for probing the energy range of $J$ and $\omega_0$ typical 
in inorganic quantum magnets are readily achievable. 

In summary, experiments of the type we discuss to establish and to control 
bulk quantum spin NESS are possible in real magnetic materials. The sole 
caveat is that it should be possible to prepare the system with a thickness in 
the range of tens of nanometres. Even at the rather modest phonon occupations 
required to observe nontrivial nonequilibrium spin states, maintaining the 
spin system at a low temperature over a long period of steady driving does 
pose a significant challenge to the cooling capacity of a conventional cold 
finger, which normally is designed to control the system temperature with 
high precision using liquid $^4$He coolant, rather than functioning as an 
optimized heat sink. We assume that both of the issues we have identified 
can be solved for a wide range of quantum magnets. However, in the event 
of a materials system that does not allow the driving energy to be removed 
quickly enough to avoid heating, one solution may lie in altering the 
experimental geometry away from laser irradiation of the entire sample, 
as we discuss in more detail in Sec.~\ref{sd}B. 

\section{Discussion}
\label{sd}

\subsection{Approximations: Time, Coupling, and Intensity Scales}

In constructing our description of the phonon-driven and dissipative 
quantum magnetic system we have appealed to a number of approximations. 
In fact establishing the validity of the framework presents an interlinked 
problem involving (i) the treatments we have adopted for the laser, for the 
spin and phonon sectors, and in the master-equation method, (ii) the fast 
and slow timescales of the spin-lattice system, (iii) the coupling constants, 
and (iv) the intensities or mode occupations. As examples, the magnetic 
interactions determine our treatment of the dimerized chain, a relatively 
weak spin-lattice coupling is intrinsic to our treatment of both sectors, 
the timescales of the dynamics in these sectors should allow the Born-Markov 
and rotating-wave approximations within the quantum master equation, and 
if mode occupations are too high anharmonic or non-linear effects can set in. 

\subsubsection{Triplons as Bosons}

A first approximation is that we treat the triplons as non-interacting bosons, 
diagonalizing them by a standard Bogoliubov transformation. The triplons in a 
system of coupled dimers are in fact hard-core bosons, because at most one may 
be present at each site, and finite inter-triplon interactions are well known 
when the quasiparticles are adjacent to each other in real space. However, for 
relatively low densities (below our threshold of $n_{\rm x}^{\rm max} = 0.2$) and 
weak inter-dimer coupling, $\lambda = J'/J \le 0.5$, approximating the triplons 
as interaction-free bosons is well justified \cite{rnr}, as discussed in 
Sec.~\ref{smm}A. To study the regime of larger inter-dimer coupling, the 
standard Bogoliubov transformation can be replaced by a unitary transformation 
controlled to high orders in $\lambda$ \cite{rku,rksu,rkdu}.

\subsubsection{Laser and mean-field decoupling}

As noted in Sec.~\ref{smm}, we have described the laser field driving the 
optical phonon as a classical oscillating field. In view of the fluences 
commonly used in experiment, which make the quantum fluctuations of the 
laser field negligible relative to its expectation value, this approximation 
is perfectly justified. The time-dependent mean-field decoupling of the 
driven phonon and the spin system [Eq.~\eqref{eq:mf-split}] is a further 
approximation, although we have demonstrated in Sec.~\ref{seh} that it is 
well justified at all relevant driving frequencies. From the definition of 
Eq.~\eqref{eq:expval-phon}, $O(10^{-2})$ values $n_\text{ph}$ on every bond 
mean that the optical phonon is macroscopically occupied (the phonon number, 
being proportional to the system size, is extensive). Thus the relative size 
of the quantum fluctations is again negligible, justifying a mean-field 
treatment of the phononic field. While we cannot exclude completely that more 
complex physics occurs for particularly large spin-phonon coupling, such as 
triplon-phonon bound-state formation, this would need to be built first into 
the ground states and then into the driven dynamics. 

\subsubsection{Lindblad damping of driven phonon}

The damping rate, $\gamma$, of the driven phonon should be significantly
smaller than its energy, and we have set it to a value of order 2\% of the 
phonon energy. The way that $\gamma$ enters, which leads to the description of 
the driven phonon as a classical damped harmonic oscillator, is well justified 
for the reasons stated in the preceding paragraph. The Lindblad framework 
treats the relaxation of a degree of freedom by using its established damping
term. More generally, fundamental theorems about the Lindblad formalism 
\cite{rbp} state that the dynamics of an open quantum system can always be
captured by decay rates for certain Lindblad operators, which we labelled 
$A_l$ in Eq.~\eqref{eaqme}; at this level, the only open issue is which 
operators in system Hamiltonian are the relevant Lindblad operators, but 
this is manifestly obvious for the driven Einsein phonon considered here. 

A more physical question concerns the microscopic mechanism of this damping.
Clearly, the only bath in an insulator at the energies we consider consists 
of the optical and acoustic phonons. Due to anharmonic effects, the driven 
phonon can decay into two (or more) other phonons. Compared to the single 
driven phonon, the large number of other phonons in a 3D material constitute 
a large bath, which is not strongly influenced by the driven phonon, except 
in cases where the driving exceeds the heat-sink capacities in the sense of 
Sec.~\ref{seh} and heating effects enter. The fact that the phonon lines 
observed in inelastic scattering studies are usually rather sharp 
demonstrates that the coupling of a given phonon to the bath represented 
by the other phonons is weak, as a result of which the Born approximation 
is fully justified. 

A further required property of the bath is to obey the Markov approximation, 
that its correlations should decay significantly faster than the decay 
dynamics of the quantum system (specified by the Hamiltonian). This property 
is difficult to verify without a detailed knowledge of all the phonons and 
their anharmonicities, but an estimate is possible. In general the spectrum 
of phonons covers the energy range from zero to the Debye energy, $\hbar 
\omega_\text{D}$, and hence $1/\omega_\text{D}$ sets the timescale for the 
decay of bath correlations. The Debye energy is typically 50-100 meV (12-24 
THz). For driving frequencies in the 1-10 THz regime and a decay rate, 
$\gamma$, which is 1/50 of these, it is clear that the correlation time 
scale, $1/\omega_\text{D}$, is indeed shorter than the time scale $1/\gamma$ 
of the phonon damping (with the possible exception of very soft materials). 
Finally, the rotating-wave approximation is the statement that one may 
neglect fast oscillations to focus only on the slow variables (as we did in 
Sec.~\ref{str}A) \cite{rbp}, and again this is clearly justified because the 
oscillatory terms for a phonon driven at $\omega_0$ are fast on the timescale 
of the damping. 

\subsubsection{Lindblad damping of triplons}

The previous arguments can be repeated to justify the use of $t_k$ as the 
Lindblad operators in the spin sector. We observed in Sec.~\ref{smm}C that 
these operators break spin conservation, so that our treatment is relevant 
for systems with finite spin-orbit coupling. If this coupling is low, spin 
conservation requires that one consider terms of the type $C_{kq} = t_k^\dag 
t_q$, which we discuss in the next subsection (Sec.~\ref{sd}B). However, weak 
spin-orbit coupling also implies that the damping of spin excitations due to 
a phononic bath is weak, and thus it is not unreasonable to treat any magnetic 
excitations, which here are the triplons, as weakly damped oscillators. Because 
the strongest effects of driving the magnetic system occur when the driving 
frequency, $\omega_0$, matches the magnetic energies, $2\omega_k$, estimates 
for the validity of the Born-Markov and rotating-wave approximations 
\cite{rbp} are the same as for the driven phonon. We defer comments on 
momentum conservation in $C_{kq}$ to Sec.~\ref{sd}B.

In summary, while the validity issue is a complex one, all of the 
approximations we have made are appropriate, and in fact for a typical 
condensed-matter system there is a reasonable amount of parameter space 
(Sec.~\ref{seh}). A broader discussion of materials and experiment may be 
found in Sec.~\ref{sd}C below. From a theory standpoint, our current approach 
is by design the simplest available, whose explicit intent is to establish the 
basic phenomenology, and a more detailed discussion of any given issue may 
require more sophisticated methodology. One example of this would be the use 
of flow-equation methods \cite{rvlbf} to extend the regimes of validity, in 
the hierarchy of timescale approximations, of the equations of motion. 

We close this part of the discussion by recalling that the intrinsic properties
of the phonon-driven spin system lead to a number of phenomena occurring over a 
range of different frequencies and times. By frequency, the key regimes of 
driving are (i) in the spin band, where the response is resonant, (ii) below 
the spin band, where it is controlled by multiphonon processes, and (iii) above 
the spin band, which is the Floquet regime, featuring weak energy absorption 
and coherently superposed phase- and frequency-shifted states. By time, 
transient phenomena at switch-on occur (mostly) on the scale of the inverse 
damping (Sec.~\ref{str}A), drive-induced heating occurs on a strongly 
$\omega_0$-dependent timescale, and relaxation phenomena at switch-off 
follow the Lindblad form to restore the starting state (Sec.~\ref{str}C). 

\subsection{Bath Models and System Heating}

The equations of motion whose solutions we have studied in Secs.~\ref{sness} 
to \ref{str} are intrinsic to one type of bath. As stated in Sec.~\ref{smm}C, 
the physical content of the Lindblad formalism is that the spin operators are 
damped by bath operators that also appear in the spin Hamiltonian. However, 
the nature of these terms must reflect the physics of the entire system, by 
which is meant the manner in which energy can be dissipated by spin and 
phononic processes. Here we comment briefly, and with one specific example, 
on how our analysis would be extended in the case of more complex bath terms. 

It is clear in Sec.~\ref{smm}C that our use of $t_k$ as the sole type of 
spin-bath operator delivers the most straightforward equations of motion, 
and we have exploited the complete independence of all $k$ states to 
explore a wide range of phenomena. However, it is also clear that damping 
processes involving a single $t_k$ operator are spin-non-conserving, 
allowing a triplon to decay into a phonon. In a truly 1D system, meaning 
a spin chain in a 1D lattice, momentum conservation would restrict the 
phase space available for such processes, raising problems with the 
applicability of the Lindblad formalism (which requires that the bath 
contain a continuum of energy states \cite{rbp}) that would at minimum 
mandate a significant $k$-dependence of the decay rate, $\gamma_\text{s}(k)$. 
However, as explained in Sec.~\ref{smm}C, such concerns are not relevant to 
the present analysis because the 1D triplons are embedded in a real lattice, 
meaning with 3D phonons. In this case, for each momentum, $\hbar k$, along 
the spin chains, there is a continuum of perpendicular momenta, $\hbar 
\vec{k}_\perp$, and hence the annihilation of the triplon can occur for a 
wide range of $\vec{k}_\perp$ values, with the bath phonons that are created 
covering a broad energy range. This energetic continuum will depend on $k$, 
but only weakly, which both justifies using the Lindblad formalism and 
indicates a constant, momentum-independent damping rate, $\gamma_\text{s}$.

Nevertheless, the conventional spin-phonon coupling in any 3$d$ 
transition-metal compound, and hence the spin-damping effect of its 
phononic modes, takes the form of $H_\text{sp}$ in Eq.~(\ref{ehp}), and the 
spin-isotropic nature of this interaction means that phonon modes cannot 
alter the spin state (the number of excited triplons) directly. The most 
straightforward spin-conserving bath operators appropriate to this situation, 
based on the operators $t_k$ in the spin-system Hamiltonian, would be of the 
form $C_{kq} = t_k^\dag t_q$, and for the reason given above need not be 
momentum-conserving. Bath operators with the form of $C_{kq}$ manifestly act 
to mix wave-vector states of the system and thus lead to a significantly more 
involved set of equations of motion, with in general $N^2$ coupled equations 
rather than only $N$. We defer a detailed analysis of this case to a follow-up 
study. 

As already noted, the type of bath studied in the present work provides a 
meaningful description of systems with spin-dependent phonon scattering 
processes. These can arise in systems with appreciable spin-orbit coupling, 
meaning 4$d$ and especially 5$d$ magnetic ions, where the resulting 
anisotropic interactions include may Dzyaloshinskii-Moriya (DM), exchange 
anisotropies (XXZ and XYZ), or even bond-selective interactions. However, 
only in rather exceptional circumstances would these dissipative channels 
be stronger than spin-conserving damping terms, and thus the consideration 
of more advanced bath operators is required to discuss real experiments. 
In addition to the question of spin conservation, it is also necessary to 
address the issue of spin-system dimensionality, which as in the present 
study may be lower than the phonon-system dimensionality (which is 3D), and 
hence to establish the level at which to enforce conservation of momentum. 

The nature of the bath reflects directly on the heating of the system, which 
was discussed for the simple, spin-non-conserving case in Sec.~\ref{seh}. In 
a more complex bath, one may expect the redistribution of energy through the 
modes of the spin system to be more efficient, although this is in general 
a small contribution to the (phonon-dominated) system temperature. Of more 
practical relevance to the issue of quantum coherence is the fact that a 
momentum-mixing bath operator would also impact the coherence of individual 
$k$-components of the spin system. From this standpoint one may consider 
``reservoir engineering'' \cite{Diehl08}, meaning influencing the form of 
$C_{kq}$ (for example by promoting forward-, backward-, or skew-scattering 
in the bath), as an alternative to controlling the system temperature only 
through the balance between the laser driving strength and the cooling 
apparatus (Sec.~\ref{seh}). 

With a view to maintaining quantum coherence in the spin sector, we turn 
to some more general considerations for controlling the temperature of the 
system. We remark that in some classes of system it is possible to decouple 
the degrees of freedom in such a way as to obtain effective electronic or 
spin temperatures different from the lattice temperature. However, this is 
not an option in a system of spins localized on the sites of a lattice, where 
the temperature controlling the response of the spin system is that of the 
lattice. Similarly, our model is also far from the paradigm of a spatially 
separate system and bath, where different effective temperatures for the 
two components are related by controllable coupling constants. Within the 
confines of the situation we consider, we mention two approaches to
temperature control, namely system geometry and the laser driving protocol. 

In the present work we have considered only bulk driving by the electric 
field of the laser, meaning that the Einstein phonon of every bond is 
stimulated. In Sec.~\ref{seh}B we showed that this ``bulk'' system should in 
fact be rather thin (tens of nanometres). However, it is certainly possible 
that a device of $\mu$ or mm length is illuminated only at one end, causing 
the phonon and spin excitations to propagate through the equilibrium material 
over a distance far larger than the nonequilibrium irradiated portion, and 
possibly larger than the penetration depth of the light. Such a situation 
would require a model for spatial gradients of heat, magnetization, and 
temperature, which would certainly be of direct interest for switching and 
transport in spintronic devices. To date some experiments already present 
this type of situation \cite{rjbasyzb}, and certain theoretical discussions 
have also invoked the framework of driving only at the ends of the system 
\cite{rp,rzpp,rbkvm}.

Finally, another means of controlling the system temperature lies in driving 
by repeated short pulses. In its simplest form, this allows the system to 
relax back to its cold state by the action of the heat sink (Sec.~\ref{str}C), 
although the required pulse separation would be a very slow timescale 
(Sec.~\ref{seh}B). At a more sophisticated level, pulsed driving processes 
allow new degrees of control over the system, including the imposition of 
dynamics on new timescales quite separate from the driving frequency, as in 
the case of Floquet engineering of the electronic band structure \cite{rok}. 
While certain types of driving protocol have already been proposed for 
controlling small numbers of quantum spins \cite{ru} and ensembles of 
effectively $S = 1/2$ quantum dots \cite{Greilich07,rkea}, for now we leave 
open the application of these ideas to the many-body spin systems considered 
here. 

\subsection{Experiment}

As stated in Sec.~\ref{sintro}, the last decade has seen an enormous expansion 
in the technological capabilities of laser sources, both in ultrafast 
timescales and in high intensities, with applications both for pump-probe 
experiments and for steady driving. Where in Sec.~\ref{sintro} the focus 
of our remarks was the new physics made possible by these new sources, it is 
also worth commenting on the new technologies that have led to such growth in 
the application of lasers to condensed matter. For decades this was limited 
by the ``Terahertz gap,'' the problem that light at the energies of most 
interest to the intrinsic processes in condensed matter was neither easily 
generated nor easily guided or focused, but was easily absorbed and scattered. 
Starting with the initial compilation of methods making it possible to 
engineer transient states of condensed matter \cite{rktn}, further  
technological solutions have been developed and applied to frontier science 
challenges \cite{rzsz}. The best review of terahertz enabling technologies, 
both for generation and for beam control, may be found in Ref.~\cite{Salen19}. 
On the generation side, one has not only new free-electron laser sources but 
also a range of new ``table-top'' techniques, including plasma-based sources 
and high-harmonic generation, many made possible by exploiting new materials.
On the control side, beam guiding, transport, focusing, and diagnostics 
have also benefited strongly from the optical properties of certain materials. 
Together this progress has led to a qualitative expansion in the type of 
physics that can be probed, or indeed created, in condensed matter, and 
the aim of the present study is to extend these capabilities to quantum 
magnetic materials. 

For the purposes of this preliminary analysis, we have focused on 
well-dimerized (and thus robustly gapped) quantum spin chains, meaning 
that the system we consider does not, either at equilibrium or in its driven 
state, approach a phase transition to a magnetically ordered, to a gapless 
quantum disordered, or to any other different magnetic state. In a 1D system 
with only Heisenberg spin interactions, the primary requirement is simply that 
the spin gap (the one-triplon band minimum, $\omega_{\rm min} = \omega_{k = 0}$) 
does not close, including on laser driving of a selected phonon. An excellent 
example of an inorganic compound realizing quasi-1D alternating $S = 1/2$ spin 
chains is Cu(NO$_3$)$_2$ \cite{rtljenmfkct}, which is thought to have no 
anisotropy, negligible second-neighbor interactions, and a gap of approximately 
0.38 meV (compared to a band width of 0.12 meV, yielding $\lambda = 0.14$). 
This material also shows no evidence of strong phonon coupling to the spin 
excitations \cite{rtljenmfkct}, which indicates that it belongs in the 
weak-coupling regime we study. However, the magnetic energy scales in 
Cu(NO$_3$)$_2$ are lower by a factor of 20 than the test-case numbers 
presented in Sec.~\ref{seh}, presenting a different balance of slower 
heating rates, slower convergence to NESS, and altered damping ratios. 

A recently discovered class of alternating spin-chain materials includes 
AgVOAsO$_4$ \cite{Ahmed17} and NaVOAsO$_4$ \cite{Arjun19}, which have 
magnetic coupling constants in the 5 meV range. Although here $\lambda$ 
is at the upper validity limit of our present simple treatment of the 
spin chain (Sec.~\ref{smm}A), as noted earlier, more sophisticated 
approaches are available for this purpose and no part of the equations 
of motion (Sec.~\ref{smm}C) is invalidated. Another class of candidate 
systems is the set of metal-organic TTF compounds \cite{Jacobs76,Cross79}, 
and even purely organic TCNQ compounds \cite{Huizinga79}, in which the 
spin-Peierls transition has been observed and the distorted (low-temperature) 
state is an alternating spin chain. A further category of interest in quantum 
magnetism has been alternating antiferromagnetic-ferromagnetic (AF-FM) $S = 
1/2$ chains, primarily because of a tendency to Haldane physics in the strong 
FM regime, but in the remainder of the parameter space, which includes the 
materials Na$_3$Cu$_2$SbO$_6$ ($\lambda = - 0.79$) \cite{Miura06} and 
(CH$_3$)$_2$NH$_2$CuCl$_3$ ($\lambda = - 0.92$) \cite{Stone07}, our analysis 
remains fully applicable regardless of the signs of the interactions. 

In addition to CuGeO$_3$, which we introduced in Sec.~\ref{seh} to consider 
its thermal properties, (VO)$_2$P$_2$O$_7$ \cite{run1,run2} constitutes a 
further system that in fact realizes alternating $S = 1/2$ spin chains 
with significant interchain interactions. However, the anomalously large $g$ 
values that made both of these compounds attractive for equilibrium experiments 
in quantum magnetism do place them outside the weak-coupling regime we analyze 
here. In a later study we will extend our considerations to the regime of 
strong spin-phonon coupling, which is also described by the equations of motion 
derived in Sec.~\ref{smm}. Here one may anticipate nonlinear driving effects 
(Figs.~\ref{fdcnph} and \ref{fdcfd}), which could allow experiments at lower 
laser intensities, stronger mixing of frequency harmonics (Figs.~\ref{fdcft} 
and \ref{fdcftc}), stronger below-band multi-phonon processes 
(Fig.~\ref{fdchpa}), more delicate driving-induced anomalous 
convergence (Sec.~\ref{str}A), and stronger transfer of spectral 
weight between different frequencies (Fig.~\ref{fdces}). 

Returning to the spin sector alone, our present formalism is readily extended 
to alternating chains with different gap-to-bandwidth ratios, although an 
accurate treatment of systems with small gaps would require a more numerical 
approach for the systematic summation of perturbative terms to high orders, 
as in the method of continuous unitary transformations (CUTs) \cite{rku,rksu,
rkdu}. Other 1D gapped spin systems to which our considerations are immediately 
applicable include even-leg $S = 1/2$ spin ladders and Haldane ($S = 1$) 
chains. A parallel class of systems that could be treated by very similar 
considerations would be that of magnetically ordered quantum spin systems, 
which includes many 2D and 3D materials of all spin quantum numbers; this 
would involve the straightforward adoption of a (constrained) spin-wave 
framework to describe the spin sector. Ordered magnetic systems also provide 
the simplest cases in which to analyze the effects of anisotropic magnetic 
interactions, such as single-ion, DM, XXZ, and other terms, which have 
recently attracted intensive interest with a view to creating topological 
magnons \cite{rsmmo,rnkkl,Wang18,Malki19}, vortices \cite{Gao17}, skyrmions 
\cite{rdkl}, and other means of encoding protected quantum information 
\cite{Chumak15}. More complex spin sectors include anisotropic systems such 
as Ising and XY models without magnetic order, gapless spin chains, and gapped 
or gapless non-ordered states in higher dimensions, meaning in the former 
category Z$_2$ quantum spin liquids and in the latter algebraic spin liquids 
and quantum critical systems. Here the challenge is not only to find a suitable 
framework in which to describe the complex correlated spin sector, especially 
if this is changed by using laser driving to push it across a magnetic quantum 
phase transition, but also to deal with the situation where the excitations 
of the spin system extend to arbitrarily low energies, thus interacting 
strongly with even the acoustic phonons. 

In addition to an adequate treatment of the spin sector, the quantitative 
analysis of real materials will require accurate lattice dynamics calculations 
to give the phonon modes and frequencies, and the corresponding oscillator 
strengths. The normal modes can be used to estimate spin-phonon coupling 
strengths and the frequencies to choose the laser excitation parameters.
Typically, the phonon spectrum in inorganic materials extends up to the 
Debye energy, which rarely exceeds 100 meV (24 THz), while a lower limit for 
optical phonons is perhaps 5-10 meV. Spin energy scales can extend up to a 
one-magnon band maximum of 300 meV (70 THz) in cuprates, and have no lower 
limit. There is no established relationship between the two, as materials 
with predominantly high-energy phonon modes can have a very low-energy spin 
sector, and those with high spin energies need have no special phonon 
properties. However, metal-organic systems do tend to have a softer phonon 
sector, due to the nature of the interactions between weakly polar organic 
groups, and very low magnetic energies due to the long paths between magnetic 
ions. Thus it is safe to say that a very wide range of frequency scales and 
phononic modulation possibilities is available for planning experiments of 
the type we discuss. 

Finally, it is also necessary to consider how to measure all of the 
physical quantities characterizing the spin system using a terahertz laser. 
In Sec.~\ref{seh}A we presented the example of absorption of the incident 
laser fluence (Fig.~\ref{fdces}), while further quantities that are also 
functions of frequency and temperature include reflection, polarization 
rotation (due to birefringence changes or the Faraday effect), and the 
two-magnon response. It is not generally possible to probe the wave-vector 
response of the spin sector, except in special cases possessing a strong 
coupling to one well-characterized ``probe phonon'' mode. In most such 
measurements, the signal arising due to the spin system will be weak in 
comparison with the many other contributions to the total response of a 
sample, and here we point to the strong frequency-selectivity allowed by 
the phonon-coupled model, and visible in Figs.~\ref{fdcef} and \ref{fdces}, 
as the primary means of ensuring that the spin signal is detectable. 

\section{Summary}
\label{ss}

We have investigated the nonequilibrium steady quantum mechanical states of 
a lattice spin system under the continuous, coherent laser excitation of 
phonons at a single frequency and in the presence of a realistic dissipation. 
In real materials, this dissipation is dominated by the many phonons of the 
lattice system, and its inclusion at the operator level, which we effect 
within the Lindblad formalism, goes beyond much of the work on driven many-body 
systems presently in the literature. We have focused for pedagogical purposes 
on a simple example of a gapped quantum mechanical spin system, the dimerized 
spin chain, and a simple example of a driving phonon, a bulk Einstein mode, 
but stress that the framework we have established can be extended, with 
additional numerics, to fully realistic examples of both systems. By 
establishing and solving the quantum master equations governing the 
time-evolution of this model system, we have demonstrated the establishment 
of quantum spin NESS and investigated their dependence on all of the parameters 
describing the lattice and spin sectors, including their driving, dissipation, 
and coupling.

We have performed a detailed analysis of the internal dynamics of the driven 
quantum spin NESS, and of the accompanying behavior of the driven lattice 
system. We find that the NESS amplitude shows a dramatic sensitivity to the 
frequency of the driving phonon, peaking strongly at the upper and lower 
edges of the band of two-triplon excitations. Beyond the frequency, we 
characterize this response as a function of the driving electric field, the 
lattice and spin damping coefficients, and the spin-phonon coupling, which 
causes a rapid onset of strong mutual feedback between sectors. We use the 
Fourier transform to analyze the components of the spin NESS appearing at 
different harmonics of the driving frequency even in the weak-coupling regime. 
By investigating the $k$-resolved response of the spin system we demonstrate 
that resonance in frequency is also strongly $k$-selective. 

Our equations of motion are valid at all times and we use them to study both 
the transient behavior of the system when driving is switched on and the 
relaxation to equilibrium when the driving is removed. At switch-on we find 
a complex phenomenology where even the weakly-coupled system can be driven 
close to thresholds, in triplon occupation and rate of excitation, at which 
its characteristic timescales are renormalized strongly. We have computed the 
energy flow through the composite spin-lattice system, from its arrival as the 
driving laser light to its dissipative loss. The energy offers a new window on 
frequency-sensitivity, allows us to gauge the self-consistency of our analysis 
by applying sum rules, and shows an unexpected ``self-blocking'' effect, 
whereby the spin system suppresses the uptake of laser power near resonance. 
Because the Lindblad formalism gives direct access to the energy flow into the 
bath, we have used our framework to estimate heating timescales and hence the 
practical requirements, in the form of driving limits, sample geometry, and 
cooling capacity, of a NESS experiment in a real material.

The framework we establish makes it possible to perform quantitative 
investigations of many different types of spin system, including those with 
magnetic order, with small or vanishing gaps, with topological properties, 
or with nontrivial quantum entanglement. It also enables the analysis of more 
complex types of bath, most notably ones describing spin-conserving dissipative 
processes, and hence the modelling of real materials with laser-driven phonons. 
With appropriate treatment of spatial gradients (in driving, magnetization, 
and temperature), one may also model real device geometries, leading to 
spintronic applications where the challenge is to preserve quantum coherence 
over the timescale required for reading and processing the quantum information 
encoded in the spin sector.

\begin{acknowledgments}
We thank D. Bossini, F. Giorgianni, S. Haas, C. Lange, Z. Lenar\v{c}i\v{c}, 
A. Rosch, Ch. R\"uegg, and B. Wehinger for helpful discussions. Research at 
TU Dortmund was supported by the Deutsche Forschungsgemeinschaft (DFG, German 
Research Foundation) through Grant No.~UH 90-13/1 and together with the Russian 
Foundation of Basic Research through project TRR 160. Research at the University
of G\"ottingen was funded by the DFG through Grant No.~217133147/SFB 1073, 
project B03. Research at LANL was supported by the U.S. DOE NNSA under Contract 
No.~89233218CNA000001 through the LANL LDRD Program.
\end{acknowledgments}

\begin{appendix}

\section{Detuned Triplon Pair Creation}
\label{app:detuned}

In Sec.~\ref{str}A we analyzed the gradual change of the diagonal triplon 
components, $u_k(t)$, in the resonant case, by which is meant for $\omega_0 
 = 2\omega_{k_\text{res}}$. Henceforth we omit the subscript $k$. If this 
resonance condition is not met, we state that there is a finite detuning, 
$\delta = 2\omega - \omega_0$. In this appendix we present the differences 
between the derivations for the detuned and resonant cases (Sec.~\ref{str}A). 
In the detuned case, Eq. \eqref{eq:z-approx} still holds, but the definition 
of Eq.~\eqref{eq:F-def} changes to 
\begin{equation}
F(t) = \int_0^t \big( {\tilde u} (t') + {\textstyle \frac{3}{2}} 
e^{\gamma_\text{s} t'} \big) e^{-i \delta t'} dt',
\end{equation}
which implies that $F(t)$ is no longer real, but complex. Equation 
\eqref{eq:q-eq1} still holds, but Eq.~\eqref{eq:q-reson} is modified to 
\begin{equation}
\label{eq:q-detuned}
{\tilde u}(t) = e^{\gamma_{\rm s} t} u(t) \; = \; \Gamma^2 \, {\rm Re} \Bigg[ 
\int_0^t e^{i\delta t'} F(t') dt' \Bigg],
\end{equation}
with $\Gamma$ as defined in Eq.~\eqref{eq:Gamma-def}. We stress that the 
validity of replacing the rapidly oscillating terms by their average, as 
performed in Eq.~\eqref{eq:bessel}, is justified if $|\delta| \ll \omega_0$. 
If the detuning becomes too large, the deviations from the full result may
become large, but comparing our analytical approximation to the results of 
a numerical integration revealed very good agreement over a broad range of 
parameter space.

Equation \eqref{eq:q-detuned} does not lead to a closed differential equation 
by double differentiation because of the restriction caused by taking the real 
part. It is necessary instead to take three derivatives of ${\tilde u}(t)$ 
and define $x(t) = d{\tilde u}/dt(t)$, for which we obtain
\begin{equation}
\label{eq:diff-detuned}
\frac{d^2x}{dt^2}(t) = (\Gamma^2 - \delta^2) x(t) + {\textstyle \frac{3}{2}} 
\Gamma^2 \gamma_\text{s} e^{\gamma_\text{s}t}.
\end{equation}
Clearly, the nature of the solutions to this differential equation depends 
crucially on the sign of the prefactor, $\Gamma^2 - \delta^2$. As in the 
resonant case (Sec.~\ref{str}A), if it is positive then exponentially 
increasing and decreasing functions appear, whereas if it is negative then 
oscillating trigonometric functions appear. Solving Eq.~\eqref{eq:diff-detuned} 
for the initial conditions $x(0) = 0$ and $dx/dt(0) = 3 \Gamma^2/2$ and then 
integrating the resulting expression yields ${\tilde u}(t)$, from which the 
expressions given for $u(t)$ in Eqs.~\eqref{eq:u-regimeI} and 
\eqref{eq:u-regimeII} follow. 

\section{Lindemann Criterion}
\label{app:lc}

Breakdown of the lattice is governed by the Lindemann criterion \cite{rfl}, 
which dates to Lindemann's introduction of the concept that a solid will 
begin to melt when $\langle q^2 \rangle \approx \rho^2 a_0^2$, meaning when 
the fluctuations, $q$, of its atoms around their equilibrium positions exceed 
a fraction, $\rho$, of the interatomic distance, $a_0$. While Lindemann used 
the concept to relate $\langle q^2 \rangle$ to the melting temperature, 
$T_{\rm m}$, for our purposes it is sufficient to relate $\langle q^2 \rangle$ 
to $n_{\rm ph}$. 

It has been established for a broad range of condensed-matter systems that 
Lindemann's proposed relation holds, with the value of $\rho$ being in the 
range 0.1-0.15 \cite{gilva56,dudow08,gross12}. To relate this result with 
$n_{\rm ph}$, for a local phononic oscillation the potential energy is half 
of the total energy,
\begin{equation}
{\textstyle \frac12} M \omega_0^2 \langle q^2 \rangle = {\textstyle \frac12} 
\hbar \omega_0 (n_\text{ph} + {\textstyle \frac12}),
\end{equation}
which implies 
\begin{equation}
\label{eq:linde2}
\frac{\langle q^2 \rangle}{q_\text{osc}^2} = n_\text{ph} + {\textstyle \frac12},
\end{equation}
where $q_\text{osc}$ is the oscillator length, $\sqrt{\hbar /(M \omega_0)}$. 
The Born-Oppenheimer approximation \cite{born27} implies that the 
characteristic lengthscale of the atomic motion, $q_\text{osc}$, is a 
fraction $(m_{\rm e}/M)^{1/4}$ of the electronic lengthscale, which we identify 
crudely with $a_0$; here $m_{\rm e}$ is the mass of an electron and $M$ the mass 
of the oscillating atom. Thus one obtains from Eq.~\eqref{eq:linde2} that 
\begin{equation}
\frac{\langle q^2\rangle}{a_0^2} \sqrt{\frac{M}{m_{\rm e}}} = n_\text{ph}
 + {\textstyle \frac12} ,
\end{equation}
and hence 
\begin{equation}
n_\text{ph} + {\textstyle \frac12} \approx \rho^2 \sqrt{\frac{M}{m_{\rm e}}}. 
\end{equation}
Using the mass of the oxygen atom as a generic value for $M$, and $\rho^2 
\approx 0.02$ for the Lindemann ratio, we conclude that the phonon number 
should not exceed 
\begin{equation}
\label{eq:nph-max}
n_\text{ph} \approx 3
\end{equation}
if the sample is to remain solid. Here $n_{\rm ph}$ is the total number per 
atom of phonons polarized in one direction and the conventional Lindemann 
criterion applies because we assume the chains to be embedded in a 3D crystal 
(no low-dimensional instability need be considered). From the estimates made 
above, it is clear that the Lindemann threshold is in no way threatened by 
the driven phonon occupations we consider, and thus that the integrity of 
the periodic lattice is not an issue. 

\end{appendix}


\begin{thebibliography}{199}


\bibitem{Baltz18}
V. Baltz, A. Manchon, M. Tsoi, T. Moriyama, T. Ono, and Y. Tserkovnyak, 
Antiferromagnetic spintronics,
Rev. Mod. Phys. {\bf 90}, 015005 (2018).

\bibitem{Chumak15}
A. V. Chumak, V. I. Vasyuchka, A. A. Serga, and B. Hillebrands, 
Magnon spintronics, 
Nature Phys. {\bf 11}, 453 (2015).


\bibitem{rjqq}
D.-Q. Jiang, M. Qian, and M.-P. Qian, 
{\sl Mathematical Theory of Nonequilibrium Steady States} 
(Springer, Heidelberg, 2004).

\bibitem{ru}
G. S. Uhrig, 
Quantum coherence from commensurate driving with laser pulses and decay,
SciPost Phys. {\bf 8}, 040 (2020).

\bibitem{Rudner13}
M. S. Rudner, N. H. Lindner, E. Berg, and M. Levin, 
Anomalous Edge States and the Bulk-Edge
Correspondence for Periodically Driven Two-Dimensional Systems, 
Phys. Rev. X {\bf 3}, 031005 (2013).

\bibitem{rdp}
L. D’Alessio and A. Polkovnikov, 
Many-body energy localization transition in periodically driven systems, 
Ann. Phys. {\bf 333}, 19 (2013).

\bibitem{rnh}
R. Nandkishore and D. A. Huse, 
Many-Body Localization and Thermalization in Quantum Statistical Mechanics, 
Annu. Rev. Condens. Matter. Phys. {\bf 6}, 15 (2015).


\bibitem{Chong18}
K.-O. Chong, J.-R. Kim, J. Kim, S. Yoon, S. Kang, and K. An, 
Observation of a non-equilibrium steady state of cold atoms in a moving 
optical lattice, 
Commun. Phys. {\bf 1}, 25 (2018).

\bibitem{Labouvie16}
R. Labouvie, B. Santra, S. Heun, and H. Ott, 
Bistability in a Driven-Dissipative Superfluid, 
Phys. Rev. Lett. {\bf 116}, 235302 (2016).

\bibitem{Schreiber15}
M. Schreiber, S. S. Hodgman, P. Bordia, H. P. L\"uschen, M. H. Fisher, R. Vosk, 
E. Altman, U. Schneider, and I. Bloch, 
Observation of many-body localization of interacting fermions in a quasirandom 
optical lattice,
Science {\bf 349}, 842 (2015).

\bibitem{rec}
A. Eckardt, 
Colloquium: Atomic quantum gases in periodically driven optical lattices, 
Rev. Mod. Phys. {\bf 89}, 011004 (2017).

\bibitem{rbdz}
I. Bloch, J. Dalibard, and W. Zwerger, 
Many-body physics with ultracold gases, 
Rev. Mod. Phys. {\bf 80}, 885 (2008).


\bibitem{rg}
I. Gierz, 
Probing carrier dynamics in photo-excited graphene with time-resolved ARPES, 
J. Electron Spectroscopy and Related Phenomena {\bf 219}, 53 (2017).

\bibitem{rcr}
A. Cavalleri, 
Photo-induced superconductivity, 
Contemp. Phys. {\bf 59}, 31 (2018).

\bibitem{Buzzi18}
M. Buzzi, M. F\"orst, R. Mankowsky, and A. Cavalleri, 
Probing dynamics in quantum materials with femtosecond X-rays, 
Nature Reviews Materials {\bf 3}, 299 (2018).

\bibitem{Zong18}
A. Zong, X. Shen, A. Kogar, L. Ye, C. Marks, D. Chowdhury, T. Rohwer, B. 
Freelon, S. Weathersby, R. Li, J. Yang, J. Checkelsky, X. Wang, and N. Gedik, 
Ultrafast manipulation of mirror domain walls in a charge density wave, 
Science Adv. {\bf 4}, eaau5501 (2018).

\bibitem{Iwai03}
S. Iwai, M. Ono, A. Maeda, H. Matsuzaki, H. Kishida, H. Okamoto, and Y. Tokura, 
Ultrafast Optical Switching to a Metallic State by Photoinduced Mott Transition 
in a Halogen-Bridged Nickel-Chain Compound, 
Phys. Rev. Lett. {\bf 91}, 057401 (2003).

\bibitem{Zong19}
A. Zong, A. Kogar, Y.-Q. Bie, T. Rohwer, C. Lee, E. Baldini, E. Erge\c{c}en, 
M. B. Yilmaz, B. Freelon, E. J. Sie, H. Zhou, J. Straquadine, P. Walmsley, 
P. E. Dolgirev, A. V. Rozhkov, I. R. Fisher, P. Jarillo-Herrero, B. V. Fine, 
and N. Gedik, 
Evidence for topological defects in a photoinduced phase transition, 
Nature Phys. {\bf 15}, 27 (2019).


\bibitem{rkobfd}
T. Kitagawa, T. Oka, A. Brataas, L. Fu, and E. Demler, 
Transport properties of nonequilibrium systems under the application of 
light: Photoinduced quantum Hall insulators without Landau levels, 
Phys. Rev. B {\bf 84}, 235108 (2011).

\bibitem{rlrg}
N. H. Lindner, G. Refael, and V. Galitski, 
Floquet topological insulator in semiconductor quantum wells,
Nature Phys. {\bf 7}, 490 (2011).

\bibitem{rrbs}
P. Rodriguez-Lopez, J. J. Betouras, and S. E. Savel'ev, 
Dirac fermion time-Floquet crystal: Manipulating Dirac points, 
Phys. Rev. B {\bf 89}, 155132 (2014).

\bibitem{Iadecola13}
T. Iadecola, D. Campbell, C. Chamon, C.-Y. Hou, R. Jackiw, S.-Y. Pi, and 
S. V. Kusminskiy, 
Materials Design from Nonequilibrium Steady States: Driven Graphene as a 
Tunable Semiconductor with Topological Properties, 
Phys. Rev. Lett. {\bf 110}, 176603 (2013).

\bibitem{Sentef15}
M. A. Sentef, M. Claassen, A. F. Kemper, B. Moritz, T. Oka, J. K. Freericks, 
and T. P. Devereaux, 
Theory of pump-probe photoemission in graphene: Ultrafast tuning of Floquet 
bands and local pseudospin textures,
Nature Commun. {\bf 6}, 7047 (2015).

\bibitem{Foerst11}
M. F\"orst, C. Manzoni, S. Kaiser, Y. Tomioka, Y. Tokura, R. Merlin, and 
A. Cavalleri, 
Nonlinear phononics as an ultrafast route to lattice control, 
Nature Phys. {\bf 7} 854 (2011).

\bibitem{rscg}
A. Subedi, A. Cavalleri, and A. Georges, 
Theory of nonlinear phononics for coherent light control of solids, 
Phys. Rev. B {\bf 89}, 220301 (2014).

\bibitem{vonHoegen18}
A. von Hoegen, R. Mankowsky, M. Fechner, M. F\"orst, and A. Cavalleri, 
Probing the Interatomic Potential of Solids with Strong-Field Nonlinear 
Phononics, Nature {\bf 555}, 79 (2018).


\bibitem{rlk}
A. L\"auchli and C. Kollath, 
Spreading of correlations and entanglement after a quench in the 
one-dimensional Bose-Hubbard model, 
J. Stat. Mech. P05018 (2008).

\bibitem{rhu}
S. A. Hamerla and G. S. Uhrig, 
One-dimensional fermionic systems after interaction quenches and their 
description by bosonic field theories,
New J. Phys. {\bf 15}, 073012 (2013).

\bibitem{rpfokmm}
S. Paeckel, B. Fauseweh, A. Osterkorn, T. K\"ohler, D. Manske, and S. R. 
Manmana, 
Detecting superconductivity out of equilibrium,
Phys. Rev. B {\bf 101}, 180507(R) (2020). 

\bibitem{Schwarz20}
L. Schwarz, B. Fauseweh, N. Tsuji, N. Cheng, N. Bittner, H. Krull, M. Berciu, 
G. S. Uhrig, A. P. Schnyder, S. Kaiser, and D. Manske, 
Classification and characterization of nonequilibrium Higgs modes in 
unconventional superconductors,
Nature Commun. {\bf 11}, 287 (2020). 

\bibitem{rfz}
B. Fauseweh and J.-X. Zhu, 
Laser pulse driven control of charge and spin order in the two-dimensional 
Kondo lattice, 
to appear in Phys. Rev. B (arXiv:2002.03023).

\bibitem{rr}
M. Rigol, 
Quantum quenches and thermalization in one-dimensional fermionic systems, 
Phys. Rev. A {\bf 80}, 053607 (2009).

\bibitem{rbbw}
J. Berges, Sz. Bors\'anyi, and C. Wetterich, 
Prethermalization, 
Phys. Rev. Lett. {\bf 93}, 142002 (2004).

\bibitem{rok}
T. Oka and S. Kitamura, 
Floquet Engineering of Quantum Materials, 
Annu. Rev. Condens. Matter Phys. {\bf 10}, 387 (2019).

\bibitem{rakl}
E. Arrigoni, M. Knap, and W. von der Linden, 
Nonequilibrium Dynamical Mean-Field Theory: An Auxiliary Quantum Master 
Equation Approach, 
Phys. Rev. Lett. {\bf 110}, 086403 (2013).

\bibitem{Aoki14}
H. Aoki, N. Tsuji, M. Eckstein, M. Kollar, T. Oka, and P. Werner, 
Nonequilibrium dynamical mean-field theory and its applications, 
Rev. Mod. Phys. {\bf 86}, 779 (2014).

\bibitem{rqh} 
T. Qin and W. Hofstetter, 
Nonequilibrium steady states and resonant tunneling in time-periodically 
driven systems with interactions, 
Phys. Rev. B {\bf 97}, 125115 (2018).

\bibitem{rew}
M. Eckstein and P. Werner, 
Ultrafast Separation of Photodoped Carriers in Mott Antiferromagnets, 
Phys. Rev. Lett. {\bf 113}, 076405 (2014).

\bibitem{rmw}
Y. Murakami and P. Werner, 
Nonequilibrium steady states of electric field-driven Mott insulators, 
Phys. Rev. B {\bf 98}, 075102 (2018).

\bibitem{rhmew}
A. Herrmann, Y. Murakami, M. Eckstein, and P. Werner, 
Floquet prethermalization in the resonantly driven Hubbard model, 
Europhys. Lett. {\bf 120}, 57001 (2018).


\bibitem{rl}
G. Lindblad, On the generators of quantum dynamical semigroups, 
Comm. Math. Phys. {\bf 48}, 119 (1976).

\bibitem{rbp}
H.-P. Breuer and F. Petruccione, {\sl The Theory of Open Quantum Systems}, 
2nd Ed. (Oxford University Press, Oxford, 2007).

\bibitem{rw}
U. Weiss, {\sl Quantum Dissipative Systems}, 2nd Ed. 
(World Scientific, Singapore, 2012).  


\bibitem{Sato19}
S. A. Sato, P. Tang, M. A. Sentef, U. De Giovannini, H. H\"ubener, and A. 
Rubio, 
Light-induced anomalous Hall effect in massless Dirac fermion systems and 
topological insulators with dissipation, 
New J. Phys. {\bf 21} 093005 (2019).

\bibitem{Babadi17}
M. Babadi, M. Knap, I. Martin, G. Refael, and E. Demler, 
Theory of parametrically amplified electron-phonon superconductivity,
Phys. Rev. B {\bf 96}, 014512 (2017).

\bibitem{rllr} 
F. Lange, Z. Lenar\v{c}i\v{c}, and A. Rosch, 
Pumping approximately integrable systems,
Nature Commun. {\bf 8}, 15767 (2017). 

\bibitem{rlar} 
Z. Lenar\v{c}i\v{c}, E. Altman, and A. Rosch, 
Activating many-body localization in solids by driving with light,
Phys. Rev. Lett. {\bf 121}, 267603 (2018).

\bibitem{Peronaci20}
F. Peronaci, O. Parcollet, and M. Schir\'o,
Enhancement of local pairing correlations in periodically driven Mott 
insulators,
Phys. Rev. B {\bf 101}, 161101(R) (2020).

\bibitem{Walldorf19}
N. Walldorf, D. M. Kennes, J. Paaske, and A. J. Millis,
The antiferromagnetic phase of the Floquet-driven Hubbard model,
Phys. Rev. B {\bf 100}, 121110(R) (2019).

\bibitem{Murakami17}
Y. Murakami, N. Tsuji, M. Eckstein, and P. Werner,
Nonequilibrium steady states and transient dynamics of conventional 
superconductors under phonon driving, 
Phys. Rev. B {\bf 96}, 045125 (2017).


\bibitem{rp}
T. Prosen, 
Open XXZ Spin Chain: Nonequilibrium Steady State and a Strict Bound on 
Ballistic Transport,
Phys. Rev. Lett. {\bf 106}, 217206 (2011).

\bibitem{rzpp}
M. \v{Z}nidari\v{c}, T. Prosen, and P. Prelovšek, 
Many-body localization in the Heisenberg XXZ magnet in a random field, 
Phys. Rev. B {\bf 77}, 064426 (2008).

\bibitem{ri}
J. Z. Imbrie, 
On Many-Body Localization for Quantum Spin Chains, 
J. Stat. Phys. {\bf 163}, 998 (2016).

\bibitem{rwk}
S. A. Weidinger and M. Knap, 
Floquet prethermalization and regimes of heating in a periodically driven,
interacting quantum system, 
Sci. Rep. {\bf 7}, 45382 (2017).

\bibitem{rzhks}
B. \v{Z}unkovi\v{c}, M. Heyl, M. Knap, and A. Silva, 
Dynamical Quantum Phase Transitions in Spin Chains with Long-Range 
Interactions: Merging Different Concepts of Nonequilibrium Criticality, 
Phys. Rev. Lett. {\bf 120}, 130601 (2018).


\bibitem{Greilich07}
A. Greilich, A. Shabaev, D. Yakovlev, Al. L. Efros, I. A. Yugova, D. Reuter, 
A. D. Wieck, and M. Bayer,
Nuclei-Induced Frequency Focusing of Electron Spin Coherence,
Science {\bf 317}, 1896 (2007).

\bibitem{rkea}
I. Kleinjohann, E. Evers, P. Schering, A. Greilich, G. S. Uhrig, M. Bayer, 
and F. B. Anders,
Magnetic field dependency of electron spin revival amplitude in periodically 
pulsed quantum dots,
Phys. Rev. B {\bf 98}, 155318 (2018).


\bibitem{rsmmo}
R. Shindou, R. Matsumoto, S. Murakami, and J.-i. Ohe, 
Topological chiral magnonic edge mode in a magnonic crystal, 
Phys. Rev. B {\bf 87}, 174427 (2013).

\bibitem{rnkkl}
K. Nakata, S. K. Kim, J. Klinovaja, and D. Loss, 
Magnonic topological insulators in antiferromagnets, 
Phys. Rev. B {\bf 96}, 224414 (2017).

\bibitem{Wang18}
X. S. Wang, H. W. Zhang, and X. R. Wang,
Topological magnonics: A Paradigm for Spin-Wave Manipulation and Device Design, 
Phys. Rev. App. {\bf 9}, 024029 (2018).

\bibitem{Malki19}
M. Malki and G. S. Uhrig, 
Topological magnon band for magnonics,
Phys. Rev. B {\bf 99}, 174412 (2019).

\bibitem{rbcknns}
C. Broholm, R. J. Cava, S. A. Kivelson, D. G. Nocera, M. R. Norman, and 
T. Senthil, 
Quantum Spin Liquids, 
Science {\bf 367}, eaay0668 (2020).

\bibitem{Gao17}
S. Gao, O. Zaharko, V. Tsurkan, Y. Su, J. S. White, G. S. Tucker, B. Roessli, 
F. Bourdarot, R. Sibille, D. Chernyshov, T. Fennell, A. Loidl, and Ch. R\"uegg, 
Spiral spin-liquid and the emergence of a vortex-like state in MnSc$_2$S$_4$, 
Nature Phys. {\bf 13}, 157 (2017).

\bibitem{rdkl}
S. A. D\'{\i}az, J. Klinovaja, and D. Loss, 
Topological Magnons and Edge States in Antiferromagnetic Skyrmion Crystals, 
Phys. Rev. Lett. {\bf 122}, 187203 (2019).


\bibitem{Wadley16}
P. Wadley, B. Howells, J. \v{Z}elezn\'{y}, C. Andrews, V. Hills, R. P. Campion, 
V. Nov\'{a}k, K. Olejn\'{i}k, F. Maccherozzi, S. S. Dhesi, S. Y. Martin, T. 
Wagner, J. Wunderlich, F. Freimuth, Y. Mokrousov, J. Kune\v{s}, J. S. Chauhan, 
M. J. Grzybowski, A. W. Rushforth, K. W. Edmonds, B. L. Gallagher, and T. 
Jungwirth, 
Electrical switching of an antiferromagnet, 
Science {\bf 351}, 587 (2016).

\bibitem{Lebrun18}
R. Lebrun, A. Ross, S. A. Bender, A. Qaiumzadeh, L. Baldrati, J. Cramer, 
A. Brataas, R. A. Duine, and M. Kl\"aui, 
Tunable long-distance spin transport in a crystalline antiferromagnetic 
iron oxide,
Nature {\bf 561}, 222 (2018).

\bibitem{ryba}
H. Yu, S. D. Brechet, and J.-P. Ansermet, 
Spin caloritronics, origin and outlook, 
Phys. Lett. A {\bf 381}, 825 (2017).

\bibitem{rgkmn}
M. Gibertini, M. Koperski, A. F. Morpurgo, and K. S. Novoselov, 
Magnetic 2D materials and heterostructures, 
Nature Nanotech. {\bf 14}, 408 (2019).

\bibitem{Li13}
T. Li, A. Patz, L. Mouchliadis, J. Yan, T. A. Lograsso, I. E. Perakis, and 
J. Wang, 
Femtosecond switching of magnetism via strongly correlated spin-charge quantum 
excitations, 
Nature {\bf 496}, 69 (2013).

\bibitem{Bossini16}
D. Bossini, S. Dal Conte, Y. Hashimoto, A. Secchi, R. V. Pisarev, Th. Rasing, 
G. Cerullo, and A. V. Kimel,
Macrospin dynamics in antiferromagnets triggered by sub-20 femtosecond 
injection of nanomagnons, 
Nature Commun. {\bf 7}, 10645 (2016).

\bibitem{Bossini19}
D. Bossini, S. Dal Conte, G. Cerullo, O. Gomonay, R. V. Pisarev, M. Borovsak, 
D. Mihailovic, J. Sinova, J. H. Mentink, Th. Rasing, and A. V. Kimel, 
Laser-driven quantum magnonics and terahertz dynamics of the order parameter 
in antiferromagnets, 
Phys. Rev. B {\bf 100}, 024428 (2019).


\bibitem{rjbasyzb} 
M. J\"ackl, V. I. Belotelov, I. A. Akimov, I. V. Savochkin, D. R. Yakovlev, 
A. K. Zvezdin, and M. Bayer, 
Excitation of magnon accumulation by laser clocking as a source of 
long-range spin waves in transparent magnetic films, 
Phys. Rev. X {\bf 7}, 021009 (2017).


\bibitem{Manipatruni19}
S. Manipatruni, D. E. Nikonov, C.-C. Lin, T. A. Gosavi, H. Liu, B. Prasad, 
Y.-L. Huang, E. Bonturim, R. Ramesh, and I. A. Young, 
Scalable energy-efficient magnetoelectric spin-orbit logic,
Nature {\bf 565}, 35 (2019). 

\bibitem{Seifert18}
T. S. Seifert, S. Jaiswal, J. Barker, S. T. Weber, I. Razdolski, J. Cramer, 
O. Gueckstock, S. F. Maehrlein, L. Nadvornik, S. Watanabe, C. Ciccarelli, A. 
Melnikov, G. Jakob, M. M\"unzenberg, S. T. B. Goennenwein, G. Woltersdorf, 
B. Rethfeld, P. W. Brouwer, M. Wolf, M. Kl\"aui, and T. Kampfrath, 
Femtosecond formation dynamics of the spin Seebeck effect revealed by 
terahertz spectroscopy,
Nature Commun. {\bf 9}, 2899 (2018).


\bibitem{rss}
B. S. Shastry and B. I. Shraiman, 
Theory of Raman Scattering in Mott-Hubbard Systems,
Phys. Rev. Lett. {\bf 65}, 1068 (1990).

\bibitem{rmbe}
J. H. Mentink, K. Balzer, and M. Eckstein,
Ultrafast and reversible control of the exchange interaction in Mott 
insulators, 
Nat. Commun. {\bf 6}, 6708 (2015).


\bibitem{Melnikov18}
A. A. Melnikov, K. N. Boldyrev, Yu. G. Selivanov, V. P. Martovitskii, S. V. 
Chekalin, and E. A. Ryabov, 
Coherent phonons in a Bi$_2$Se$_3$ film generated by an intense single-cycle 
THz pulse, 
Phys. Rev. B {\bf 97}, 214304 (2018).

\bibitem{Giorgianni20}
F. Giorgianni, B. Wehinger, S. Allenspach, N. Colonna, C. Vicario, P. Puphal, 
E. Pomjakushina, B. Normand, and Ch. R\"uegg,
Nonlinear Quantum Magnetophononics in SrCu$_2$(BO$_3$)$_2$, 
unpublished (arXiv:2101.01189).


\bibitem{Singla15}
R. Singla, G. Cotugno, S. Kaiser, M. F\"orst, M. Mitrano, H. Y. Liu, A. 
Cartella, C. Manzoni, H. Okamoto, T. Hasegawa, S. R. Clark, D. Jaksch, and 
A. Cavalleri,
THz-Frequency Modulation of the Hubbard $U$ in an Organic Mott Insulator,
Phys. Rev. Lett. {\bf 115}, 187401 (2015).

\bibitem{Kennes17}
D. M. Kennes, E. Y. Wilner, D. R. Reichman, and A. J. Millis,
Transient superconductivity from electronic squeezing of optically pumped 
phonons,
Nature Physics {\bf 13}, 479 (2017).

\bibitem{Sentef17}
M. A. Sentef, 
Light-enhanced electron-phonon coupling from nonlinear electron-phonon coupling,
Phys. Rev. B {\bf 95}, 205111 (2017).

\bibitem{Grandi20}
F. Grandi, J. Li, and M. Eckstein,
Ultrafast Mott transition driven by nonlinear electron-phonon interaction,
Phys. Rev. B {\bf 103}, L041110 (2021).


\bibitem{Nova17}
T. F. Nova, A. Cartella, A. Cantaluppi, M. F\"orst, D. Bossini, R. V. 
Mikhaylovskiy, A. V. Kimel, R. Merlin, and A. Cavalleri, 
An effective magnetic field from optically driven phonons, 
Nature Phys. {\bf 13}, 132 (2017).

\bibitem{Fechner18}
M. Fechner, A. Sukhov, L. Chotorlishvili, C. Kenel, J. Berakdar, and 
N. A. Spaldin, 
Magnetophononics: Ultrafast spin control through the lattice, 
Phys. Rev. Mater. {\bf 2}, 064401 (2018).


\bibitem{rmg}
G. Mazza and A. Georges, 
Superradiant Quantum Materials, 
Phys. Rev. Lett. {\bf 122}, 017401 (2019).

\bibitem{rsrr}
M. A. Sentef, M. Ruggenthaler, and A. Rubio, 
Cavity quantum-electrodynamical polaritonically enhanced electron-phonon 
coupling and its influence on superconductivity, 
Sci. Adv. {\bf 4}, eaau6969 (2018).


\bibitem{rsb} S. Sachdev and R. Bhatt, 
Bond-operator representation of quantum spins: Mean-field theory of frustrated 
quantum Heisenberg antiferromagnets,
Phys. Rev. B {\bf 41}, 9323 (1990). 

\bibitem{rgrs} S. Gopalan, T. M. Rice, and M. Sigrist, 
Spin ladders with spin gaps: A description of a class of cuprates,
Phys. Rev. B {\bf 49}, 8901 (1994). 
 
\bibitem{rmnrs} M. Matsumoto, B. Normand, T. M. Rice, and M. Sigrist, 
Field- and pressure-induced magnetic quantum phase transitions in TlCuCl$_3$,
Phys. Rev. B {\bf 69}, 054423 (2004). 

\bibitem{rsu} K. P. Schmidt and G. S. Uhrig, 
Excitations in one-dimensional $S = 1/2$ quantum antiferromagnets,
Phys. Rev. Lett. {\bf 90}, 227204 (2003). 
 
\bibitem{rrnmnfkgsm} Ch. R\"uegg, B. Normand, M. Matsumoto, Ch. Niedermayer, A. 
Furrer, K. W. Kr\"amer, H. U. G\"udel, Ph. Bourges, Y. Sidis, and H. Mutka,
Quantum Statistics of Interacting Dimer Spin Systems,
Phys. Rev. Lett. {\bf 95}, 267201 (2005).

\bibitem{rnr} B. Normand and Ch. R\"uegg, 
Complete bond-operator theory of the two-chain spin ladder,
Phys. Rev. B {\bf 83}, 054415 (2011).

\bibitem{rfu}
B. Fauseweh and G. S. Uhrig, 
Low-temperature thermodynamics of multiflavored hardcore bosons by the 
Br\"uckner approach,
Phys. Rev. B {\bf 92}, 214417 (2015).


\bibitem{rhtu}
M. Hase, I. Terasaki, and K Uchinokura, 
Observation of the spin-Peierls transition in linear Cu$^{2+}$ (spin-1/2) 
chains in an inorganic compound CuGeO$_3$, 
Phys. Rev. Lett. {\bf 70}, 3651 (1993).

\bibitem{rwgb}
R. Werner, C. Gros, and M. Braden, 
Microscopic spin-phonon coupling constants in CuGeO$_3$, 
Phys. Rev. B {\bf 59}, 14356 (1999).

\bibitem{uhrig97a}
G. S. Uhrig,
Symmetry and Dimension of the Dispersion of Inorganic Spin-Peierls Systems, 
Phys. Rev. Lett. {\bf 79}, 163 (1997).

\bibitem{Popovic95}
Z. V. Popovi\'c, S. D. Devi\'c, V. N. Popov, G. Dhalenne, and A. Revcolevschi, 
Phonons in CuGeO$_3$ studied using polarized far-infrared and Raman-scattering 
spectroscopies,
Phys. Rev. B {\bf 52}, 4185 (1995).

\bibitem{Lorenz97}
T. Lorenz, H. Kierspel, S. Kleefisch, B. B\"uchner, E. Gamper, A. Revcolevschi, 
and G. Dhalenne,
Specific heat, thermal expansion, and pressure dependencies of the transition
temperatures of doped CuGeO$_3$,
Phys. Rev. B {\bf 56}, R501 (1997). 

\bibitem{Ecolivet99}
C. Ecolivet, M. Saint-Paul, G. Dhalenne, and A. Revcolevschi,
Brillouin scattering and ultrasonic measurements of the elastic constants 
of CuGeO$_3$,
J. Phys. Condens. Matter, {\bf 11}, 4157 (1999). 

\bibitem{Hofmann02}
M. Hofmann, T. Lorenz, A. Freimuth, G. S. Uhrig, H. Kageyama, Y. Ueda, 
G. Dhalenne, and A. Revcolevschi, 
Heat transport in SrCu$_2$(BO$_3$)$_2$ and CuGeO$_3$,
Physica B {\bf 312-313}, 597 (2002).

\bibitem{Duthil14}
P. Duthil, 
Materials Properties at Low Temperature,
CERN Yellow Report {\bf 005}, 77 (2014).


\bibitem{rku}
C. Knetter and G. S. Uhrig, 
Perturbation theory by flow equations: dimerized and frustrated $S=1/2$ chain
Eur. Phys. J. B {\bf 13}, 209 (2000).

\bibitem{rksu}
C. Knetter, K. P. Schmidt and G. S. Uhrig, 
The structure of operators in effective particle-conserving models
J. Phys. A: Math. Gen. {\bf 36}, 7889 (2003).

\bibitem{rkdu}
H. Krull, N. A. Drescher, and G. S. Uhrig, 
Enhanced perturbative continuous unitary transformations,
Phys. Rev. B {\bf 86}, 125113 (2012).

\bibitem{rvlbf}
M. Vogl, P. Laurell, A. D. Barr and G. A. Fiete, 
Flow Equation Approach to Periodically Driven Quantum Systems, 
Phys. Rev. X {\bf 9}, 021037 (2019).


\bibitem{Diehl08}
S. Diehl, A. Micheli, A. Kantian, H. P. B\"uchler and P. Zoller, 
Quantum states and phases in driven open quantum systems with cold atoms, 
Nature Phys. {\bf 4}, 878 (2008).

\bibitem{rbkvm}
W. Berdanier, M. Kolodrubetz, R. Vasseur, and J. E. Moore,
Floquet Dynamics of Boundary-Driven Systems at Criticality,
Phys. Rev. Lett. {\bf 118}, 260602 (2017).


\bibitem{rktn}
T. Kampfrath, K. Tanaka, and K. A. Nelson, 
Resonant and nonresonant control over matter and light by intense terahertz 
transients,
Nature Photonics {\bf 7}, 680 (2013).

\bibitem{rzsz}
X. C. Zhang, A. Shkurinov, and Y. Zhang
Extreme terahertz science,
Nature Photonics {\bf 11}, 16 (2017).

\bibitem{Salen19}
P. Sal\'en, M. Basini, S. Bonetti, J. Hebling, M. Krasilnikov, A. Y. Nikitin, 
G. Shamuilov, Z. Tibai, V. Zhaunerchyk, and V. Goryashko,
Matter manipulation with extreme terahertz light: Progress in
the enabling THz technology,
Phys. Rep. {\bf 836-837}, 1 (2019).

\bibitem{rtljenmfkct}
D. A. Tennant, B. Lake, A. J. A. James, F. H. L. Essler, S. Notbohm, H.-J. 
Mikeska, J. Fielden, P. K\"ogerler, P. C. Canfield, and M. T. F. Telling, 
Anomalous dynamical line shapes in a quantum magnet at finite temperature, 
Phys. Rev. B {\bf 85}, 014402 (2012).

\bibitem{Ahmed17}
N. Ahmed, P. Khuntia, K. M. Ranjith, H. Rosner, M. Baenitz, A. A. Tsirlin, 
and R. Nath,
Alternating spin chain compound AgVOAsO$_4$ probed by $^{75}$As NMR,
Phys. Rev. B {\bf 96}, 224423 (2017).

\bibitem{Arjun19}
U. Arjun, K. M. Ranjith, B. Koo, J. Sichelschmidt, Y. Skourski, M. Baenitz, 
A. A. Tsirlin, and R. Nath,
Singlet ground state in the alternating spin-1/2 chain compound NaVOAsO$_4$,
Phys. Rev. B {\bf 99}, 014421 (2019).

\bibitem{Jacobs76}
I. S. Jacobs, J. W. Bray, H. R. Hart, L. V. Interrante, J. S. Kasper, G. D. 
Watkins, D. E. Prober, and J. C. Bonner,
Spin-Peierls transitions in magnetic donor-acceptor compounds of 
tetrathiafulvalene (TTF) with bisdithiolene metal complexes,
Phys. Rev. B {\bf 14}, 3036 (1976).

\bibitem{Cross79}
M. C. Cross and D. S. Fisher,
A new theory of the spin-Peierls transition with special relevance to the 
experiments on TTFCuBDT,
Phys. Rev. B {\bf 19}, 402 (1979).

\bibitem{Huizinga79}
S. Huizinga, J. Kommandeur, G. A. Sawatzky, B. T. Thole, K. Kopinga, 
J. M. de Jonge, and J. Roos,
Spin-Peierls transition in 
N-methyl-N-ethylmorpholinium-ditetracyanoquinodimethanide [MEM-(TCNQ)$_2$],
Phys. Rev. B {\bf 19}, 4723 (1979).

\bibitem{Miura06}
Y. Miura, R. Hirai, Y. Kobayashi, and M. Sato,
Spin-Gap Behavior of Na$_3$Cu$_2$SbO$_6$ with Distorted Honeycomb Structure,
J. Phys. Soc. Jpn. {\bf 75}, 084707 (2006).

\bibitem{Stone07}
M. B. Stone, W. Tian, M. D. Lumsden, G. E. Granroth, D. Mandrus, J.-H. Chung, 
N. Harrison, and S. E. Nagler,
Quantum Spin Correlations in an Organometallic Alternating-Sign Chain,
Phys. Rev. Lett. {\bf 99}, 087204 (2007).

\bibitem{run1} 
G. S. Uhrig and B. Normand, 
Magnetic properties of (VO)$_2$P$_2$O$_7$ from frustrated interchain coupling, 
Phys. Rev. B {\bf 58}, R14705 (1998).

\bibitem{run2} 
G. S. Uhrig and B. Normand, 
Magnetic properties of (VO)$_2$P$_2$O$_7$: two-plane structure and spin-phonon
interactions,
Phys. Rev. B {\bf 63}, 134418 (2001). 

\bibitem{rfl}
F. A. Lindemann, 
\"Uber die Berechnung molekularer Eigenfrequenzen,
Phys. Z. {\bf 11}, 609 (1910).

\bibitem{gilva56}
J. J. Gilvarry, 
The Lindemann and Gr\"uneisen Laws
Phys. Rev. {\bf 102}, 308 (1956). 

\bibitem{dudow08}
J. Dudowicz, K. F. Freed, and J. F. Douglas,
Generalized entropy theory of polymer glass formation, 
Adv. Chem. Phys. {\bf 137}, 125 (2008).

\bibitem{gross12} 
R. Gross and A. Marx, {\sl Festk\"orperphysik} (Oldenbourg, Munich, 2012).

\bibitem{born27}
M. Born and J. R. Oppenheimer, 
Zur Quantentheorie der Molek\"ulen
Ann. Phys. {\bf 84}, 457 (1927). 

\end{thebibliography}
\end{document}